\RequirePackage{silence}
\documentclass[fleqn,usenatbib,useAMS]{mnras} %Class file altered on lines 116, 119, 122 and 123.
\usepackage{graphicx}
\usepackage{amsmath}

\DeclareMathOperator{\sech}{sech}
\usepackage{amsfonts}
\usepackage{float}
\usepackage{bm}
\usepackage{cancel}
\usepackage[perpage,symbol*]{footmisc}%At end of square bracket: ,symbol* \usepackage[T1]{fontenc}
\usepackage{ae,aecompl}
\usepackage{array}
\usepackage{soul}
\usepackage{mathtools}
\usepackage{multirow}
\usepackage[utf8]{inputenc}
\usepackage{booktabs}
\usepackage{graphicx}
\usepackage{makecell}

\DeclareRobustCommand{\appropto}{\mathrel{\vcenter{
		\offinterlineskip\halign{\hfil$##$\cr %The $##$ is not a mistake!
			\propto\cr\noalign{\kern2pt}\sim\cr\noalign{\kern-2pt}}}}}

\DeclareRobustCommand{\VAN}[3]{#2} % set up for citation
\DeclareRobustCommand{\DE}[3]{#2} % set up for citation

\defcitealias{Banik_Ryan_2018}{BRZ18}
\title[Formation of the Local Group satellite planes]{3D hydrodynamic simulations for the formation of the Local Group satellite planes} %Use [Short title]{Full title} if the two are inequivalent, former should be <=45 characters unless you want to explicitly make it multi-line (not allowed in MNRAS), must not include \& in short title.

\author[I. Banik, I. Thies, R. Truelove, G. Candlish, B. Famaey, M. S. Pawlowski, R. Ibata, P. Kroupa]{\parbox[t]{\textwidth} {Indranil Banik$^{1, 2}$\thanks{Email:
\href{mailto:ib45@st-andrews.ac.uk}{ib45@st-andrews.ac.uk} (Indranil Banik)\newline $~~~~~~~~~~~~~~~\,$ \href{mailto:ithies@astro.uni-bonn.de}{ithies@astro.uni-bonn.de} (Ingo Thies)}, Ingo Thies$^2$, Roy Truelove$^1$, Graeme Candlish$^3$, Benoit Famaey$^4$, Marcel S. Pawlowski$^5$, Rodrigo Ibata$^4$ \& Pavel Kroupa$^{2, 6}$} \vspace{10pt} \\
$^{1}$Scottish Universities Physics Alliance, University of Saint Andrews, North Haugh, Saint Andrews, Fife, KY16 9SS, UK \\
$^{2}$Helmholtz-Institut f\"ur Strahlen und Kernphysik (HISKP), University of Bonn, Nussallee 14$-$16, D-53115 Bonn, Germany \\
$^3$Instituto de F\'{i}sica y Astronom\'{i}a, Universidad de Valpara\'{i}so, Gran Breta\~{n}a 1111, Valpara\'{i}so, Chile\\
$^4$Universit\'{e} de Strasbourg, CNRS UMR 7550, Observatoire astronomique de Strasbourg, 11 rue de l'Universit\'{e}, 67000 Strasbourg, France \\
$^5$Leibniz-Institut f\"ur Astrophysik Potsdam (AIP), An der Sternwarte 16, D-14482 Potsdam, Germany \\
$^6$Astronomical Institute, Faculty of Mathematics and Physics, Charles University, V Hole\v{s}ovi\v{c}k\'ach 2, CZ-180 00 Praha 8, Czech Republic}

\pubyear{2022}
\begin{document}
\label{firstpage}
\pagerange{\pageref{firstpage}--\pageref{lastpage}}

\maketitle

\begin{abstract} %No paragraphs. Do not mention LMC.

The existence of mutually correlated thin and rotating planes of satellite galaxies around both the Milky Way (MW) and Andromeda (M31) calls for an explanation. Previous work in Milgromian dynamics (MOND) indicated that a past MW-M31 encounter might have led to the formation of these satellite planes. We perform the first-ever hydrodynamical MOND simulation of the Local Group using \textsc{phantom of ramses}. We show that an MW-M31 encounter at $z \approx 1$, with a perigalactic distance of about 80~kpc, can yield two disc galaxies at $z=0$ oriented similarly to the observed galactic discs and separated similarly to the observed M31 distance. Importantly, the tidal debris are distributed in phase space similarly to the observed MW and M31 satellite planes, with the correct preferred orbital pole for both. The MW-M31 orbital geometry is consistent with the presently observed M31 proper motion despite this not being considered as a constraint when exploring the parameter space. The mass of the tidal debris around the MW and M31 at $z=0$ compare well with the mass observed in their satellite systems. The remnant discs of the two galaxies have realistic radial scale lengths and velocity dispersions, and the simulation naturally produces a much hotter stellar disc in M31 than in the MW. However, reconciling this scenario with the ages of stellar populations in satellite galaxies would require that a higher fraction of stars previously formed in the outskirts of the progenitors ended up within the tidal debris, or that the MW-M31 interaction occurred at $z>1$.

\end{abstract}


\section{Introduction}
\label{Introduction}

It has been known since the early work of \citet{Kunkel_1976}, \citet{Lynden_Bell_1976}, and \citet{Lynden_Bell_1982} that dwarf spheroidal galaxies around the Milky Way (MW) have an anisotropic spatial distribution, as later confirmed by \citet*{Kroupa_2005} and \citet*{Metz_2007}. The orbital poles that can be measured indicate that this Vast Polar Structure (VPOS) is corotating \citep{Metz_2008, Pawlowski_2013_LG, Pawlowski_2017, Pawlowski_2020, Li_2021}. This is in stark contrast with expectations based on the standard $\Lambda$ cold dark matter ($\Lambda$CDM) cosmological model \citep*{Efstathiou_1990, Ostriker_Steinhardt_1995} because the existence of such a corotating plane would {\it a priori} imply a significant amount of dissipation. The issue is compounded by a similar satellite plane (SP) discovered around Andromeda \citep[M31;][]{Ibata_2013, Ibata_2014} and Centaurus A \citep[Cen A;][]{Muller_2018, Muller_2021}, with radial velocities (RVs) suggestive of corotation in both cases.\footnote{Hints of the M31 SP were already evident in \citet{Metz_2007} and \citet*{Metz_2009}. Hints of the Cen A SP were evident in \citet{Tully_2015_Cen_A}, but the two planes they identified were later revealed to be part of one thicker plane \citep{Muller_2016}.} Proper motions (PMs) have recently become known for two M31 SP members, and indeed indicate that it too is likely corotating within its plane \citep{Sohn_2020}.

Such structures are difficult to explain using dissipationless haloes of CDM, as first pointed out by \citet{Kroupa_2005} and more recently by \citet{Pawlowski_2014}, who considered and excluded a wide range of different proposed explanations within the $\Lambda$CDM context. For instance, accreting most satellites as a single group \citep{Metz_2009_group} or along a single filament would yield some anisotropy, but not enough to explain the very thin MW SP \citep{Shao_2018}. The two available PMs of M31 SP members are also in tension with $\Lambda$CDM expectations because they indicate motion nearly within the M31 SP \citep{Pawlowski_Sohn_2021}. Dissipationless collapse of CDM halos is therefore insufficient to account for the observed SPs if they are composed of primordial dwarfs. Including baryonic physics does not change this picture very much \citep{Ahmed_2017, Pawlowski_2020, Samuel_2021}.

Since dissipation in the CDM component would cause a significant Galactic DM disc that is in tension with observations \citep{Buch_2019, Widmark_2021}, an obvious possibility is that the necessary dissipation occurred in baryons. Given that the MW SP is almost polar with respect to its disc, this would require a tidal interaction with another galaxy \citep*{Pawlowski_2011, Pawlowski_2012}. In this scenario, the MW SP would consist of tidal dwarf galaxies (TDGs) that condensed out of gas-rich tidal debris expelled during a past interaction. A similar scenario could have occurred around M31. The formation of TDGs has been observed outside the Local Group (LG), for instance around the Antennae \citep*{Mirabel_1992} and the Seashell Galaxy \citep{Bournaud_2004}. The LG SPs may have formed analogously. A common origin for both LG SPs is possible as a result of a major merger experienced by M31 \citep{Hammer_2010} expelling tidal debris towards the MW \citep{Hammer_2013}, perhaps explaining the high Galactocentric velocities of MW satellites \citep{Hammer_2021}. Since the M31 SP is viewed close to edge-on from our vantage point in the MW \citep{Conn_2013, Ibata_2013, Isabel_2020}, it is possible that if M31 experienced a major merger, then some of the tidal debris expelled from M31 formed its SP while some reached a much larger distance and is now close to the MW.

In a $\Lambda$CDM context, TDGs would be free of DM due to its dissipationless nature and its initial distribution in a dispersion-supported near-spherical halo \citep*{Barnes_1992, Wetzstein_2007}. During a tidal interaction, DM of this form is clearly incapable of forming into a thin dense tidal tail which might undergo Jeans collapse into TDGs. As a result, any TDGs would be purely baryonic and thus have only a very small escape velocity, preventing them from subsequently accreting DM out of the Galactic halo. For this reason, TDGs formed in the $\Lambda$CDM framework cannot explain the high observed internal velocity dispersions ($\sigma_{\text{int}}$) of the MW satellites if they are in equilibrium. Non-equilibrium solutions were considered by \citet{Kroupa_1997}, \citet{Klessen_1998}, and \citet{Casas_2012}. While their proposed solutions match many properties of the observed MW satellites, this scenario would not yield elevated $\sigma_{\text{int}}$ values at Galactocentric distances $\ga 150$~kpc. Satellites in such a non-equilibrium phase would also be very fragile, requiring us to be observing them at just the right epoch prior to total disruption but after significant tidal perturbation. These issues are undoubtedly a major reason why CDM-free galaxies are very rare in the latest $\Lambda$CDM simulations \citep{Haslbauer_2019}, making it very unlikely that so many TDGs are in the LG right now around both the MW and M31. Moreover, postulating that most of their satellites are TDGs would mean that they have very few primordial satellites.

This conundrum presents an open invitation to consider a different theoretical framework in which DM is dynamically irrelevant to holding galaxies together against their high $\sigma_{\text{int}}$, removing the problem that TDGs are expected to be free of DM. In this case, the effects conventionally attributed to CDM must instead be due to a non-Newtonian gravity law on galactic scales. The best developed such proposal is Milgromian dynamics \citep[MOND;][]{Milgrom_1983}. In MOND, galaxies lack CDM but the gravitational field strength $g$ at distance $d$ from an isolated point mass $M$ transitions from the Newtonian $g_{_N} = {GM/d^2}$ law at short range to
\begin{eqnarray}
	g ~=~ \sqrt{g_{_N} a_{_0}} ~~~\text{for } ~ g_{_N} \ll a_{_0} \, .
	\label{Deep_MOND_limit}
\end{eqnarray}
MOND introduces $a_{_0}$ as a fundamental acceleration scale of nature below which the deviation from Newtonian dynamics becomes significant. Empirically, $a_{_0} \approx 1.2 \times {10}^{-10}$ m/s$^2$ to match galaxy rotation curves \citep*[RCs;][]{Begeman_1991, Gentile_2011}. With this value of $a_{_0}$, MOND predicts the detailed shape of galaxy RCs very well using only their directly observed baryonic matter \citep[e.g.][]{Kroupa_2018, Li_2018, Sanders_2019}. In particular, observations confirm the prior MOND prediction of very large departures from Newtonian dynamics in low surface brightness galaxies \citep[LSBs;][and references therein]{McGaugh_2020}. More generally, there is a very tight empirical `radial acceleration relation' (RAR) between the gravity inferred from RCs and that expected from the baryons alone in Newtonian dynamics \citep{McGaugh_Lelli_2016, Lelli_2017}, with RCs asymptotically reaching a flatline level $\propto \sqrt[^4]{M}$ as required by Equation \ref{Deep_MOND_limit} \citep{McGaugh_2012, Lelli_2019}. The observed phenomenology of galactic RCs confirm all the central predictions of \citet{Milgrom_1983}, as reviewed in e.g. \citet{Famaey_McGaugh_2012}.

MOND can also explain the X-ray temperature profile \citep{Milgrom_2012} and internal dynamics of elliptical galaxies \citep[figure 8 of][]{Lelli_2017}, which reveal a similar characteristic acceleration scale to spirals \citep{Chae_2020_elliptical, Shelest_2020}. At the low-mass end, MOND is consistent with the $\sigma_{\text{int}}$ of pressure-supported systems like the satellites of the MW \citep{McGaugh_2010}, M31 \citep{McGaugh_2013a, McGaugh_2013b}, non-satellite LG dwarfs \citep{McGaugh_2021}, Dragonfly 2 \citep[DF2;][]{Famaey_2018, Kroupa_2018_Nature}, and DF4 \citep{Haghi_2019}. These predictions rely on correctly including the external field effect (EFE) arising from the non-linearity of MOND \citep{Milgrom_1986}, which we discuss further in Section \ref{Including_the_EFE}. For the more isolated galaxy DF44, the MOND prediction without the EFE is consistent with the observed $\sigma_{\text{int}}$ profile \citep{Bilek_2019, Haghi_2019_DF44}. Note that the situation with the EFE is however less clear for ultra-diffuse galaxies located deep in the potential well of galaxy clusters \citep{Freundlich_2022}. The EFE also plays a role in accounting for the observed weak bar of M33 \citep{Banik_2020_M33}, which is quite difficult to understand in the presence of a live DM halo \citep*{Sellwood_2019} due to bar-halo angular momentum exchange \citep[e.g.][]{Debattista_2000, Athanassoula_2002}. This problem is related to the bar pattern speeds in galaxies, which seem to be too slow in $\Lambda$CDM cosmological simulations due to dynamical friction on the bar exerted by the DM halo (\citealt{Algorry_2017, Peschken_2019, Roshan_2021_disc_stability, Roshan_2021_bar_speed}; though see \citealt{Fragkoudi_2021}). In addition, the morphological properties of MW satellites are more easily understood in MOND as a consequence of differing levels of tidal stability \citep{McGaugh_2010}. This is because in MOND, the lack of CDM haloes and the EFE render the satellites much more susceptible to Galactic tides, which are not so relevant in $\Lambda$CDM (see their figure 6). Neglecting tides leads to erroneous conclusions regarding the viability of MOND since it predicts that much fewer satellites $-$ especially at the ultra-faint end $-$ are amenable to equilibrium virial analysis \citep{Fattahi_2018}.

Further work will however be necessary to make rigorous MOND predictions in a cosmological context, which needs a relativistic theory. A few such theories exist, with one promising proof-of-concept being the relativistic MOND theory of \citet{Skordis_2019} in which gravitational waves travel at the speed of light. Such theories allow MOND calculations of weak gravitational lensing by foreground galaxies in stacked analyses \citep[e.g.][]{Brimioulle_2013}, which so far seems to agree with expectations \citep{Milgrom_2013, Brouwer_2017, Brouwer_2021}. At larger distances from the central galaxy, the EFE from surrounding structures would cause the gravity law to become inverse square and to depart from spherical symmetry \citep{Banik_2015}, which may explain some recent observations \citep{Schrabback_2021}. More detailed calculations would require knowledge of how large-scale structure forms in a Milgromian framework and the resulting EFE on galaxies. This is a critical next step for MOND, though care is required when comparing with observations as these could have a rather different interpretation to what is usually assumed. One possible smoking gun signature of MOND in weak lensing convergence maps would be the discovery that the convergence parameter is negative in some regions \citep{Oria_2021}. This cannot arise in GR and is possible only in gravitational theories that are non-linear in the weak-field regime.

At this stage, general statements can however already be made about a MONDian cosmology, in which the key difference with $\Lambda$CDM would be faster structure formation \citep{Sanders_1998}. This might be relevant for the so-called Hubble tension, i.e. the fact that, to fit the anisotropies in the cosmic microwave background \citep[CMB;][]{Aiola_2020, Planck_2020}, $\Lambda$CDM requires a local Hubble constant $H_0$ below the directly measured value at high significance based on multiple independent techniques \citep[e.g.][]{Riess_2020, Valentino_2021, Riess_2022}. This tension could be due to our position within a large local supervoid underdense by $\approx 30\%$ out to a radius of $\approx 300$~Mpc. Such a large and deep underdensity has indeed been observed in multiple surveys and is called the KBC void after its discoverers \citep*{Keenan_2013}. This is incompatible with $\Lambda$CDM at $6.04\sigma$, one major reason for the high significance of the tension being that the relevant observations cover 90\% of the sky \citep*{Haslbauer_2020}. However, a KBC-like void could arise naturally in a MOND cosmology supplemented by light sterile neutrinos playing the role of hot DM (HDM), as proposed by \citet{Angus_2009}. The main feature in the \citet{Haslbauer_2020} void scenario is faster structure formation than in $\Lambda$CDM. From an observational point of view, this is also evident in the properties of the high-redshift interacting galaxy cluster El Gordo \citep*{Asencio_2021}. The lack of analogues to El Gordo at low redshift could well be due to our location within the KBC void, which might also explain why the MOND simulations of \citet{Angus_2013} seemingly overproduced massive clusters when comparing their whole simulation volume with low-redshift datasets. Therefore, MOND with HDM could potentially account for astronomical observations ranging from the kpc scales of galaxies \citep[where HDM would play no role;][]{Angus_2010_mass} all the way to the Gpc scale of the local supervoid, without causing any obvious problems in the early Universe or in galaxy clusters \citep*{Angus_2010}. It is also possible to fit the CMB in MOND without any sterile neutrino component \citep{Skordis_2021}, though it is unclear whether this approach can explain the properties of galaxy clusters. For a recent review of MOND that considers evidence from a wide range of scales and discusses the cosmological aspects in some detail, we refer the reader to \citet{Banik_2022}.

Concerning the history of the well-observed LG, MOND implies a very strong mutual attraction between the MW and M31. Acting on their almost radial orbit \citep{Van_der_Marel_2012}, this leads to a close encounter ${9 \pm 2}$ Gyr ago \citep{Zhao_2013}, consistent with the timing of a few other events putatively linked to the flyby (Section \ref{Other_evidence_flyby}). An $N$-body simulation of this interaction showed that it is likely to yield anisotropically distributed tidal debris reminiscent of an SP \citep{Bilek_2018}. Around the same time, \citet*[][hereafter \citetalias{Banik_Ryan_2018}]{Banik_Ryan_2018} considered a wider range of orbital geometries using a less computationally intensive restricted $N$-body approach in which the MW and M31 were treated as point masses surrounded by test particle discs.\footnote{This leads to a numerically more tractable axisymmetric potential.} Their section~2 demonstrated that the MW-M31 trajectory is consistent with negligible peculiar velocity in the early universe, a constraint known as the timing argument \citep{Kahn_Woltjer_1959}. Despite lacking hydrodynamics or disc self-gravity, the initial setup of each galaxy as a rotating disc was sufficient to cause significant clustering of the tidal debri orbital poles. In some models, the preferred directions aligned with the actually observed orbital poles of the LG SPs. Therefore, these structures could well have formed as TDGs that condensed out of tidal debris orbiting in the correct plane. Indeed, \citet{Bilek_2021} conclude that all satellite galaxies in the LG SPs are TDGs based on the earlier simulations of \citet{Bilek_2018}.

TDGs are expected to be more resilient in MOND due to their enhanced self-gravity, as explored with earlier high-resolution hydrodynamical MOND simulations \citep*{Renaud_2016}. Besides helping them to survive, this would also explain the high observed $\sigma_{\text{int}}$ of satellite galaxies around the MW \citep{McGaugh_2010} and M31 \citep{McGaugh_2013a, McGaugh_2013b}. However, the enhancement to the self-gravity would typically be less than provided by a CDM halo, leading to a greater degree of tidal susceptibility in MOND. This is more in line with the observed morphologies of LG satellites \citep{McGaugh_2010}.

A key distinguishing characteristic of more recently formed TDGs is their high metallicity for their mass \citep{Duc_2014}. However, this relies on a long process of metal enrichment in the disc of the progenitor galaxy. When considering a very ancient interaction, there might not have been enough time for such an enrichment \citep*{Recchi_2015}, especially as TDGs are expected to form out of material that was initially several disc scale lengths out \citepalias{Banik_Ryan_2018}. This could explain why the M31 satellites within and outside its SP have similar properties, including in terms of metallicity \citep{Collins_2015}.%It is also possible that many of the M31 satellites outside the SP considered here \citep{Ibata_2013} are part of another SP formed in a different interaction \citep{Isabel_2020}.

A past MW-M31 flyby also has interesting consequences for the rest of the LG. Due to the high MW-M31 relative velocity around the time of their flyby, they would likely have gravitationally slingshot several LG dwarfs out at high speed. As discussed further in Section~\ref{Other_evidence_flyby_NGC3109}, this could lead to the existence of LG dwarfs with an unusually high RV in a $\Lambda$CDM context, such as the dwarfs in the NGC 3109 association \citep{Pawlowski_McGaugh_2014, Peebles_2017}. These could be backsplash from the MW-M31 flyby, a scenario that was considered in detail by \citet{Banik_2018_anisotropy}. Backsplash galaxies also exist in $\Lambda$CDM, but very rarely have properties resembling NGC 3109 \citep{Banik_2021}.

In this contribution, we build on the earlier studies of \citet{Bilek_2018} and \citetalias{Banik_Ryan_2018} by conducting 3D hydrodynamical MOND simulations of the flyby using \textsc{phantom of ramses} \citep[\textsc{por};][]{Lughausen_2015, Nagesh_2021}. Our main objective is to vary the MW-M31 orbital pole and pericentre distance to find if there are models where the tidal debris around each galaxy aligns with its observed SP. Achieving this simultaneously for both the MW and M31 is a highly non-trivial test of the past flyby scenario. We also check if their discs are preserved and end up with realistic properties.

In the following, we describe the initial conditions and setup of our simulations (Section \ref{Setup}). We then present our results and analyses regarding the MW-M31 trajectory (Section \ref{Section_MW_M31_orbit}) and proper motion (PM; Section \ref{Proper_motion_section}), the tidal debris (Section \ref{Tidal_debris}), and the MW and M31 disc remnants (Section \ref{Disc_remnants}). We discuss our results in Section~\ref{Discussion} and conclude in Section \ref{Conclusions}. Videos of the LG in our best-fitting simulation with frames every 10 Myr are publicly available.\footnote{\href{https://seafile.unistra.fr/d/6bb8e94212764324868e/}{https://seafile.unistra.fr/d/6bb8e94212764324868e/}}

\section{Methods}
\label{Setup}

\subsection{Poisson equation}

The simulations presented in this paper are conducted with \textsc{por}, which solves the governing equation of QUMOND:
\begin{eqnarray}
	\overbrace{\nabla \cdot \bm{g}}^{\propto \rho_{\text{eff}} \, \equiv \, \rho_{_{\text{PDM}}} + \rho_{_b}} ~=~ \nabla \cdot \left[ \bm{g}_{_N} \nu \left( g_{_N} \right) \right] \, ,
	\label{QUMOND_equation}
\end{eqnarray}
where $\bm{g}_{_N}$ is the Newtonian gravity determined from the baryonic density $\rho_{_b}$ using standard techniques, $v \equiv \left| \bm{v} \right|$ for any vector $\bm{v}$, and $\bm{g}$ is the true gravity. It is often helpful to think of what density distribution would lead to this $\bm{g}$ under Newtonian gravity. The required effective density $\rho_{_{\text{eff}}} \equiv \rho_{_{\text{PDM}}} + \rho_{_b}$, where $\rho_{_{\text{PDM}}}$ is the phantom dark matter (PDM) density which captures the MOND corrections. Equation \ref{QUMOND_equation} is derived from a non-relativistic Lagrangian \citep{QUMOND}, so QUMOND obeys the usual conservation laws regarding energy and momentum. This is also true of an earlier version \citep{Bekenstein_Milgrom_1984}, though we do not consider it here as it is computationally less efficient due to a non-linear grid relaxation stage.

To solve Equation \ref{QUMOND_equation}, we must assume an interpolating function $\nu$ between the Newtonian and Milgromian regimes. In spherical symmetry, this has the effect that $\bm{g} = \nu \bm{g}_{_N}$, softening the transition between the Newtonian inverse square law and Equation \ref{Deep_MOND_limit}. In this work, we use the `simple' form of the interpolating function \citep{Famaey_Binney_2005}:
\begin{eqnarray}
	\nu \left( g_{_N} \right) ~=~ \frac{1}{2} ~+~ \sqrt{\frac{1}{4} + \frac{a_{_0}}{g_{_N}}} \, .
	\label{Simple_interpolating_function}
\end{eqnarray}
This is numerically rather similar to the function used by \citet{McGaugh_Lelli_2016} and \citet{Lelli_2017} to fit galaxy RCs, but can be inverted analytically. Other reasons for using this function were discussed in section 7.1 of \citet{Banik_2018_Centauri} in preference to functions with a sharper transition. Equation \ref{Simple_interpolating_function} is quite accurate for $g_{_N} \approx \left( 0.1 - 10 \right) a_{_0}$ as relevant for the MW-M31 flyby problem, but Solar system constraints imply that a more rapid convergence to the Newtonian result is required for $g_{_N} \gg a_{_0}$ \citep{Hees_2014, Hees_2016}. However, the precise nature of this convergence is not important if the relevant quantity is $\bm{g}$ rather than merely its deviation from $\bm{g}_{_N}$.

For the computation of $\bm{g}_{_N}$, we use the standard boundary condition that the Newtonian potential at large distance is
\begin{eqnarray}
	\Phi_N ~=~ -\frac{GM}{r} \, ,
	\label{Phi_N}
\end{eqnarray}
where $G$ is the Newtonian gravitational constant, $M$ is the total mass within the simulation volume, and $\bm{r}$ is the position relative to its barycentre. The boundary condition for the true potential $\Phi$ will be discussed in Section~\ref{New_features_por} based on Equations \ref{Potential_adjustment_dark_energy} and \ref{Phi_g_ext_domination}.

\subsection{Treatment of the MW-M31 orbit}
\label{MW_M31_orbit}

An important aspect of our simulations is choosing an appropriate initial position and velocity for the MW and M31 disc templates to be discussed in Section \ref{Disc_templates}. We do this using a semi-analytic backwards integration (hereafter SAM) very similar to that used in section 2 of \citetalias{Banik_Ryan_2018}, to which we refer the reader for a detailed discussion of the time evolution of the MW-M31 separation $\bm{d} \left( t \right)$. Briefly, their separation $d$ in physical coordinates (used throughout this paper) is governed by
\begin{eqnarray}
	\ddot{d} ~=~ \frac{\ddot{a}}{a}d - g + \frac{h^2}{d^3} \, ,
	\label{Timing_argument_basic}
\end{eqnarray}
where $a$ is the cosmic scale factor, $\bm{h}$ is the angular momentum with magnitude $h$ and direction $\widehat{\bm{h}}$ (the orbital pole), $g$ is the radially inward component of the mutual gravity between the MW and M31, and an overdot denotes a time derivative. The term involving $a$ is present in a homogeneously expanding universe, but $g$ and $h$ are zero in this case. As discussed in section 3.1.1 of \citet{Haslbauer_2020}, we adopt a standard Planck cosmology \citep{Planck_2014_cosmology} at the background level (Table \ref{Cosmological_parameters}). MOND can in principle explain the high locally measured $H_0$ \citep[e.g.,][]{Riess_2021, Valentino_2021} as arising from outwards peculiar velocities induced by the observed KBC void \citep{Keenan_2013}, so our choice of cosmological parameters should be consistent with both early and late Universe probes of the expansion rate.

\begin{table}
	\centering
	\caption{Our adopted cosmological parameters \citep{Planck_2014_cosmology}, with 0 subscripts denoting present values. We assume a standard flat background cosmology and neglect other components as we are not considering the very early universe.}
	\begin{tabular}{cc}
		\hline
		Parameter & Value \\
		\hline
		$H_0$ & 67.3 km/s/Mpc \\
		$\Omega_{m, 0}$ & 0.315 \\
		$\Omega_{\Lambda, 0}$ & 0.685 \\
		\hline
	\end{tabular}
	\label{Cosmological_parameters}
\end{table}

The angular momentum barrier $h^2/d^3$ is necessary to prevent an unrealistic direct collision between the MW and M31. However, SAM is a timing argument analysis constrained to give zero peculiar velocity when $a = 0.1$, which implies $h = 0$ then. We square this circle by assuming $h = 0$ prior to first apocentre, after which $\bm{h}$ instantaneously jumps to a particular value that remains fixed until today. The discontinuous behaviour of $\bm{h}$ occurs at a time when the angular momentum barrier is least important to the trajectory, minimizing numerical effects. Physically, it would be reasonable if $\bm{h}$ was mostly gained from tidal torques around the time of apocentre, but the more recent apocentre would be less relevant due to cosmic expansion driving external perturbers much further from the LG (Section~\ref{Tidal_torques}). Importantly, our primary objective in this contribution is to conduct \textsc{por} simulations of the flyby. These are initialized 1 Gyr before the flyby, which is safely after the jump in $\bm{h}$.

The calculation of $g$ is rather complicated $-$ we summarize only the main points here, and refer the reader to \citetalias{Banik_Ryan_2018} for a detailed discussion. In the absence of any other bodies and assuming the deep-MOND limit (gravitational fields $\ll a_{_0}$), we get a mutual gravity of
\begin{eqnarray}
	g_{\text{iso}} ~&=&~ \frac{Q\sqrt{GMa_{_0}}}{d} \, , \\
	Q ~&=&~ \frac{2 \left( 1 - {q_{_{\text{MW}}}}^\frac{3}{2} - {q_{_{\text{M31}}}}^\frac{3}{2}\right)}{3 \, q_{_{\text{MW}}} q_{_{\text{M31}}}} \, ,
	\label{g_iso}
\end{eqnarray}
where $M$ is the total mass of the MW and M31, of which a fraction $q_{_{i}}$ resides in galaxy $i$. The parameter $Q$ accounts for the finite mass ratio between the MW and M31 \citep*{Zhao_2010}. Roughly speaking, it is caused by the PDM halo of one galaxy being reduced in mass due to gravity from the other galaxy, which is a manifestation of the EFE. We assume that $q_{_{\text{MW}}} = 0.3$ and $q_{_{\text{M31}}} = 0.7$ \citepalias[consistently with][]{Banik_Ryan_2018}, so $Q = 0.7937$.

When the MW and M31 are near pericentre, it is less accurate to assume the deep-MOND limit. However, the higher $g$ allows us to neglect the relatively much weaker EFE on the whole LG, so we consider it as an isolated two-body problem where we get $g$ numerically. Such Newtonian corrections are expected to have only a minor impact on our results because only a small portion of the SAM trajectory is subject to them.

When the MW and M31 are near apocentre, we expect the EFE from large-scale structure to be important (Section \ref{Including_the_EFE}). In the limit where $g_{\text{ext}} \gg g_{\text{iso}}$, we can use the external field (EF)-dominated analytic solution found by \citet{Banik_2015} to get that
\begin{eqnarray}
	g_{_{\text{EFE}}} ~&=&~ \frac{GMa_{_0}}{d^2 g_{\text{ext}}} \left( \frac{3 + \cos^2 \theta}{4} \right) \, ,
	\label{g_EFE}
\end{eqnarray}
where $\theta$ is the angle between $\bm{d}$ and the EF $\bm{g}_{\text{ext}}$. In reality, $d$ is never large enough for $g_{\text{ext}}$ to dominate, requiring an interpolation between the isolated and EF-dominated regimes. We achieve this using equation 14 of \citetalias{Banik_Ryan_2018}, which is a fit to numerical results for the case $\theta = 0$.

In addition to considering a uniform $\bm{g}_{\text{ext}}$, we also follow the \citetalias{Banik_Ryan_2018} approach to include the tidal effect of M81, IC 342, and Cen A. Since in each case their gravity on the LG is $\ll g_{\text{ext}}$ from more distant structures (see their section 2.3.1), we can superpose the gravitational field of each perturber on the LG.\footnote{MOND becomes a linear gravity theory if the EFE dominates \citep{Banik_2015}.}

For simplicity, we assume that $\bm{h}$ is constant in SAM except for the above-mentioned discontinuity at first turnaround, which occurs well before our \textsc{por} simulation starts. Our \textsc{por} calculations would miss changes in $\bm{h}$ due to tidal torques from perturbers beyond the LG, but we will show later that such effects are quite small (Section \ref{Tidal_torques}). We therefore expect any changes in $\bm{h}$ to arise mostly from the EFE (Section \ref{External_field_torque}) and from torques on the MW-M31 orbit during the flyby. Neither effect is included in SAM, but both are included in \textsc{por}. Since the MW-M31 orbit is nearly radial, changes in $\bm{h}$ should have little effect on the timing argument, which is the main purpose of SAM.

In summary, we will hereafter use the present MW-M31 separation, direction, and mutual RV as present-day constraints on the SAM. We vary their orbital pole and mutual two-body angular momentum magnitude to match the observed SP orbital poles. Constraints on the PM of M31 will not be taken into account in this process.

\subsection{Disc templates}
\label{Disc_templates}

SAM can only tell us the initial position and velocity of the MW and M31 discs. We therefore complement SAM with a system to generate a Milgromian disc template, to which we then apply the appropriate rotation and Galilean transformation. We generate two stable isolated Milgromian discs using the procedures described in \citet{Banik_2020_M33}, i.e. by using our adapted version of the Newtonian code \textsc{disk initial conditions environment} \citep[\textsc{dice};][]{Perret_2014}. The MW and M31 are assumed to have exponential surface density profiles, motivated by the fact that disc galaxies usually have an exponential radial profile \citep{Freeman_1970}, which arises naturally with about the right mass-size relation when spherical gas clouds collapse under MOND gravity \citep*{Wittenburg_2020}. We implicitly assume that at the start time of our simulations $\approx 8$ Gyr ago (redshift $z \approx 1$), the MW and M31 discs had already formed with nearly their present masses. Thin rotationally supported disc galaxies do exist at even higher redshift \citep{Lelli_2018, Neeleman_2020, Rizzo_2020, Lelli_2021}. The relatively isolated nature of the LG \citep[e.g.][]{Banik_2021} suggests that its major galaxies might well have attained nearly their present mass rather early in cosmic history, especially in a framework with enhanced long-range gravity \citep{Peebles_2010} where mergers are less common due to the lack of dynamical friction between extended CDM haloes \citep{Kroupa_2015, Renaud_2016}. This is quite plausible given the difficulty faced by the $\Lambda$CDM paradigm in explaining the high observed fraction of thin disc galaxies, which could be due to mergers being too frequent in this framework \citep{Peebles_2020, Haslbauer_2022}.

The orientation, barycentre position, and velocity of each disc are set to the desired initial values by applying a rotation and Galilean transformation to its particles using an adapted version of the \textsc{ramses} patch known as \texttt{condinit}. This also assigns the density and velocity of each gas cell \citep*{Teyssier_2010}. To avoid severe thermal effects when the MW and M31 encounter each other, we use the same gas temperature of $T2\_ISM = 4.65 \times {10^5}$~K (465 kK) for both galaxies. We set a temperature floor of $T2\_star = 0.8 \, T2\_ISM$ and disable star formation and metallicity-dependent cooling, since at this exploratory stage we are mainly trying to reproduce the observed orientations of the LG SPs. If a suitable encounter geometry can be found, then it would be worthwhile to conduct a more detailed simulation with realistic star formation and stellar feedback prescriptions. However, this is beyond the scope of this work.

Unlike the \textsc{por} simulations of M33 in \citet{Banik_2020_M33}, an important aspect of the present \textsc{por} simulations is that we need to consider the outer parts of the simulated discs in much greater detail because we expect these regions to be the original source material for the SPs. Indeed, the restricted $N$-body models of \citetalias{Banik_Ryan_2018} showed that the SPs mainly consist of material at an initial galactocentric distance of ${\approx 50}$~kpc (see their figures 6 and 7). Material at such a large distance would be very poorly resolved with a computationally feasible number of equal mass particles. Therefore, we devise a procedure to vary the stellar particle mass in our disc templates so as to maintain a good resolution in the outer parts (see Appendix \ref{Particle_mass_setup}). Near the disc centre, the particle mass is approximately constant at $6 \times 10^5 M_\odot$ ($1.4 \times 10^6 M_\odot$) for the MW (M31). Each disc template consists of $5 \times 10^5$ particles. The spatial resolution of the \textsc{por} gravity solver also needs to be sufficient $-$ this is discussed further in Section \ref{Simulation_setup}. The resolution is 1.5 kpc in the best-resolved regions, though we show that improving this to 0.75 kpc has little effect on our results (Appendix \ref{Higher_resolution_simulation}).

\subsubsection{Initial disc parameters}
\label{Disc_parameters_section}

%M31 advised initial spin axis:
%238.3848
%-33.7111

%M31 advised initial spin axis, Cartesian:
%-0.4361
%-0.7084
%-0.5550

%MW advised initial spin axis:
%55.1106
%-84.8875

%MW advised initial spin axis, Cartesian:
%0.0510
%0.0731
%-0.9960

The rotation curve of each disc template is calculated using MOND gravity with the present value of $a_{_0}$, since throughout this work we assume that $a_{_0}$ remains constant with time. Therefore, the initial MW and M31 discs would lie on the RAR defined by nearby galaxies. Possible consequences of a time-varying $a_{_0}$ were discussed in \citet{Milgrom_2015_a0_variation} and section 5.2.3 of \citet{Haslbauer_2020}, though we do not consider this here for simplicity. Constraining the evolution is challenging observationally because, amongst other issues, it is not yet possible to use 21-cm observations of neutral hydrogen to obtain rotation curves in the distant Universe, so the H$\alpha$ line is typically used instead. The high-$z$ rotation curve data of \citet{Genzel_2017} are quite consistent with a time-independent $a_{_0}$, which sets some limits on its evolution \citep{Milgrom_2017}. Other works also suggest that high-$z$ galaxies are consistent with MOND \citep{Stott_2016, Harrison_2017, Genzel_2020, Sharma_2021}. Tighter constraints on this issue would be valuable to better understand the possible theoretical underpinnings of MOND, and more generally to test its prediction that isolated galaxies in dynamical equilibrium at any fixed redshift lie on a tight RAR.

%\citep{Milgrom_2015_a0_variation}
%\citep{Milgrom_2017}
%\citep{Stott_2016, Harrison_2017, Genzel_2020}

We assume that the present MW and M31 disc scale lengths are as given in table 3 of \citetalias{Banik_Ryan_2018}. When setting up their discs at the start of our simulation, we scale their present lengths by 0.8 for M31 and 0.6 for the MW. Starting with smaller discs is required to allow them to expand after the flyby to reach a realistic present-day configuration, and is also in line with the observed expansion of the stellar component of galaxies over cosmic time \citep{Sharma_2021}. We use the \textsc{dice} setting $Q\_lim = 1.25$ to ensure that all disc components have an initial Toomre parameter $Q \geq 1.25$, with the MOND generalization of the Toomre condition \citep{Toomre_1964} discussed further in section 2.2 of \citet{Banik_2020_M33}. The initial parameters of each disc are summarized in Table \ref{Disc_parameters}.

\renewcommand{\arraystretch}{1.1}
\begin{table}
	\centering
	\caption{Parameters of the MW and M31 discs, each consisting of $5 \times 10^5$ particles. Both galaxies have an initial gas temperature of 465 kK and a gas fraction of 0.5, which in the MW case is achieved by converting all of its more extended component and part of its less extended component into gas. Thus, the Galactic gas disc has a double exponential profile, while the stellar disc is a single exponential. We adopt an outer limit of 100 kpc for the M31 discs and for the more extended MW component, while 40 kpc is used for its less extended component. This is $\approx 25$ scale lengths in all cases. The initial spin vector of each disc is given in Galactic coordinates $\left(l, b \right)$.}
	\begin{tabular}{ccccc}
		\hline
		Galaxy and & MW & MW & M31 & M31 \\
		component & inner & outer & stars & gas \\
		\hline
		Total mass & \multicolumn{2}{c}{$9.15 \times 10^{10} \, M_\odot$} & \multicolumn{2}{c}{$2.135 \times 10^{11} \, M_\odot$} \\
		Fraction of mass & 0.8236 & 0.1764 & 0.5 & 0.5 \\
		Gas fraction & 0.3929 & 1 & 0 & 1 \\
		Scale length (kpc) & 1.29 & 4.2 & 4.24 & 4.24 \\
		Aspect ratio & 0.15 & 0.0461 & 0.15 & 0.15 \\
		Disc spin vector & \multicolumn{2}{c}{$\left( 55.11^\circ, \, -84.89^\circ \right)$} & \multicolumn{2}{c}{$\left( 238.38^\circ, \, -33.71^\circ \right)$} \\
		\hline
	\end{tabular}
	\label{Disc_parameters}
\end{table}
\renewcommand{\arraystretch}{1}

Gas dissipation in the tidal tails is likely important to obtaining thin SPs. While we cannot include this process as rigorously as we would like due to numerical limitations, it is certainly not appropriate to assume that the pre-flyby MW and M31 had a similar gas fraction to what we observe. Both discs are assigned an initial gas fraction of 0.5 because the flyby was $\approx 7-9$ Gyr ago \citep{Zhao_2013}, when the gas fractions were likely much higher than today \citep{Stott_2016}. Chemical evolution modelling of the MW indicates a gas fraction of $\approx 0.5$ at that time \citep[figure 8 of][]{Snaith_2015}. For simplicity, we use the same gas fraction for M31.%Another argument is that the simulation duration of 8.2~Gyr (Section~\ref{Results}) combined with a present MW gas fraction of 17.64\% \citep{Banik_2018_escape} implies an average star formation rate of $\dot{M}_{\star} = 3.6 \, M_\odot$/yr. This is somewhat higher than the presently observed $\dot{M}_{\star} \approx 2 \, M_\odot$/yr \citep{Licquia_2015, Mor_2019}, which is to be expected as the flyby would have enhanced $\dot{M}_{\star}$ \citep{Renaud_2016}.

The stellar and gas discs of M31 are assumed to have the same scale length of 4.24 kpc, so we request only one disc component in the M31 namelist for the hydrodynamical version of \textsc{dice}. For the MW, a double exponential profile is assumed, so two components are defined at that stage. The outer (more extended) component corresponds to its gas disc today, so we assume this was entirely gas at the start of our simulation. However, it only comprises 17.64\% of the total MW mass \citep{Banik_2018_escape}, so we also convert a substantial fraction of the inner (less extended) MW component into gas to get a total gas fraction of 0.5.

The aspect ratios of the M31 disc and the inner MW disc are set to 0.15, so the inner MW component has a vertical density profile with a characteristic $\sech^2$ scale height of 193.5 pc. The outer MW disc component is set to an aspect ratio of 0.0461 so that both components have the same scale height in pc. For both galaxies, the radial run of the gas disc scale height is then found by \textsc{dice} to ensure it is as close to equilibrium as practicable following section 2.3 of \citet{Banik_2020_M33}. They also described how particles in the \textsc{dice} template are written out with a reduced mass or not at all, with the removed mass put back in as gas through the \texttt{condinit} routine in \textsc{por} to ensure the correct gas fraction. Therefore, our hydrodynamical MOND version of \textsc{dice} does not yield an equilibrium disc template by itself $-$ it must be carefully combined with a modified version of \textsc{por}. All the algorithms we use are publicly available \footnote{\url{https://bitbucket.org/SrikanthTN/bonnpor/src/master/}} for reproducibility, with an accompanying user manual \citep{Nagesh_2021}.

Since it is not possible to guarantee that our disc templates are exactly in equilibrium initially, we start our simulation 1 Gyr before the expected time of the flyby that we found semi-analytically in Section \ref{MW_M31_orbit}. The initial thickness profiles of the MW and M31 gas discs (shown in Figure~\ref{Initial_gas_thickness}) are similar to that used by \citet{Banik_2020_M33} in their 100 kK model of M33.

We expect the MW and M31 discs to precess slightly from their initial orientations \citepalias[see section 3.1 of][]{Banik_Ryan_2018}. Therefore, the observed orientations of the MW and M31 discs differ slightly from our adopted initial orientations, which are given in {\it final} Galactic coordinates $\left(l, b\right)$ in Table \ref{Disc_parameters}. We use this system throughout this article when specifying directions of vectors $-$ it is the system used in the simulation. To iteratively correct for disc precession, we find the rotation matrix between the final simulated and observed orientation of each disc, and then apply the inverse of this rotation to the observed orientation to initialize the next simulation. We will see later that the final simulated orientation of each disc agrees quite closely with observations after just one such iterative correction (Section \ref{Disc_remnants}).

\subsection{Adding features to \normalfont \scshape \textsc{por}}
\label{New_features_por}

The SAM procedure discussed in Section \ref{MW_M31_orbit} is very accurate for following the overall behaviour of $d \left( t \right)$, but insufficient to model tidal debris generated by the interaction. This is the main purpose of the \textsc{por} simulations we will conduct in the present paper. There are some slight differences between the physics considered in SAM and in \textsc{por}, which we try to rectify by adding features to \textsc{por} and adjusting the initial conditions.

\subsubsection{An allowance for tides}
\label{Impulse}

Tides from objects outside the LG are not directly included in the \textsc{por} simulations, but are considered in SAM to estimate the flyby time as accurately as possible. To approximately include tides in \textsc{por}, we estimate the amount of energy gained by the MW-M31 system due to tidal compression in the 1 Gyr preceding the flyby, which we estimate directly from SAM using information on the forces caused by each perturber. This energy is put into the radial component of $\dot{\bm{d}}$ at the start of our \textsc{por} simulation, which has the effect of slightly increasing how quickly the MW and M31 are approaching each other at that time. Our \textsc{por} models neglect the impact of tides after the flyby, which is justified as the perturbers are much further apart then due to cosmic expansion.

\subsubsection{Dark energy}
\label{Including_dark_energy}

The present MW-M31 separation of 783 kpc \citep{McConnachie_2012} is large enough that our \textsc{por} model should include the cosmological term in Equation \ref{Timing_argument_basic}. This partly consists of a decelerating term due to matter, which is included automatically because the LG mass mainly resides in the MW and M31, which we directly include. At late times, there is also an outwards repulsion from dark energy. We include this while operating \textsc{ramses} in non-cosmological mode, since this is required by the \textsc{por} patch. For some dark energy parameter $\Omega_{\Lambda,0}$, the idea is to create an extra repulsive force
\begin{eqnarray}
	\Delta \bm{g} ~=~ {H_0}^2 \Omega_{\Lambda,0} \, \bm{r} \, ,
	\label{Delta_g_dark_energy}
\end{eqnarray}
where $\bm{r}$ is the position relative to the barycentre.

We can reproduce Equation \ref{Delta_g_dark_energy} with a standard Poisson solver if we adjust the density and boundary condition. Since we want $\bm{g}_{_N}$ to be calculated in a standard way so that it is correctly MONDified, we apply the density increment only to the PDM.
\begin{eqnarray}
	\Delta \rho_{_{\text{PDM}}} ~=~ -\frac{3 {H_0}^2 \Omega_{\Lambda,0}}{4 \mathrm{\pi} G} \, .
	\label{Density_adjustment_dark_energy}
\end{eqnarray}
For consistency, we also adjust the boundary potential for only the MOND stage by
\begin{eqnarray}
	\Delta \Phi ~=~ -\frac{{H_0}^2 \, \Omega_{\Lambda,0}}{2} \, .
	\label{Potential_adjustment_dark_energy}
\end{eqnarray}
Equations \ref{Density_adjustment_dark_energy} and \ref{Potential_adjustment_dark_energy} are implemented by appropriate adjustments to the \textsc{por} algorithm, thereby yielding Equation \ref{Delta_g_dark_energy} in the interior.
%IB: the potential needs to decline outwards in order to create a repulsive force.

The impact of dark energy on the MW-M31 dynamics is quite small in MOND as their mutual gravity is $\approx 30\times$ stronger than the cosmological acceleration term in Equation \ref{Timing_argument_basic} \citepalias[table 10 of][]{Banik_Ryan_2018}. We nonetheless include it for completeness. The fact that the cosmological acceleration is small compared to the internal gravity means that our results are robust with respect to uncertainty regarding how MOND should be applied in a cosmological context \citep[the `Hubble field effect' discussed in section 5.2.3 of][]{Haslbauer_2020}. It is however possible to make some plausible assumptions and simulate systems where the average enclosed density differs only slightly from the cosmic mean, as done in that work and in several others \citep[e.g.][]{Katz_2013, Candlish_2016}.

\subsubsection{The external field effect (EFE)}
\label{Including_the_EFE}

As discussed in section 2 of \citetalias{Banik_Ryan_2018}, the EFE from large-scale structure has a significant effect on the LG gravitational field when the MW and M31 are close to apocentre, where they spend a significant amount of time. The EFE is a non-standard phenomenon caused by the non-linearity of MOND (Equation \ref{Deep_MOND_limit}). It leads to the internal gravitational dynamics of a system being affected by $\bm{g}_{\text{ext}}$ even in the absence of tidal effects \citep{Milgrom_1986}. Strong evidence for the EFE in field galaxies was recently reported by \citet{Chae_2020_EFE, Chae_2021} by comparing the RCs of galaxies in isolated and more crowded environments \citep[building on similar earlier work;][]{Haghi_2016, Hees_2016}.

%This is evident from Equation \ref{QUMOND_equation} if we consider an isolated dwarf galaxy with very low internal acceleration. It lies deep in the MOND regime, so $\sigma_{\text{int}}$ greatly exceeds the Newtonian expectation. Suppose the dwarf moves closer to a massive galaxy cluster such that the total (internal + external) gravity now exceeds $a_{_0}$. This will reduce $\nu$ to order unity. Neglecting the effect of tides, $\sigma_{\text{int}}$ would now be significantly smaller if the dwarf remains gravitationally bound. In this way, the inherent non-linearity of MOND causes it to break the strong equivalence principle. \citet{Chae_2020_EFE} recently reported a highly significant correlation between the large-scale environment of a galaxy and its internal gravity for the same baryonic distribution, which seemingly represents a detection of the predicted EFE at $8-11\sigma$. The most convincing aspect is that the outer RC is flat for more isolated galaxies, but is declining in more crowded environments.

%Notice that in the above example, we could simply consider both the dwarf and its host cluster, avoiding the concept of an `external' field $\bm{g}_{\text{ext}}$. Though perfectly valid, this approach is often very computationally intensive since it entails a significant increase in the simulated volume. Thus, another approach is to consider a much smaller box around the dwarf, but with appropriately adjusted boundary conditions. This is valid if tidal effects can be neglected and we only care about the dwarf's internal properties. 

In the QUMOND approach where we must first get $\bm{g}_{_N}$, the main change is to add the Newtonian-equivalent external field to the Newtonian gravity sourced by the system under study.
\begin{eqnarray}
	\bm{g}_{_N} ~\to~ \bm{g}_{_N} + \bm{g}_{_{N, \text{ext}}} \, .
	\label{g_N_adjustment}
\end{eqnarray}
Assuming that $\bm{g}_{\text{ext}}$ is sourced by a distant point-like object, we get that
\begin{eqnarray}
	\bm{g}_{_{N, \text{ext}}} \overbrace{\nu \left( g_{_{N, \text{ext}}} \right)}^{\nu_{\text{ext}}} ~=~ \bm{g}_{\text{ext}} \, .
\end{eqnarray}
If we know the EF $\bm{g}_{\text{ext}}$, this can be inverted to obtain $\bm{g}_{_{N, \text{ext}}}$. 

%This is possible analytically for our adopted simple $\nu$ function (Equation \ref{Simple_interpolating_function}).
%\begin{eqnarray}
%	g_{_{N, \text{ext}}} ~=~ \frac{{g_{\text{ext}}}^2}{g_{\text{ext}} + a_{_0}} \, .
%\end{eqnarray}
Once we have adjusted $\bm{g}_{_N}$ according to Equation \ref{g_N_adjustment}, we use it to find $\nabla \cdot \bm{g}$ using Equation \ref{QUMOND_equation} as before. The reason is that this applies to the Newtonian gravity sourced by matter both within and beyond the simulated domain, the extent of which is an arbitrary decision that should have no bearing on the result. 

Including the EFE also requires us to change the boundary condition, but only for the MOND stage since the internal dynamics and external field are fully separable in Newtonian gravity. For simplicity, the boundary should be in some asymptotic regime where $\bm{g}$ has a well-understood analytic behaviour. Normally, it is sufficient for the boundary to be distant enough that the simulated system can be approximated as a point mass. If there is also a non-negligible EFE, then the simplest option is to choose a boundary where $\bm{g}_{\text{ext}}$ is much stronger than the internal gravity of the system. Its internal potential then becomes EF-dominated \citep[e.g.,][]{Banik_2015}:
\begin{eqnarray}
	\label{Phi_g_ext_domination}
	\Phi &=& -\frac{GM \nu_{\text{ext}}}{r}\left(1 + \frac{K_0}{2} \sin^2 \theta \right) \, , \\
	K_0 ~&\equiv &~ \frac{\partial \ln \nu_{\text{ext}}}{\partial \ln g_{_{N,ext}}} \, ,
\end{eqnarray}
where $M$ is the total mass within the simulation volume, $\bm{r}$ is the position relative to its barycentre, and $\theta$ is the angle between $\bm{r}$ and $\bm{g}_{\text{ext}}$. Due to the $1/r$ dependence and the fact that potentials from different sources can be superposed in this perturbative framework, the result is reminiscent of standard Newtonian mechanics, so the EF-dominated regime is also known as the quasi-Newtonian regime. Note that Equation \ref{Phi_g_ext_domination} alone is not our boundary condition because we also include a dark energy adjustment (Equation \ref{Potential_adjustment_dark_energy}).

For the Newtonian stage of solving the QUMOND Poisson equation, we continue to use a boundary potential of $-GM/r$, ignoring the constant $\bm{g}_{\text{ext}}$. This means that our simulations consider the internal dynamics in a freely falling reference frame accelerating at $\bm{g}_{\text{ext}}$, whose relevance to the internal dynamics is a violation of the strong equivalence principle.

For simplicity, we keep $\bm{g}_{\text{ext}}$ fixed over the course of our \textsc{por} simulation. However, to maximize the accuracy of the overall MW-M31 trajectory, SAM uses a time-dependent $g_{\text{ext}}$ as described in section 2.2 of \citetalias{Banik_Ryan_2018}. In both cases, we follow the approach in that paper of assuming that today, $\bm{g}_{\text{ext}} = 0.022 \, a_{_0}$ directed towards Galactic coordinates $\left( 276^\circ, \, -30^\circ \right)$, the direction in which the LG presently moves relative to the CMB \citep{Kogut_1993}. This direction is assumed fixed throughout cosmic history.

\subsection{Iterative orbit adjustment}
\label{Iterative_orbit_adjustment}

Despite our best attempts to ensure SAM and \textsc{por} handle the flyby problem as similarly as possible, the two algorithms nonetheless use very different techniques. Thus, advancing the SAM-generated initial conditions using \textsc{por} does not quite yield the presently observed $\bm{d}$ or its direction $\widehat{\bm{d}}$. In the presence of an EFE, the late-time $\widehat{\bm{d}}$ affects the torque exerted by the EFE on the MW-M31 system (Section \ref{External_field_torque}), which in turn influences the present PM of M31. Moreover, an incorrect final $\widehat{\bm{d}}$ suggests that the flyby took place in a different orientation to how it occurred in the simulation, which would influence the SPs.

To ensure the final $\widehat{\bm{d}}$ matches the observed sky position of M31 as accurately as possible, we find the rotation matrix that takes the observed $\widehat{\bm{d}}$ to the simulated final value. We then rerun SAM with the inverse of this rotation applied to the presently observed $\widehat{\bm{d}}$ and the relative velocity $\dot{\bm{d}}$ used in the previous SAM simulation. The idea is that if $\widehat{\bm{d}}$ ends up $10^\circ$ further north than observed, then we can get approximately the correct final $\widehat{\bm{d}}$ by running SAM with a present $\widehat{\bm{d}}$ that lies $10^\circ$ south of the actually observed direction towards M31, because there is some additional physical effect in \textsc{por} but not in SAM that pushes $\widehat{\bm{d}}$ further north by $10^\circ$. The initial conditions generated in this way are used to rerun the \textsc{por} simulation. We find that just one such iteration allows us to match the observed $\widehat{\bm{d}}$ to within a few degrees, which we consider sufficient.

\subsection{Simulation setup and initial conditions}
\label{Simulation_setup}

\subsubsection{SAM}
\label{SAM_setup}

We begin by running SAM with the constraints given in Table \ref{Initial_conditions_SAM}. The main model parameters that we vary are:
\begin{enumerate}
	\item the MW-M31 orbital pole $\widehat{\bm{h}}$, and
	\item the magnitude $h$ of their mutual two-body angular momentum. 
\end{enumerate}
As discussed in section 5.1.3 of \citetalias{Banik_Ryan_2018}, we account for reflex motion induced by the Large Magellanic Cloud (LMC) on the MW and by M33 on M31, leading to a present RV of $\dot{d} = -93.4$ km/s.\footnote{Although the LMC should form out of tidal debris expelled during the flyby, our model does not form individual TDGs. The simulated Galactic disc thus does not experience recoil from an orbiting massive satellite, as it might do in a more advanced model. Since the timing argument is sensitive to the gravity between the MW and M31, any massive satellites should be included in the mass and velocity of each galaxy.} We keep $\dot{d}$ fixed, but vary the present tangential velocity $\bm{v}_{\text{tan}}$ used in SAM, which does a backwards integration. Changing $h$ mainly affects the perigalacticon distance, but does not much alter the relative speed then. Varying $\widehat{\bm{h}}$ allows us to consider a wide range of possible orbital geometries. While our priority is to match the SP orientations, we subsequently compare our best-fitting model with the observed PM of M31 (Section \ref{Proper_motion_section}). Interestingly, if we neglect both considerations, the timing argument alone sets some constraints on $\widehat{\bm{h}}$ because perturbers like Cen A have a different influence on the internal dynamics of the LG depending on $\widehat{\bm{d}}$ prior to the flyby. As a result, the timing argument mass of the LG is too high for a large range of $\widehat{\bm{h}}$, which renders the models unlikely as the MOND timing argument mass should correspond to the baryonic mass \citepalias[section 5.1.3 of][]{Banik_Ryan_2018}.

\begin{table}
	\centering
	\caption{Present-day constraints imposed on SAM, which does a backwards integration. The tangential velocity of M31 is varied to best match the observed SP orbital poles.}
	\begin{tabular}{cc}
		\hline
		Initial MW-M31 $\ldots$ & Value \\
		\hline
		distance & 783 kpc \\
		direction & $\left( 121.17^\circ, -21.57^\circ \right)$ \\
		RV & $-93.4$ km/s \\
		\hline
	\end{tabular}
	\label{Initial_conditions_SAM}
\end{table}

The model parameters which we explore are the same as in \citetalias{Banik_Ryan_2018}, except that we do not vary the EF. It was indeed shown that varying the EF within the plausible range has only a small impact on the final results, which is likely due to the well-known LG velocity in the CMB frame setting some constraint on the history of $\bm{g}_{\text{ext}}$. We therefore use the best-fitting model of \citetalias{Banik_Ryan_2018} in which $\bm{g}_{\text{ext}} = 0.022 \, a_{_0}$ directed towards Galactic coordinates $\left( 276^\circ, \, -30^\circ \right)$ at the present time. To make the MW-M31 trajectory as realistic as possible, the $g_{\text{ext}}$ assumed in SAM varies with time, as discussed in their section 2.2. However, implementing a time-varying $\bm{g}_{\text{ext}}$ in \textsc{por} would involve significant complications, so this uses the present $\bm{g}_{\text{ext}}$ for the full duration of the simulation. The EFE is not expected to influence details of the MW-M31 interaction, but causes them to reach a larger apocentre at late times by weakening their mutual gravity. As a result, the main role of the EFE is to alter the timing of the MW-M31 flyby, which sets the time available for the tidal debris to settle down after the flyby. We therefore prioritize making the flyby time as accurate as possible in our \textsc{por} models, which requires a carefully prepared SAM. Due to these slight differences and the more rigorous treatment of the flyby in \textsc{por}, the MW-M31 trajectory is expected to differ somewhat compared to SAM. We do not run the \textsc{por} simulations for a different length of time than that indicated by SAM in order to obtain a better agreement with the observed distance to M31, so this is expected to differ somewhat from the observed $783 \pm 25$~kpc \citep{McConnachie_2012}.

It is important to realize that although the final M31 distance in \textsc{por} need not match observations exactly, requiring even an approximate match places non-trivial constraints on our model. This is because a very close encounter would lead to significant dynamical friction between the baryonic discs, causing a very low apocentre and a subsequent merger within a few Gyr. This possibility was neglected in the models of \citetalias{Banik_Ryan_2018}, where the $d \left( t \right)$ returned by SAM was assumed to be exactly correct. In general, a hydrodynamical model of the interaction is obviously a significant advance on the previous restricted $N$-body models, even if the much higher computational cost reduces the scope for fine-tuning to match certain observables to high precision. We tried $\approx 30$ \textsc{por} models to obtain a good fit to the SP orientations, which we judged visually.\footnote{We also ran a few more models to fine-tune the agreement with some observables like the present disc orientations and the MW-M31 direction (Section \ref{Iterative_orbit_adjustment}).}

\begin{table}
	\centering
	\caption{The initial position in kpc and velocity in km/s of the MW and M31 disc centres at the start of our best-fitting \textsc{por} simulation. The combined barycentre is at the origin because the MW:M31 mass ratio is 3:7 (Table \ref{Disc_parameters}), consistently with \citetalias{Banik_Ryan_2018}. SAM indicates that these initial conditions are valid 8.1 Gyr ago.}
	\begin{tabular}{ccccc}
		\hline
		Galaxy & \multicolumn{2}{c}{MW} & \multicolumn{2}{c}{M31} \\
		Direction & Position & Velocity & Position & Velocity \\
		\hline
		$x$ & -56.0 & 5.2 & 24.0 & -2.2 \\
		$y$ & 254.2 & -177.8 & -109.0 & 76.2 \\
		$z$ & 44.1 & -86.7 & -18.9 & 37.2 \\
		\hline
	\end{tabular}
	\label{Initial_conditions}
\end{table}

%gal_center2=-56.0212184815010,254.186762629426,44.0656113946021
%Vgal2=5.19272453497556,-177.819419917975,-86.7363387262098
%gal_center1=24.0090936349290,-108.937183984040,-18.8852620262580
%Vgal1=-2.22545337214473,76.2083228220094,37.1727165969756

\subsubsection{\sc por}
\label{PoR_settings}

Our \textsc{por} simulations use very similar settings to those described in section 2.4 of \citet{Banik_2020_M33}, so we briefly mention the main points here. We use the non-cosmological particle-in-cell mode, activate MOND and the EFE, and use the refinement conditions $m\_refine = 10^3 M_\odot$ and $n\_subcycle = \left( 1, 1, 2, 2 \right)$. Since we are simulating the whole LG, we use a much larger box size of 6144 kpc, with $7-12$ levels of refinement. The most poorly resolved regions thus have a resolution of $6144/2^7 = 48$ kpc, which improves to $6144/2^{12} = 1.5$ kpc for the best resolved regions. \citet{Teyssier_2002} provides a more detailed description of \textsc{ramses}, including default values of parameters that we do not alter. Perhaps the most important of these is the $gravity\_type$, whose default setting of 0 represents self-gravity. We also use the default Poisson convergence parameter $epsilon = 10^{-4}$.

The results of our best-fitting model are discussed next. In this model, the initial position and velocity of the MW and M31 are as given in Table~\ref{Initial_conditions} for the centres of two galaxies each consisting of $5 \times 10^5$ particles and a gas fraction of 50\%, with the disc orientations given in Table~\ref{Disc_parameters}. The initial gas temperature is $T2\_ISM = 465$~kK, with the temperature floor $T2\_star$ being 20\% lower. According to SAM, the initial MW and M31 positions correspond to 1 Gyr before the flyby, so our \textsc{por} simulations can be considered to start 5.72 Gyr after the Big Bang (8.1 Gyr ago). Starting the simulations 1 Gyr before the flyby gives the discs some time to settle down first, but avoids the need to simulate very early epochs at which the cosmological term in Equation \ref{Timing_argument_basic} would be significant.

\section{Results}
\label{Results}

In this section, we present the results of our best-fitting model and compare it with relevant observations. We use the simulation snapshot after 8.2 Gyr because this corresponds as closely as possible to the lookback time estimated by SAM for the initial conditions of Table~\ref{Initial_conditions} .

\subsection{The MW-M31 orbit}
\label{Section_MW_M31_orbit}

The initial conditions are obtained by running SAM, which gives a timing argument mass of $3.457 \times 10^{11} \, M_\odot$ for the whole LG. This is somewhat higher than the sum of the disc masses used in \textsc{por} (Table \ref{Disc_parameters}), which we consider acceptable as there would also be some mass in, e.g., a halo of gas around each galaxy, though we do not include a halo explicitly. This issue was discussed further in section 5.1.1 of \citetalias{Banik_Ryan_2018}, who argued that although the observed MW and M31 RCs suggest a combined disc mass of $2.3 \times 10^{11} \, M_\odot$, a modestly higher timing argument mass for the whole LG is quite feasible due also to satellite galaxies like the LMC and M33. Equation \ref{Deep_MOND_limit} implies that a 50\% increase in the mass of a galaxy increases the flatline level of its RC by just 11\%. Such a small increase in the MW and M31 RCs at large ($\ga 100$~kpc) distances is difficult to rule out at present, especially for M31 \citep{Corbelli_2010, Sofue_2015}.

\begin{figure*}
	\centering
	\includegraphics[width = 17.5cm] {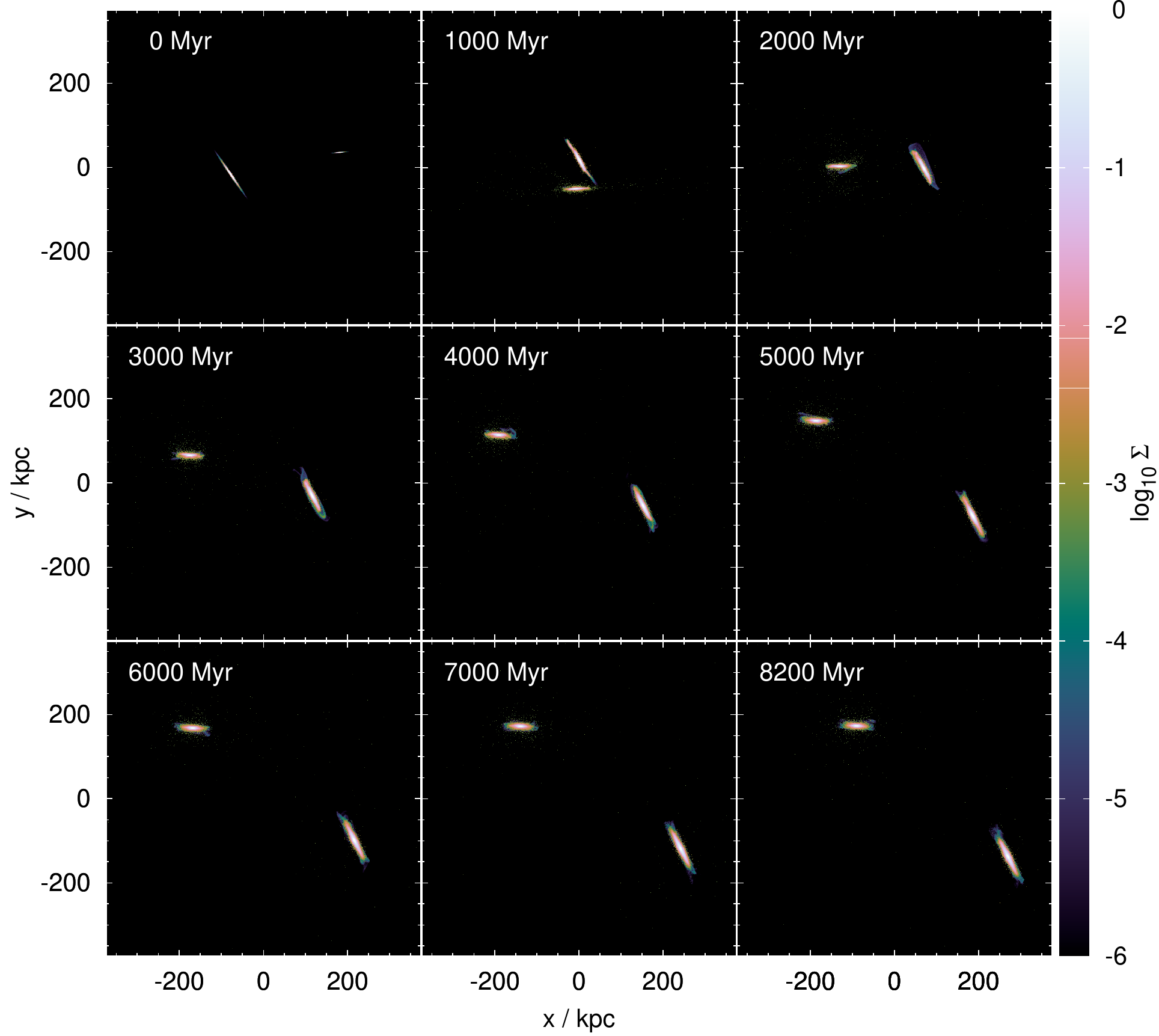}
	\caption{The stellar particles in the central 500 kpc square of our \textsc{por} simulation, viewed along the direction which would make both discs appear edge-on in actual observations at $z=0$ (Section \ref{Disc_remnants}). Since the disc orientations change somewhat during the flyby and we have not perfectly adjusted for this, the discs are not perfectly edge-on in the last snapshots. The projected surface density is shown in units of $1000 \, M_\odot$/pc$^2$. The time since the start of the simulation is indicated at the top left of each panel. The MW disc appears almost horizontal in this view. Notice that the MW and M31 retain thin discs, outside of which there is very little material.}
	\label{LG_view_part}
\end{figure*}

\begin{figure*}
	\centering
	\includegraphics[width = 17.5cm] {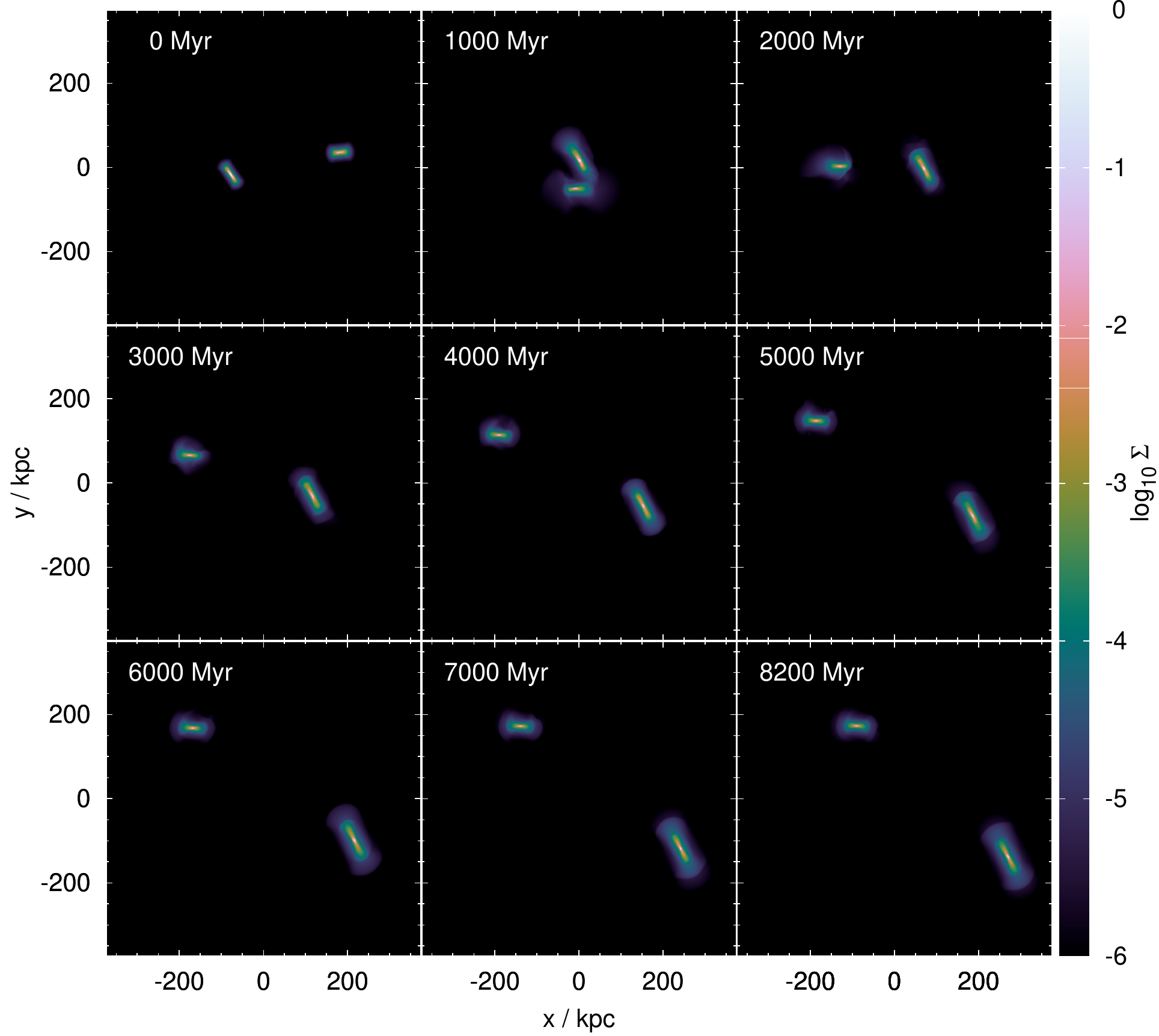}
	\caption{Similar to Figure \ref{LG_view_part}, but now showing the gas. The smallest simulated gas cell has sides of length 1.5 kpc.}
	\label{LG_view_gas}
\end{figure*}

Before conducting more detailed analyses, we show the central 500 kpc of the \textsc{por} simulation as viewed from the direction which would make both discs appear perfectly edge-on if their orientations are as observed $-$ this is very nearly the case (Section \ref{Disc_remnants}). Figure \ref{LG_view_part} shows the stellar particles, while Figure \ref{LG_view_gas} shows the gas. The discs undergo closest approach $\approx 1$ Gyr into the simulation. Dynamical friction during the flyby is small, allowing the galaxies to reach a large post-encounter separation. Moreover, the flyby does not disrupt the MW and M31 discs too severely, allowing them to retain a thin disc by the end of our simulation $-$ we will investigate this in more detail in Section \ref{Disc_remnants}. There is also a small numerical drift of the MW-M31 barycentre, which can be understood using much less computationally intensive methods \citep[see footnote 14 of][]{Banik_2020_M33}.

\begin{figure}
	\centering
	\includegraphics[width = 8.5cm] {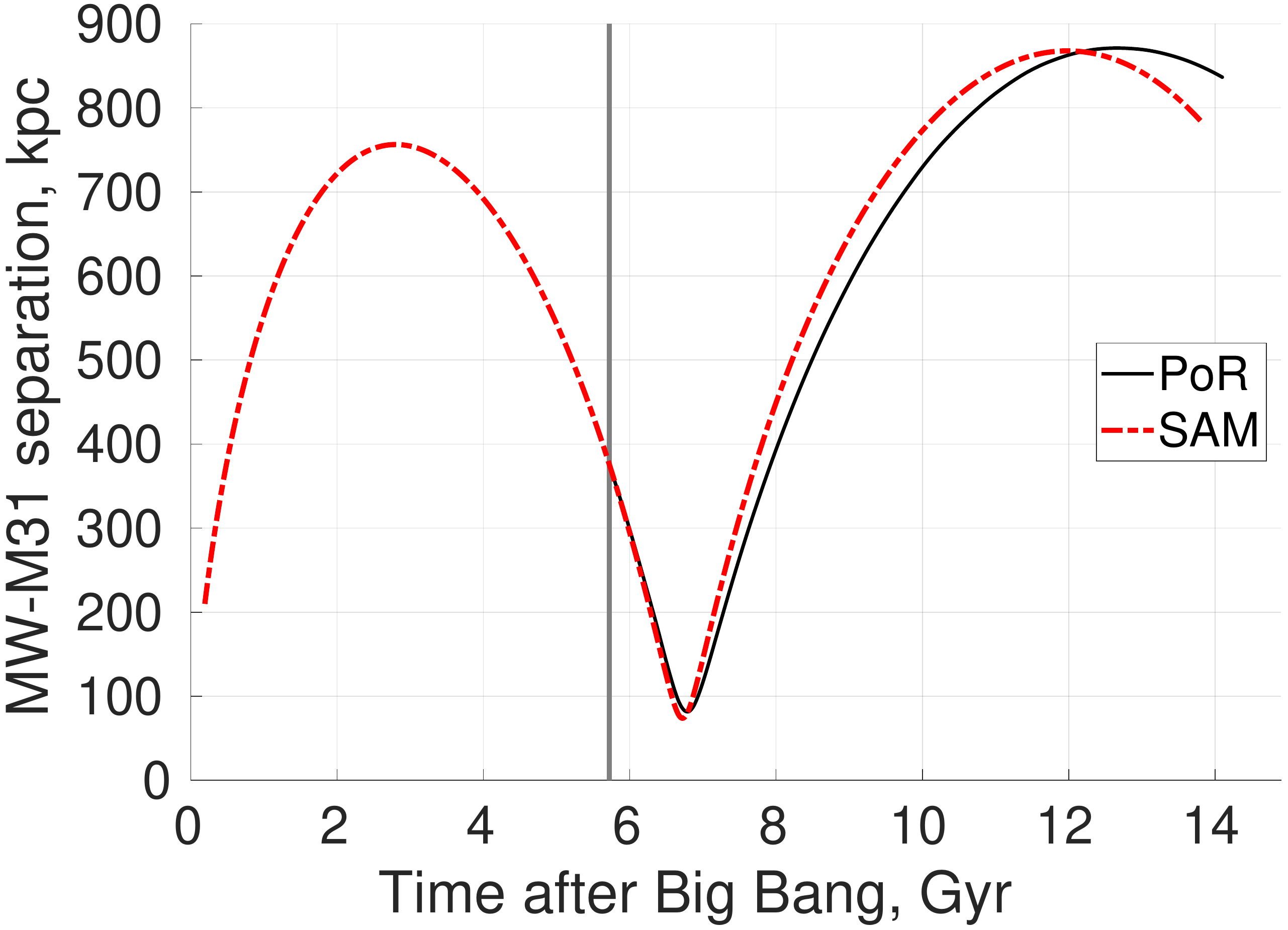}
	\includegraphics[width = 8.5cm] {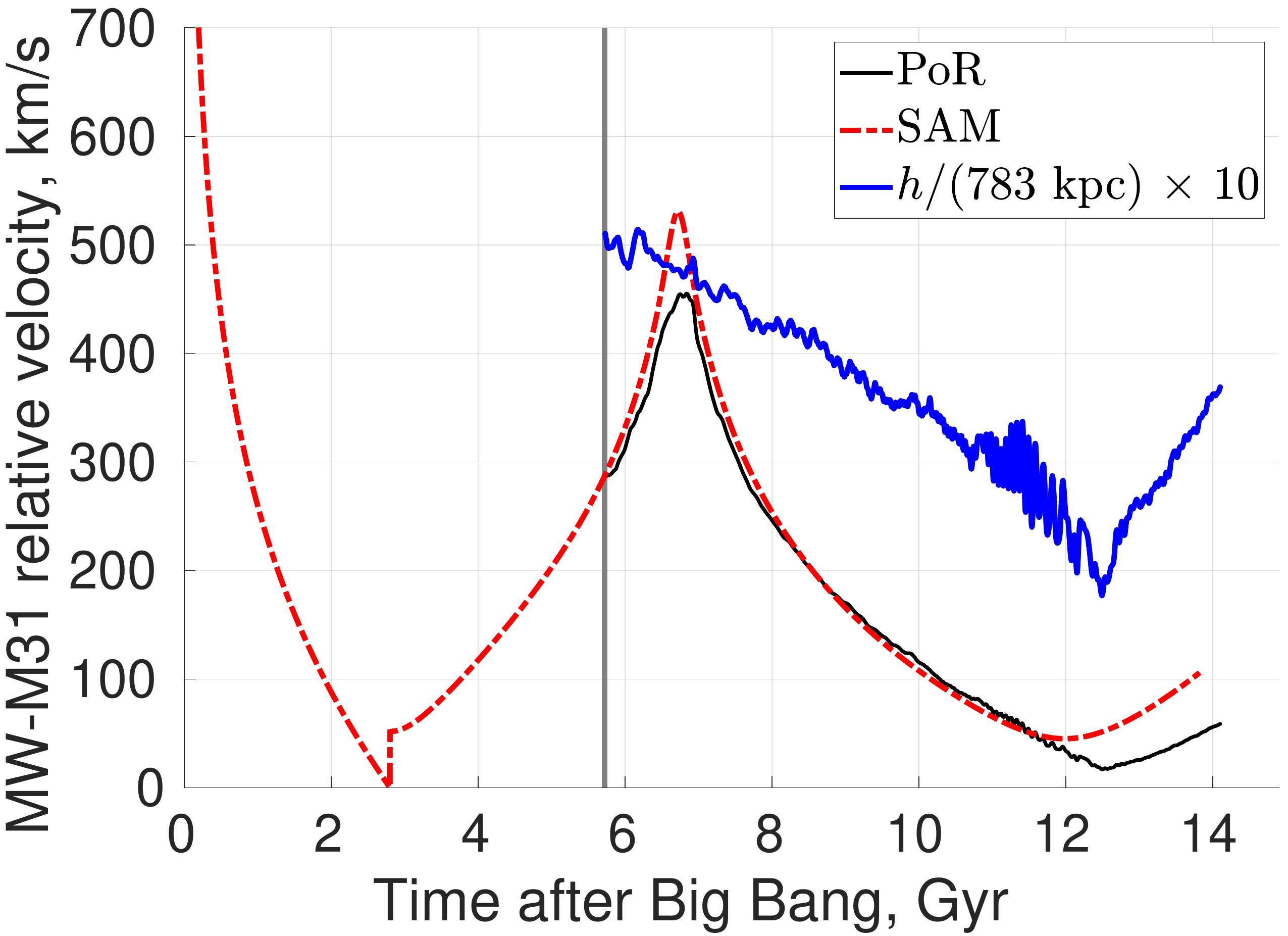}
	\caption{\emph{Top}: The MW-M31 separation as a function of time in our \textsc{por} (solid black) and SAM (dot-dashed red) models. Although our \textsc{por} simulation is not cosmological, its initial conditions are obtained from SAM 1 Gyr before the flyby using a total mass of $3.457 \times 10^{11} \, M_\odot$, thus placing the \textsc{por} model in the cosmological context of a trajectory consistent with the timing argument (Section \ref{MW_M31_orbit}). The \textsc{por} model (with a slightly lower total mass as it considers only the discs) is advanced slightly into the future, but our detailed analyses focus on the snapshot 8.2 Gyr after the start of the simulation to most closely match the post-flyby time estimated by SAM. The vertical grey line in each panel shows the start of our \textsc{por} simulation. \emph{Bottom}: The MW-M31 relative velocity in SAM (dot-dashed red) and \textsc{por} (solid black). The solid blue line with more scatter shows their relative angular momentum as an equivalent tangential velocity at their present separation of 783 kpc \citep{McConnachie_2012}, multiplied by 10 for clarity. The discontinuity in the SAM velocity around 3 Gyr arises from assuming no angular momentum prior to first apocentre, to allow the MW-M31 trajectory to reach zero peculiar velocity at very early times (Section \ref{MW_M31_orbit}). Notice the lack of a significant change in the angular momentum due to the MW-M31 flyby, indicating that it hardly alters their orbit. The gradual change over many Gyr is due to torque from the EFE (Section~\ref{External_field_torque}) and due to tidal debris falling back onto each galaxy, especially the MW (Section~\ref{Eccentricity}).}
	\label{MW_M31_trajectory}
\end{figure}

We extract the MW-M31 trajectory from our \textsc{por} simulation while it is running, without extracting every simulation output. We briefly describe this technique in Appendix \ref{Barycentre_extraction_on_the_fly} as it could also be useful for other projects and is part of the publicly available version of \textsc{por} used here. In this way, we obtain the trajectory shown in Figure \ref{MW_M31_trajectory}, with the top and bottom panel used to show the MW-M31 separation and relative velocity, respectively. The expected result using SAM is shown as a dotted red line in each panel. Both methods give a rather similar overall trajectory, indicating little dynamical friction during the encounter. This is due to the fairly large pericentre distance of 81 kpc. Table \ref{Pericentre_apocentre_SAM_PoR} summarizes information about the pericentre and apocentre in each trajectory. The higher second apocentre in SAM was discussed in section 5.1.2 of \citetalias{Banik_Ryan_2018}, who concluded that it is mainly driven by tides from Cen A due to its relatively high mass and close alignment with the MW-M31 line after but not before the flyby. Tides are not explicitly included in our \textsc{por} simulation, which moreover starts shortly before the flyby and thus includes only the most recent MW-M31 apocentre, when the perturbers are rather distant.

\begin{table}
	\centering
	\caption{Information about the MW-M31 pericentre and apocentre in our SAM and \textsc{por} simulations, based on Figure \ref{MW_M31_trajectory}. Times are relative to the Big Bang.}
	\begin{tabular}{ccccc}
		\hline
		Orbital phase & \multicolumn{2}{c}{Perigalacticon} & \multicolumn{2}{c}{Apogalacticon} \\
		Algorithm & SAM & \textsc{por} & SAM & \textsc{por} \\
		\hline
		Time (Gyr) & 6.72 & 6.79 & 12.0 & 12.6 \\
		Distance (kpc) & 73.8 & 81.5 & 867.8 & 871.2 \\
		Velocity (km/s) & 530.9 & 452.1 & 45.1 & 18.6 \\
		\hline
	\end{tabular}
	\label{Pericentre_apocentre_SAM_PoR}
\end{table}

The simulation snapshot that we analyse (8.2 Gyr after the start) has $d = 846$~kpc, which slightly exceeds the observed $783 \pm 25$ kpc distance to M31 \citep{McConnachie_2012}. As explained previously, we do not rectify this issue by running the simulation for longer. Doing so should have only a small effect on the SPs, the main focus of this work. Importantly for the overall geometry, the simulated $\widehat{\bm{d}} = \left( 121.11 ^\circ, \, -21.99^\circ \right)$, which differs by only $0.42^\circ$ from the observed sky position of M31 $\left( 121.17 ^\circ, \, -21.57^\circ \right)$.\footnote{The simulated $\widehat{\bm{d}}$ is actually found using the method discussed in Section \ref{Proper_motion_section} to identify the MW and M31 barycentres, but the results are almost identical.} Therefore, the overall MW-M31 trajectory in our \textsc{por} simulation is quite reasonable, and should be consistent with cosmological initial conditions (the timing argument) at much earlier times.

%Distance to M31 = 845.6175 kpc.
%M31 direction:
%121.1007
%-21.9879
%Offset from observed direction: 0.42053 degrees.
%Observed M31_dir = Sph_xy(-21.5729606/Radian, 121.1743170/Radian).

\subsection{Final tangential velocity and PM}
\label{Proper_motion_section}

The PM of M31 provides an important constraint on the orbital geometry of our best-fitting model. When scanning the parameter space (Section \ref{SAM_setup}), we did not consider this constraint, though we did not consider all possible orbital geometries either as some are highly disfavoured by the timing argument alone (Section \ref{MW_M31_orbit}). In the following, we describe how we obtain the predicted PM of M31, and compare this to the latest observational constraints.

To find the separation and relative velocity of the MW and M31, we first have to identify each galaxy's centre of mass. We obtain an initial guess using the iterative on-the-fly method described in Appendix \ref{Barycentre_extraction_on_the_fly} based on the particles alone. Much more detailed analyses are possible using a simulation snapshot because we also use our modified version of \textsc{rdramses} \citep[section 3 of][]{Banik_2020_M33} to obtain a list of all gas cells, treating them as particles at the cell centres. This allows the stars and gas to be analysed on an equal footing. We therefore find the barycentre of all material whose position and velocity lies within 250 kpc and 500 km/s, respectively, of our initial guess for the barycentre. These thresholds are deliberately set quite wide to reduce the risk of converging on the wrong density maximum. This process is repeated iteratively until the barycentre position shifts by $<1$ kpc and its velocity shifts by $<1$ km/s between successive iterations. The process converges very rapidly because the initial guess for the barycentre based on particles alone (Appendix \ref{Barycentre_extraction_on_the_fly}) is already highly accurate.

\begin{table}
	\centering
	\caption{The final position in kpc and velocity in km/s of the MW and M31 barycentres 8.2 Gyr into our best-fitting \textsc{por} simulation. According to SAM, this is the \textsc{por} simulation snapshot temporally closest to the present epoch. The combined barycentre is slightly offset from the origin due to numerical drift, as discussed further in footnote 14 of \citet{Banik_2020_M33}.}
	\begin{tabular}{ccccc}
		\hline
		Galaxy & \multicolumn{2}{c}{MW} & \multicolumn{2}{c}{M31} \\
		Direction & Position & Velocity & Position & Velocity \\
		\hline
		$x$ & 232.9 & -31.0 & -172.1 & -6.6 \\
		$y$ & -248.8 & 73.5 & 422.6 & 29.4 \\
		$z$ & 176.4 & -0.7 & -140.2 & -14.4 \\
		\hline
	\end{tabular}
	\label{Present_conditions}
\end{table}

%MW position in kpc:
%  232.8987
% -248.8351
%  176.3665

%MW velocity in km/s:
%  -30.9907
%   73.5017
%   -0.7365

%M31 position in kpc:
% -172.1283
%  422.5675
% -140.2416

%M31 velocity in km/s:
%   -6.6447
%   29.3561
%  -14.3980

Table~\ref{Present_conditions} shows the position and velocity of the MW and M31 in the simulation reference frame, from which we get the MW-M31 relative separation and velocity. We take the cross product of these vectors to get their orbital angular momentum $\bm{h}$ and thus the orbital pole $\widehat{\bm{h}}$. We also use $h$ to obtain the tangential velocity $v_{t, 783}$ at a distance of 783 kpc, which slightly exceeds the simulated tangential velocity by a factor of (846/783) because the final distance is slightly larger than observed. We then find what PM components of M31 on the night sky would most closely mimic $\bm{d}$ and $\bm{v}_{\text{tan}}$ in the analysed \textsc{por} simulation output. We do this by adjusting the assumed M31 PM components $\left( \mu_{\alpha, \star}, \mu_\delta \right)$ in SAM to find the combination which best matches the $\widehat{\bm{h}} = \left( 209.77^\circ, \, 3.29^\circ \right)$ and $v_{t, 783} = 34.12$ km/s found from \textsc{por}, with $\mu_{\alpha, \star}$ and $\mu_\delta$ being the angular velocity in the directions which most quickly increase the right ascension and declination, respectively. In this calculation, the present M31 direction and RV in SAM are fixed to the values in Table \ref{Initial_conditions_SAM}. The Galactic circular velocity at the Solar circle is assumed to be 239 km/s \citep{McMillan_2011}, while the non-circular motion of the Sun is taken from \citet{Francis_2014} $-$ uncertainties in these parameters and in the distance to M31 are much smaller than in its PM. We vary $\left( \mu_{\alpha, \star}, \mu_\delta \right)$ using a gradient descent algorithm \citep{Fletcher_1963} to minimize the sum of squared errors in $\widehat{\bm{h}}$ and $v_{t, 783}$, with each error scaled to an uncertainty of $1^\circ$ and 1 km/s, respectively. While it is possible to match $v_{t, 783}$ exactly, the best-fitting $\left( \mu_{\alpha, \star}, \mu_\delta \right)$ still gives a small error in $\widehat{\bm{h}}$ because of a slight difference in the final $\widehat{\bm{d}}$ between \textsc{por} and SAM. This is reduced by an iterative rerun of the simulation. Since the mismatch is then only $0.42^\circ$, our approach gives a good idea of what our simulation implies for the present M31 PM.%BF: One could also take exactly the same values as used in Salomon et al? Namely Usun=8.88, Vsun=2.91 if you take V0=239 and Wsun=3.08. Probably wont change much, but might be worth for consistency. IB: our analysis is internally consistent, we have done corrections for these ourselves. Changing the values can of course impact the results, but the effect would be much smaller than other uncertainties.

\begin{figure}
	\centering
	\includegraphics[width = 8.5cm] {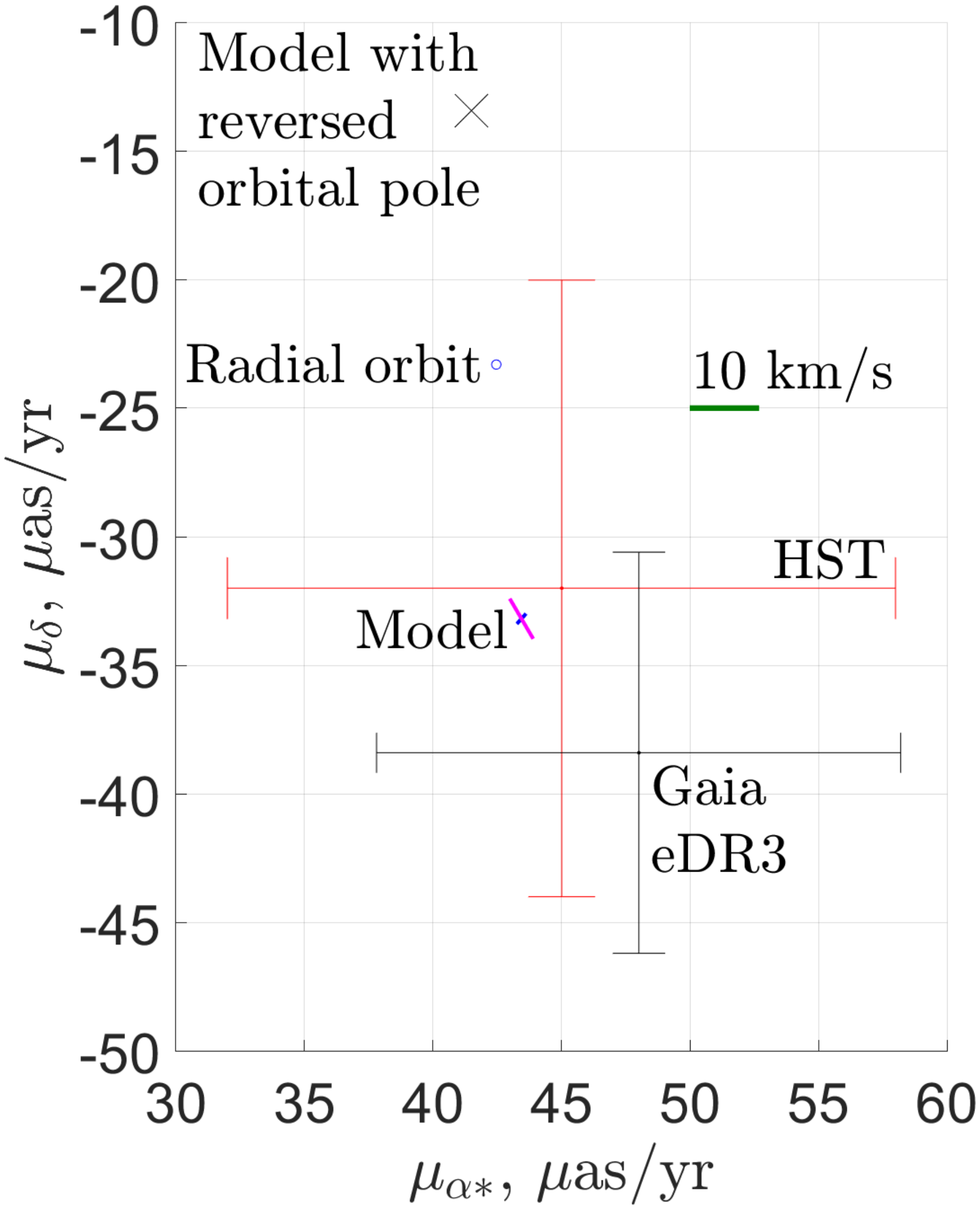}
	\caption{The heliocentric PM of M31 in our best-fitting \textsc{por} model (intersection of blue and pink lines), found using the method described in Section \ref{Proper_motion_section}. The short blue (pink) line indicates the effect of varying the LMC (M33) mass by $\pm 20\%$. In both cases, a heavier satellite increases the predicted M31 $\mu_{\alpha, \star}$. The open blue circle shows the M31 PM for a purely radial MW-M31 orbit, while the black cross shows the result of reversing the Galactocentric tangential velocity of M31 given by our \textsc{por} model. The large red error bars show the observed M31 PM using HST \citep{Van_der_Marel_2012}, while the black error bars use instead Gaia eDR3 \citep{Salomon_2021}. The measurement of \citet{Van_der_Marel_2019} is not shown here for clarity, but all three PMs are listed in Table \ref{Proper_motion_observed}. The horizontal green line represents 10 km/s at the M31 distance of 783 kpc.}
	\label{Proper_motion}
\end{figure}

Our result is shown in Figure \ref{Proper_motion}. As discussed in section 5.1.3 of \citetalias{Banik_Ryan_2018}, this includes a correction for the LMC and M33 altering the barycentric velocity of the MW and M31, respectively. Though the LMC and M33 positions and velocities are known fairly well, there is some uncertainty regarding their mass. This affects the results slightly because our model does not form individual TDGs, so it does not properly capture the induced recoil on the MW from the LMC, which ideally should form in a simulation of the flyby. The uncertainties are rather small in a MOND context because all galaxies are purely baryonic and we know the position and velocity of M33 and the LMC rather well. We therefore vary the mass of each satellite by $\pm 20\%$ and repeat the above-mentioned PM calculation, which assumes the MW (M31) position and velocity in SAM refers to the barycentre of the MW-LMC (M31-M33) system. As discussed in section 4.4 of \citet{Banik_Zhao_2016}, the PM correction due to the LMC is not as significant because its Galactocentric velocity is mostly away from M31. This is evident in the very short length of the dark blue line through the model PM in Figure \ref{Proper_motion}, which indicates the impact of a 20\% uncertainty in the LMC mass. The longer pink line shows the same for M33, whose effect is somewhat larger. Even so, the uncertainty on the M31 PM is $<1 \, \mu$as/yr in both cases, which is much smaller than the difference between its predicted PM and that for a purely radial MW-M31 orbit (open blue circle).

\begin{table}
	\centering
	\caption{The PM of M31 in $\mu$as/yr according to different investigations, which in order are: HST alone \citep{Van_der_Marel_2012}, HST and Gaia DR2 \citep{Van_der_Marel_2019}, and Gaia eDR3 \citep{Salomon_2021} using the $B_{\text{pm}}$ sample in their table 4. The latter result is in a heliocentric frame, including instrumental corrections for PMs of background quasars and using an outlier rejection system. Astrophysical motions beyond the Solar System are not accounted for in these results, since we do the corrections ourselves. The final column gives the PM in our best-fitting model.}
	\begin{tabular}{ccccc}
		\hline
		& & HST $+$ & & \\
		Component & HST & Gaia DR2 & Gaia eDR3 & Model \\
		\hline
		$\mu_{\alpha, \star}$ & $~45 \pm 13$ & $~49 \pm 11$ & $~48.0 \pm 10.2$ & 43.4 \\
		$\mu_\delta$ & $-32 \pm 12$ & $-38 \pm 11$ & $-38.4 \pm 7.8$ & $-33.2$ \\
		\hline
	\end{tabular}
	\label{Proper_motion_observed}
\end{table}

%mu_RA_star_best_fit = 43.4349;
%mu_delta_best_fit = -33.1888;

We compare the predicted M31 PM with the measurements in Table \ref{Proper_motion_observed}. For clarity, we only plot the observed values from the Hubble Space Telescope (HST) prior to Gaia \citep{Van_der_Marel_2012} and the recent result from Gaia early data release 3 \citep[eDR3;][]{Salomon_2021}. In the latter case, we use their so-called $B_{\text{pm}}$ sample in their table 4 to include corrections for instrumental effects as estimated from PMs of background quasars, but without including corrections for astrophysical motions beyond the Solar System $-$ these are handled in our analysis \citep[see, e.g., section 2.3 of][]{Banik_Zhao_2016}. The combination of HST results with Gaia data release 2 \citep[DR2;][]{Van_der_Marel_2019} is not shown in Figure \ref{Proper_motion} as it gives a similar but somewhat less precise result to the $B_{\text{pm}}$ sample in Gaia eDR3. All three PM estimates are mutually consistent within uncertainties. They also agree very well with the PM of our \textsc{por} model (Figure \ref{Proper_motion}). The PM uncertainties are now small enough that this is not guaranteed. We demonstrate this by showing the expected PM of M31 if the \textsc{por}-determined Galactocentric tangential velocity of M31 is reversed, which reverses $\bm{h}$. The result is the black cross towards the upper left of Figure \ref{Proper_motion}, which is now inconsistent with the latest determination of the M31 PM \citep{Salomon_2021}.

\begin{figure}
	\centering
	\includegraphics[width = 8.5cm] {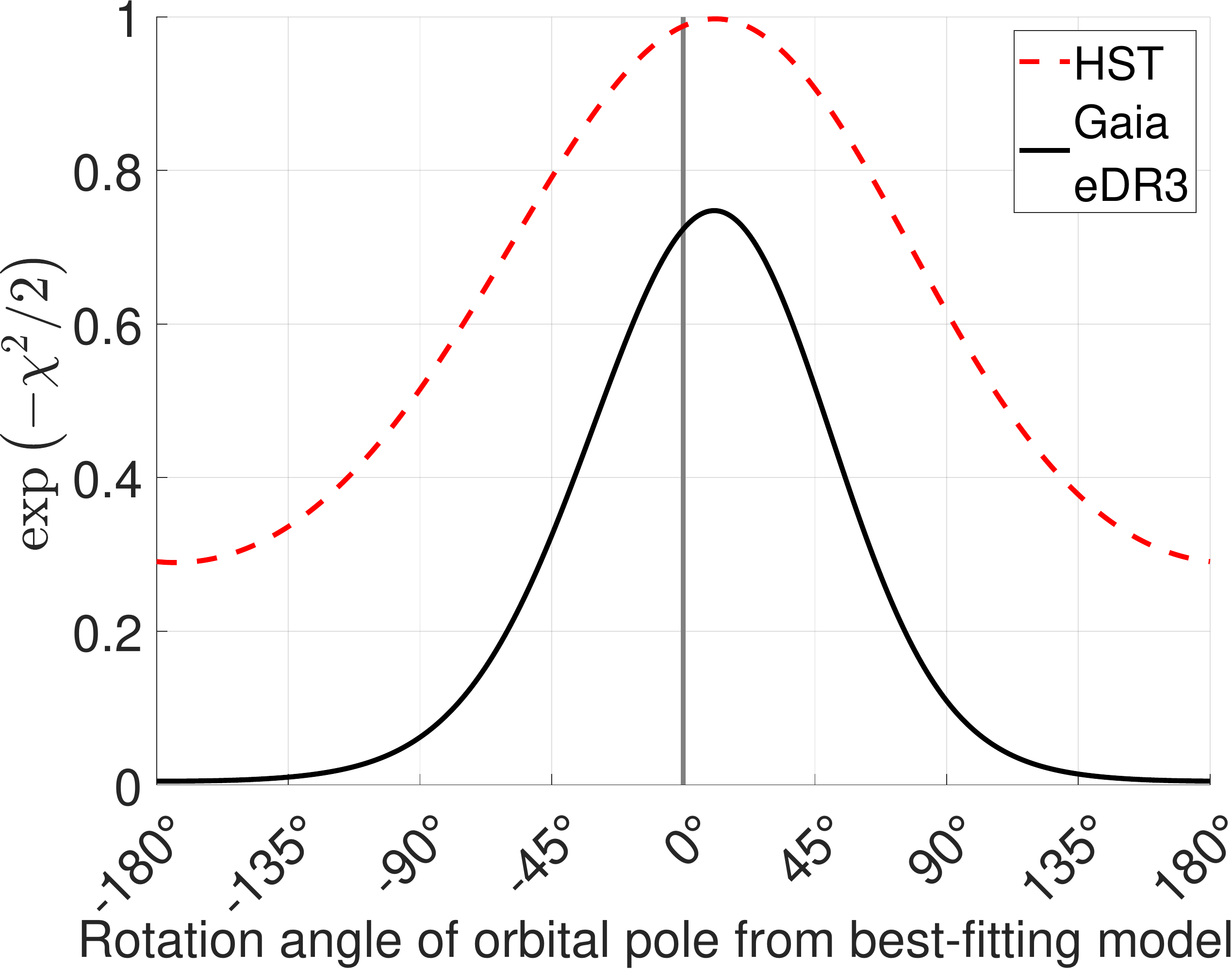}
	\caption{The probability of a higher $\chi^2$ between the observed PM of M31 and in our \textsc{por} model if the predicted Galactocentric tangential velocity of M31 is rotated around the simulated direction towards M31 by the angle shown on the $x$-axis. There is a clear preference for rotation angles close to zero, i.e. for an MW-M31 orbital pole similar to that in the model.}
	\label{Proper_motion_loop}
\end{figure}

To explore this further, we rotate the \textsc{por}-determined $\widehat{\bm{h}}$ around the M31 direction $\widehat{\bm{d}}$ by all possible angles between $-180^\circ$ and $+180^\circ$ in steps of $1^\circ$. We then determine the $\chi^2$ statistic relative to the results of \citet{Van_der_Marel_2012} and \citet{Salomon_2021}. Since this $\chi^2$ is based on two parameters (the PM components), the probability of a higher $\chi^2$ arising by chance if the model were correct is $\exp \left( -\chi^2/2 \right)$. Figure \ref{Proper_motion_loop} shows this probability for all possible rotation angles, with $0^\circ$ corresponding to the actual $\widehat{\bm{h}}$ of our \textsc{por} model. It is clear that for fixed $h$, the direction $\widehat{\bm{h}}$ preferred by the observed M31 PM (Table \ref{Proper_motion_observed}) corresponds quite closely to that of our best-fitting \textsc{por} model.

At this point, it is worth emphasizing that we did not consider the PM of M31 when selecting the best-fitting \textsc{por} simulation $-$ this was based purely on the phase space distribution of the tidal debris. The agreement between the observed M31 PM and in this model constitutes a non-trivial success thereof.

\subsubsection{Tidal torques}
\label{Tidal_torques}

Our \textsc{por} model neglects the late-time effect of tidal torques from perturbers outside the LG, which would also affect the PM of M31. To estimate the tidal torque from each perturber, we assume the deep-MOND limit and that the EFE dominates. The former assumption is clear because of the large distances to the perturbers, while the latter assumption was justified in section 2.3.1 of \citetalias{Banik_Ryan_2018}. As we are only interested in a rough estimate, we approximate that the gravity towards any object of mass $M$ is $GMa_{_0}/\left( r^2 g_{\text{ext}} \right)$ at distance $r$ (Equation \ref{g_EFE}), neglecting the small angular dependence. Using also the distant tide approximation and assuming that $g_{\text{ext}} = 0.022 \, a_{_0}$, the relative MW-M31 acceleration in the direction orthogonal to their separation has a magnitude
\begin{eqnarray}
	g_{\text{tan}} ~=~ \frac{3 \, G M a_{_0} d \cos \theta \sin \theta}{r^3 g_{\text{ext}}} \, ,
	\label{g_tan_estimation}
\end{eqnarray}
where $r$ is the distance from the MW-M31 mid-point to the perturber, $\cos \theta$ is the angle between this direction and $\widehat{\bm{d}}$, and $M$ is the mass of the perturber. $r$ is measured from the MW-M31 mid-point because the MW and M31 are treated as independent test particles freely falling in the gravitational field of the perturber, so the individual MW and M31 masses are irrelevant. Thus, the accuracy of our calculations would be maximized if using the MW-M31 mid-point. For a perturber with $r \gg d$, it does not matter exactly which point we use as the `centre' of the LG.

\begin{table}
	\centering
	\caption{Our estimated tidal torque due to each perturber, defined as the change it causes in the MW-M31 relative tangential velocity (Equation \ref{g_tan_estimation}). The assumed perturber properties are as listed in table 2 of \citetalias{Banik_Ryan_2018}. Despite the high mass of Cen A, it has only a small effect because it is almost on the MW-M31 line. The torquing effect of a perturber is maximized if $\left| \cos \theta \right| \approx 1/\sqrt{2}$, which is almost the case for IC 342. However, even its effect is substantially smaller than the PM uncertainty of $\approx 50$ km/s (Figure \ref{Proper_motion}).}%Can use sqrt(2)*3.7123*11, though the 1 sigma error is more rigorously defined with a factor of 1.515172903961339 rather than sqrt(2).
	\begin{tabular}{ccc}
		\hline
		\multirow{2}{*}{Perturber} & \multirow{2}{*}{$\left| \cos \theta \right|$} & Change in tangential \\
		 &  & velocity over 5 Gyr (km/s) \\
		\hline
		Cen A & 0.99 & 2.43 \\
		M81 & 0.32 & 3.79 \\ %3.7947
		IC 342 & 0.76 & 8.22 \\
		\hline
	\end{tabular}
	\label{Tidal_torque_estimates}
\end{table}

Assuming the same perturber properties as listed in table 2 of \citetalias{Banik_Ryan_2018} and that $d = 783$ kpc, we obtain the tidal torque estimates given in Table \ref{Tidal_torque_estimates} over a period of 5 Gyr, which is roughly the amount of time for which $d$ is similar to its maximum value (Figure \ref{MW_M31_trajectory}). Tidal torques would be much less significant around the time of the flyby. As shown by the values in Table \ref{Tidal_torque_estimates} and the green line in Figure \ref{Proper_motion} representing 10 km/s, it is clear that tidal torques from perturbers hardly affect the present PM of M31. Its measurement accuracy would need to improve another order of magnitude for such subtle effects to be discernible, which would then necessitate more detailed modelling. The small effect of tidal torques arises mainly because of the isolated nature of the LG, but also because one of the most massive external galaxies which could tidally affect it (Cen A) lies almost on the MW-M31 line \citep{Ma_1998}, limiting its tidal torque on the LG.

\subsubsection{The external field torque (EFT)}
\label{External_field_torque}

In MOND, the mutual gravity between two masses does not necessarily align with their separation if there is an external field \citep[e.g.,][]{Banik_2015}. This creates a non-tidal torque on the MW-M31 system due to objects outside it. Since this is induced by the EFE, we call this the EFT. As described in appendix A of \citetalias{Banik_Ryan_2018}, the tangential component of the relative gravity between two masses due to the EFT can be estimated as
\begin{eqnarray}
	\label{F_tan}
	g_{\text{tan}} ~=~&& \sin \theta \left[ \cos \theta + \frac{4}{5} \sin^2 \theta \cos \left( \mathrm{\pi} q_{_{\text{MW}}} \right) \frac{\tilde{r}}{1 + \tilde{r}^2}\right] \nonumber \\
	&&\times \frac{GMa_{_0}r}{2g_{\text{ext}}\left( r^3 + {r_t}^3\right)} \, , 
\end{eqnarray}
with the ``$-$'' sign changed to ``$+$'' in front of the $4/5 \ldots$ \citep[see section 4.4 of][]{Banik_2020_M33}. $\widetilde{r}$ is the separation relative to that at which the problem becomes EF-dominated:
\begin{eqnarray}
	\tilde{r} ~\equiv~ \frac{r}{r_t} \, , \quad r_t ~=~ \frac{\sqrt{G M a_{_0}}}{Q \, g_{\text{ext}}} \, .
\end{eqnarray}
$\theta$ is the angle between $\bm{d}$ and $\bm{g}_{\text{ext}}$, while $Q = 0.7937$ for an MW mass fraction of $q_{_{\text{MW}}} = 0.3$ (Equation \ref{g_iso}). Note that in appendix A of \citetalias{Banik_Ryan_2018}, the two masses considered were an external perturber (e.g. Cen A) and the whole LG, with the latter placed at the position of the MW or M31 depending on the circumstances. Their derivation leads directly to Equation \ref{F_tan} if instead we take the two masses to be the MW and M31.

The present work improves significantly upon \citetalias{Banik_Ryan_2018} in that the EFT is directly included in \textsc{por} (Section \ref{Including_the_EFE}). However, this still relies on knowing the appropriate value of $\bm{g}_{\text{ext}}$ and its even more uncertain time dependence. We can use Equation \ref{F_tan} to estimate the significance of these factors. Since the torque is small for an isolated system ($\widetilde{r} \ll 1$), it is mainly important when the MW and M31 are close to apocentre. If we use $r = 900$ kpc and integrate Equation \ref{F_tan} over a 5 Gyr period (as in Section \ref{Tidal_torques}) assuming constant $g_{\text{tan}}$, we get a tangential impulse of $\Delta v_{t, 783} = 33.6$ km/s for $g_{\text{ext}} = 0.03 \, a_{_0}$, including also a factor of 900/783 to account for changes in angular momentum due to the EFT having a greater impact on the present tangential velocity of M31 if it is currently closer to us. If instead we assume that $g_{\text{ext}} = 0.02 \, a_{_0}$, then $\Delta v_{t, 783}$ is only 25.9 km/s. This difference of $\approx 8$ km/s is also apparent at other separations, e.g. if we use $r = 600$ kpc, we get that $\Delta v_{t, 783}$ differs by 11 km/s depending on whether $g_{\text{ext}}$ is 2\% or 3\% of $a_{_0}$.

In principle, the functional form of the dependency on $q_{_{\text{MW}}}$ is not known as \citetalias{Banik_Ryan_2018} only had numerical results for the case $q_{_{\text{MW}}} = 0$. This means we can replace the factor of $\cos \left( \mathrm{\pi} q_{_{\text{MW}}} \right)$ by any function with value 1 when $q_{_{\text{MW}}} = 0$ and which is anti-symmetric with respect to $q_{_{\text{MW}}} \to 1 - q_{_{\text{MW}}}$. However, due to the low value of $\sin^2 \theta = 0.167$, these results hardly change if instead we use a mass ratio dependency of $\left(1 - 2 q_{_{\text{MW}}} \right)$ or $\cos^3 \left( \mathrm{\pi} q_{_{\text{MW}}} \right)$ instead of $\cos \left( \mathrm{\pi} q_{_{\text{MW}}} \right)$.

\begin{figure}
	\centering
	\includegraphics[width = 8.5cm] {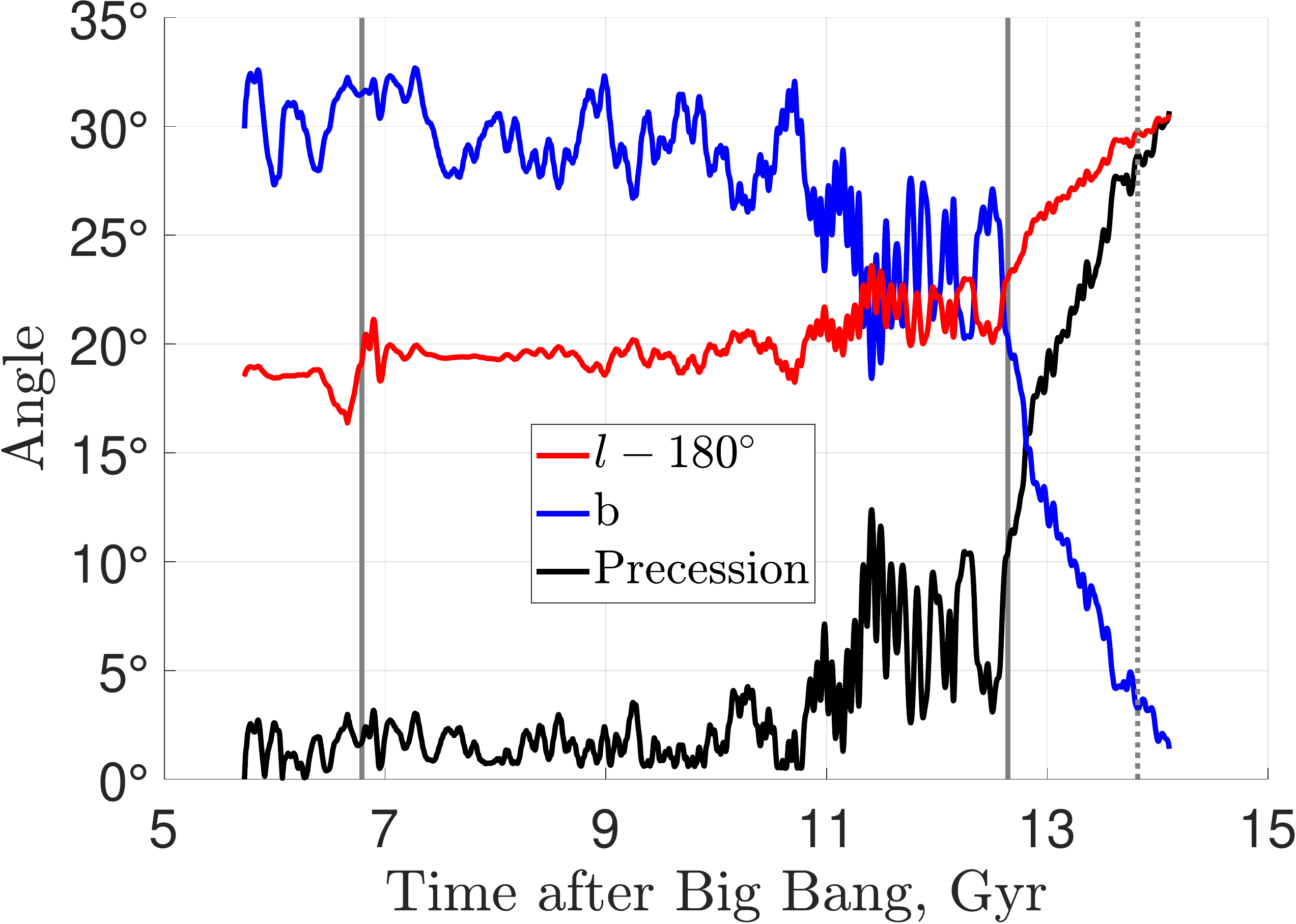}
	\caption{The MW-M31 orbital pole in our \textsc{por} simulation, shown as Galactic latitude (blue) and Galactic longitude less $180^\circ$ (red). The deviation from the initial orbital pole is shown using a solid black line. The vertical solid grey lines around 7 Gyr and 13 Gyr show the times of perigalacticon and the most recent apogalacticon, respectively, while the dotted grey line shows the present time of 13.82 Gyr. Notice the lack of precession around pericentre. Combined with the magnitude of the MW-M31 angular momentum hardly changing then (bottom panel of Figure \ref{MW_M31_trajectory}), it is clear that the MW-M31 interaction has little torquing effect on their orbit. Their orbital pole precesses more significantly around apocentre due to the EFT (Section \ref{External_field_torque}). The present orbital pole direction $\left( 209.77^\circ, \, 3.29^\circ \right)$ is very similar to that of the M31 SP.}
	\label{MW_M31_hdir}
\end{figure}
%Present time is 13.818582189959558 Gyr after the Big Bang.

Since $r$ around the time of apocentre typically has some intermediate value between 600 kpc and 900 kpc (Figure \ref{MW_M31_trajectory}), we estimate that uncertainty in the behaviour of $g_{\text{ext}}$ causes an $\approx 10$ km/s uncertainty in the PM of M31. This would not much alter the present MW-M31 orbital pole because $v_{t, 783} = 34.12$ km/s in our best-fitting \textsc{por} model. Indeed, Figure \ref{MW_M31_hdir} shows that $\widehat{\bm{h}}$ does not change much due to the EFT already included in \textsc{por}, so uncertainties in the EFT should have only a small effect. Interestingly, $\widehat{\bm{h}}$ also hardly precesses due to the flyby, which is indicated with a vertical solid grey line at pericentre (the later solid grey line represents apocentre). As a result, the final $\widehat{\bm{h}}$ is fairly similar to its initial orientation. The more rapid precession of $\widehat{\bm{h}}$ after the most recent apocentre is caused by the MW-M31 orbit becoming almost radial (notice the low value of $h$ at that time in Figure~\ref{MW_M31_trajectory}). It could also be related to tidal debris falling back onto the MW (Section~\ref{Eccentricity}).

Uncertainty in the EFT is mitigated by the geometric factors being well known: the MW-M31 line at apocentre must be quite close to its presently observed direction as they are on a nearly radial orbit \citep{Van_der_Marel_2012, Van_der_Marel_2019, Salomon_2021}. $\bm{g}_{\text{ext}}$ is also constrained by the observed motion of the LG with respect to the CMB/surface of last scattering \citepalias[section 2.2 of][]{Banik_Ryan_2018}. Since the EF on the LG mostly arises from rather distant sources, the direction of $\bm{g}_{\text{ext}}$ would not have changed much in the last 5 Gyr. As a result, the EFT at late times can only affect the PM of M31 along one particular direction. However, it is not too useful to speculate further about this because the PM of M31 still has an uncertainty $\gg 10$ km/s (Figure \ref{Proper_motion}).

\subsection{Tidal debris}
\label{Tidal_debris}

We now turn to the distribution of tidal debris around the MW and M31 disc remnants. One of the main goals of our \textsc{por} simulations is to check whether the tidal debris around each galaxy prefers a particular orbital pole, and if so, to compare this with the observed orbital pole of its SP. This requires us to define a `satellite region' around each disc, which we take to be within 250 kpc of the barycentre found in Section \ref{Proper_motion_section} and $> z_{\text{max}} = 50$ kpc from the disc plane. For simplicity, we assume that the orientation of each disc matches observations $-$ we show later that this is a fairly good assumption (Section \ref{Disc_remnants}) due to iterative adjustments to the initial orientation of each disc (Section~\ref{Disc_parameters_section}). Even if the actual disc orientations differ somewhat, our choice of $z_{\text{max}}$ should completely exclude the disc. However, only a small portion of the SPs would be lost as the satellite galaxies of the MW and M31 go out much further \citep[see, e.g., figure 1 of][]{Pawlowski_2018}. We note that using a spherical excluded region for the disc does not work very well since e.g. excluding the central 30 kpc still leaves a considerable amount of material close to the disc plane at larger radii. This makes it very difficult to clearly disentangle the disc and SP, at least unless a much larger inner radius is considered $-$ which then loses much of the satellite region. We therefore consider only a disc-shaped excluded region in what follows, or alternatively focus on precisely this region when analysing the disc remnants (Section \ref{Disc_remnants}). Appendix \ref{Appendix_zmax} shows the effect of varying $z_{\text{max}}$.

\subsubsection{Orbital pole distribution and mass}
\label{Debri_h_hat_distribution}

\begin{figure*} %17.5cm to go across page.
	\centering
	\includegraphics[width = 8.5cm] {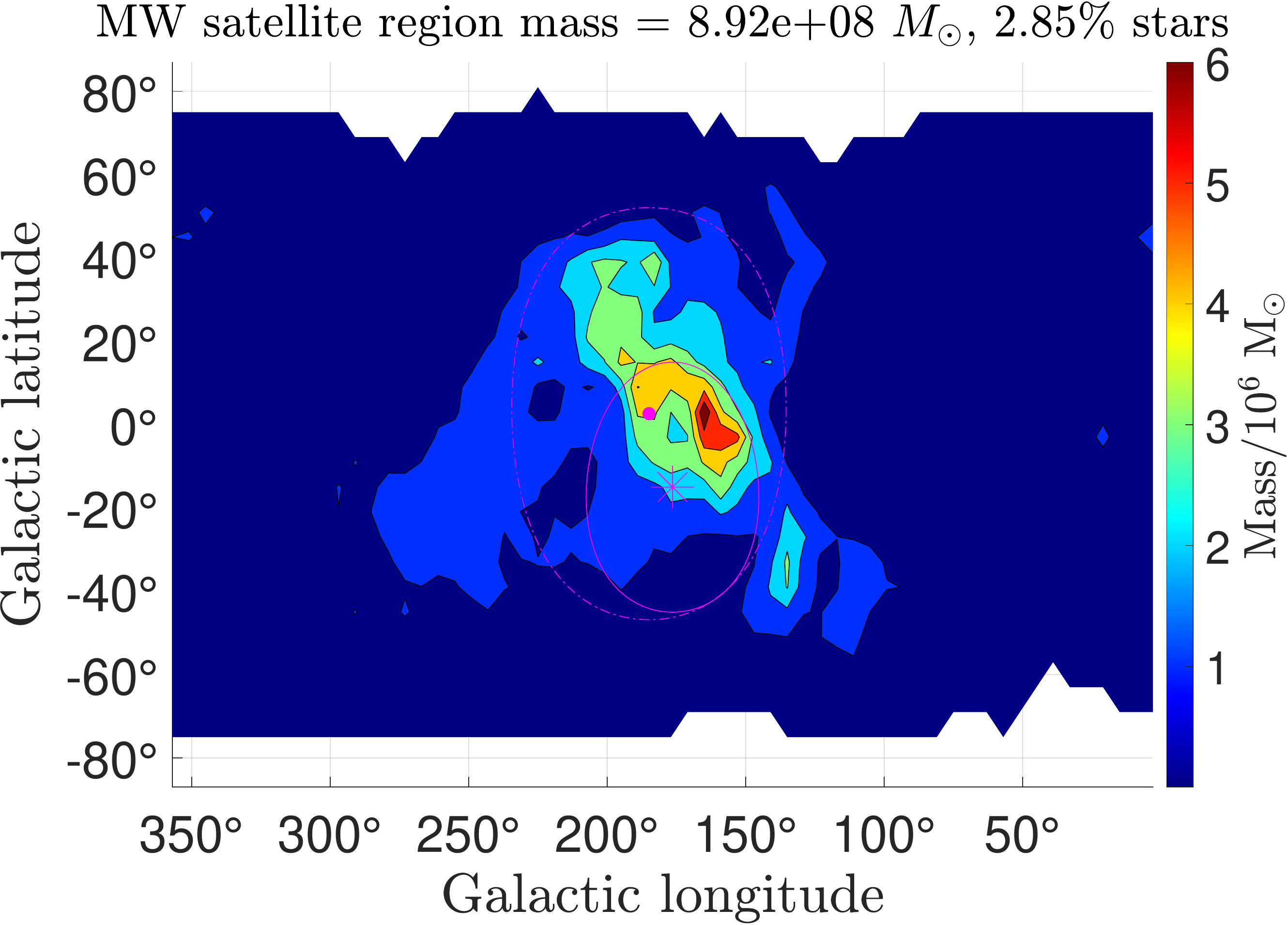}
	\hfill
	\includegraphics[width = 8.5cm] {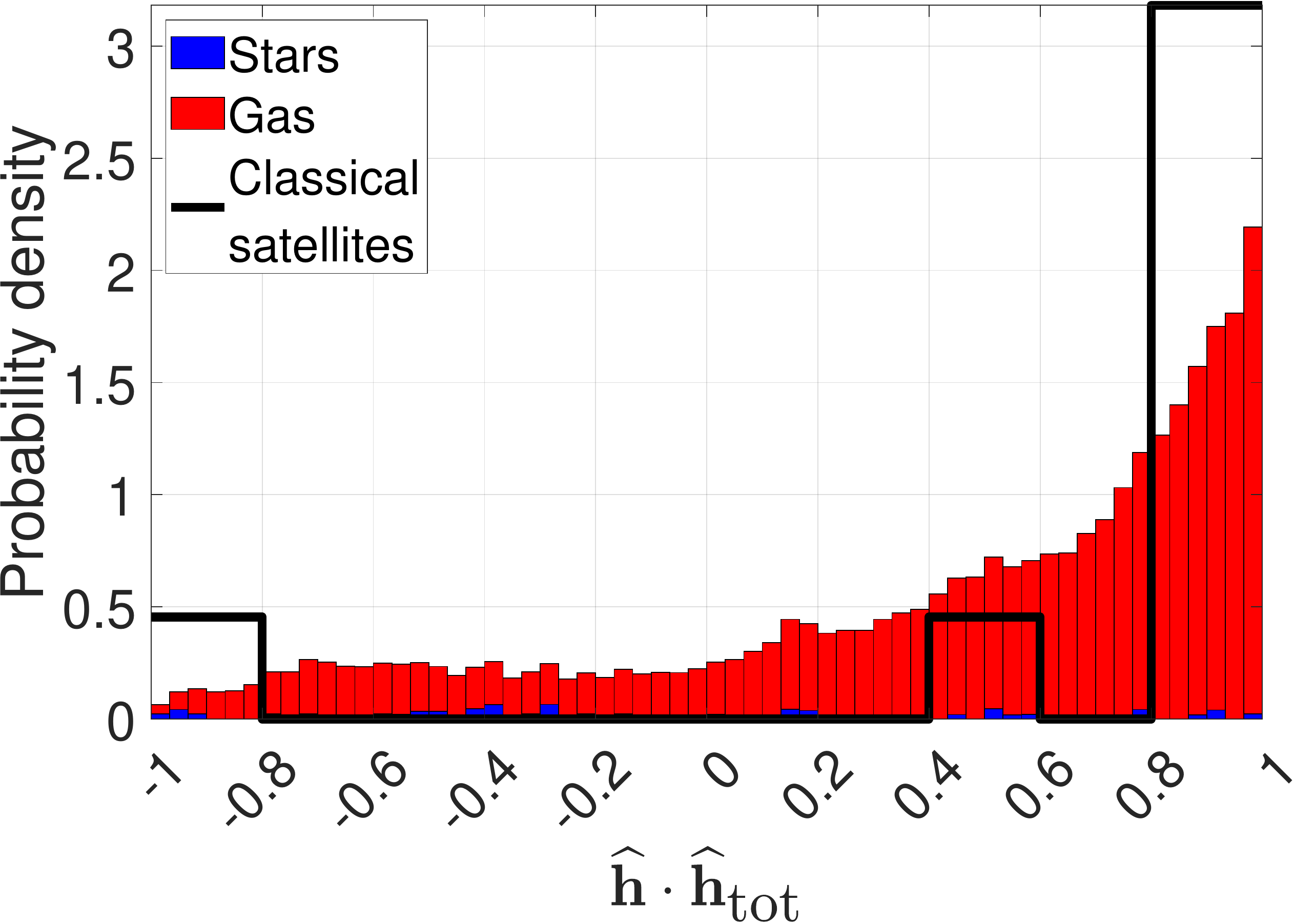}
	\caption{\emph{Left}: The orbital pole distribution of material in the MW satellite region in the 8.2 Gyr snapshot of our \textsc{por} simulation, shown in Galactic coordinates. The title shows the total mass in the satellite region and the fraction of this in stars. The colour of each pixel indicates the mass in a square of $\left(l, b \right)$ with sides of length $6^\circ$. The pink star shows the observed orbital pole of the Galactic SP (Table \ref{SP_observed_orientations}). The solid pink curve shows a cone around this direction with opening angle of $30^\circ$. The pink dot shows the preferred orbital pole in the simulation (Equation~\ref{h_tot_SP}), with the dot-dashed pink curve showing the estimated orbital pole dispersion (Equation~\ref{theta_rms}). The lack of material at very low and high Galactic latitudes is caused by excluding material within $z_{\text{max}} = 50$ kpc of the disc plane. Results appear similar for a different $z_{\text{max}}$ (Appendix \ref{Appendix_zmax}). \emph{Right}: The red bars show the cosine distribution of the angle between the orbital pole $\widehat{\bm{h}}$ and $\widehat{\bm{h}}_{\text{tot}}$, the preferred orbital pole of the simulated SP (Equation \ref{h_tot_SP}). The open black bars show the orbital pole distribution of classical MW satellites at Galactocentric distances of $50-250$~kpc based on the positions and orbital poles listed in table~2 of \citet{Pawlowski_2020}. The angles shown here are relative to the observationally preferred direction listed in Table~\ref{SP_observed_orientations}.}
	\label{MW_SP_50}
\end{figure*}

\begin{figure*}
	\centering
	\includegraphics[width = 8.5cm] {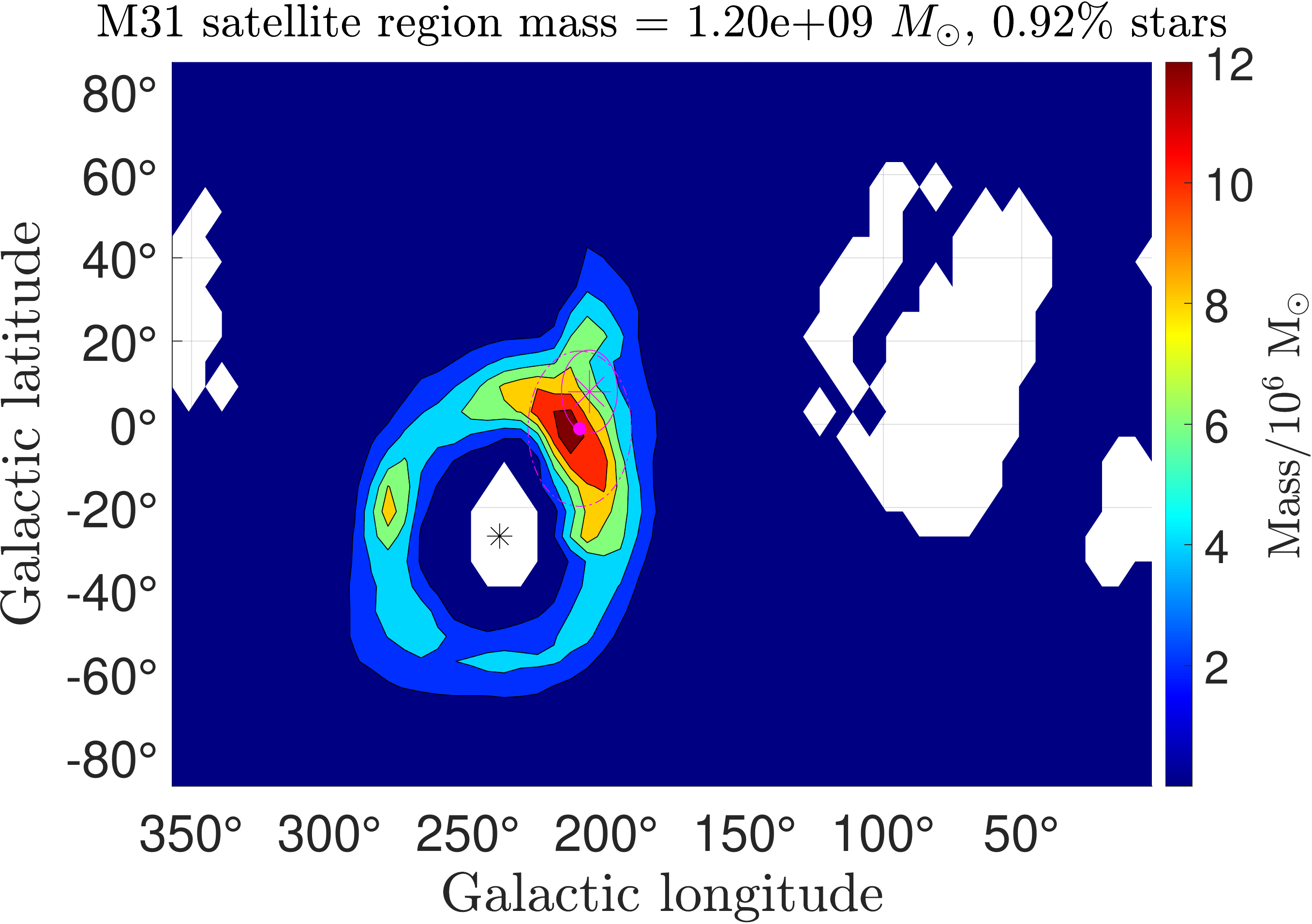}
	\hfill
	\includegraphics[width = 8.5cm] {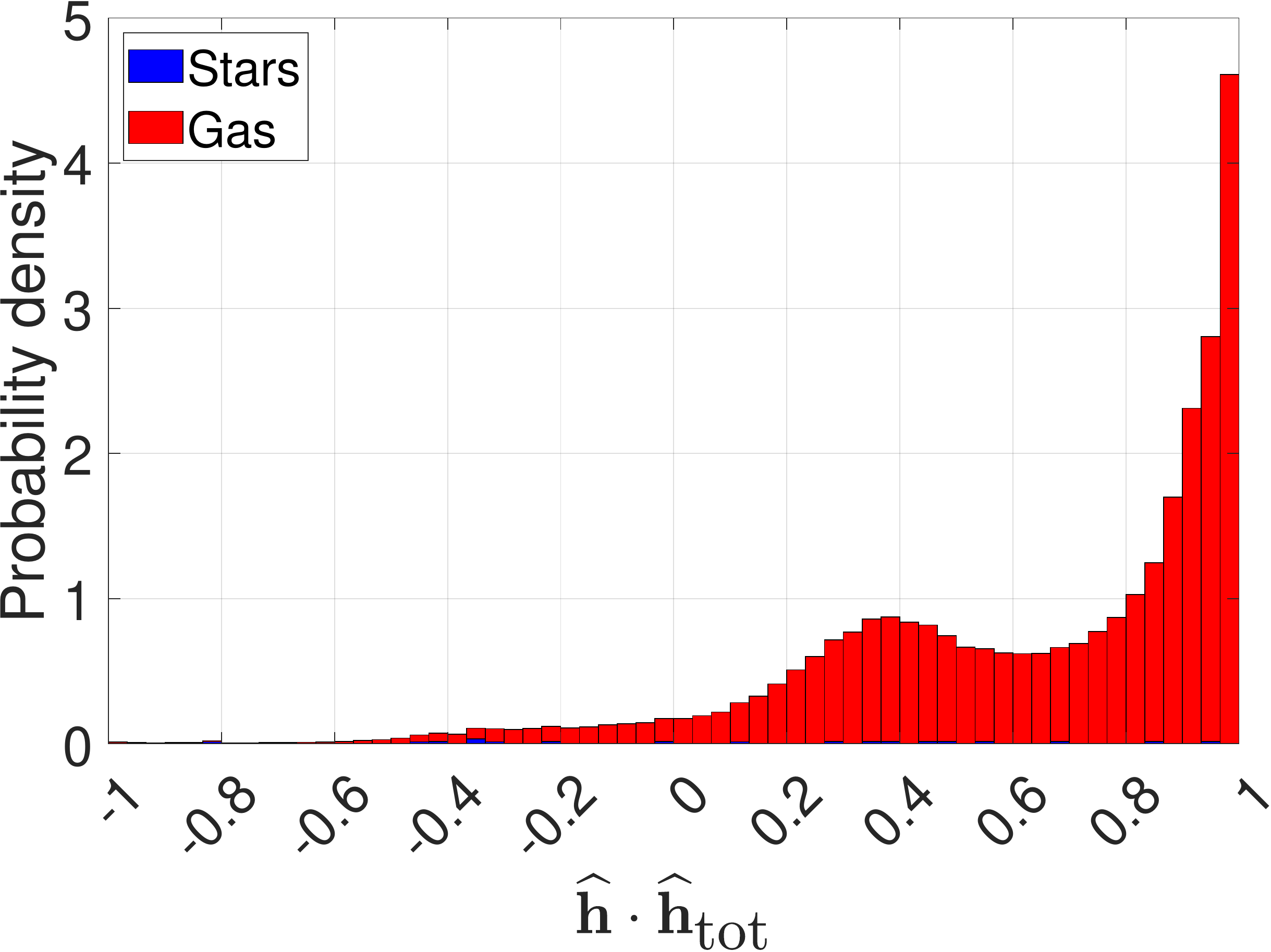}
	\caption{Similar to Figure \ref{MW_SP_50}, but showing results for M31. Its disc spin vector is shown as a black star. The lack of material with an orbital pole close to this direction or its opposite is caused by the exclusion of material close to the M31 disc plane. Notice the smaller simulated disc-SP misalignment compared to the MW, which is likely due to M31 having a higher mass and a shorter scale length for the outer disc (Table \ref{Disc_parameters}). The simulated and observed orbital pole dispersions are also much smaller than for the MW. Results appear similar for a different $z_{\text{max}}$ (Appendix \ref{Appendix_zmax}).}
	\label{M31_SP_50}
\end{figure*}

\begin{table}
	\centering
	\caption{The observed orbital pole of the MW SP \citep[section 3 of][]{Pawlowski_2013_VPOS} and of the M31 SP \citep[section 4 of][]{Pawlowski_2013_LG} in Galactic coordinates. We also show the angle between the spin vector of each SP and that of its parent disc, and the angle between the SP spin vectors. The MW-M31 orbital pole (shown in Figure \ref{MW_M31_hdir}) aligns fairly closely with that of the simulated M31 SP (shown in Figure~\ref{M31_SP_50}).}
	\begin{tabular}{ccc}
		\hline
		Galaxy & MW & M31 \\
		\hline
		SP spin vector & $\left(176.4^\circ, \, -15.0^\circ \right)$ & $\left(206.2^\circ, \, 7.8^\circ \right)$ \\
		Disc-SP misalignment & $75^\circ$ & $47^\circ$ \\
		Angle between SP spins & \multicolumn{2}{c}{$37^\circ$} \\
		\hline
	\end{tabular}
	\label{SP_observed_orientations}
\end{table}

We find the angular momentum of all particles and gas cells in the satellite region of each galaxy relative to the galaxy's barycentre. The important quantity for us is the direction of this angular momentum, which we use to build up a distribution in Galactic coordinates. The procedure is the same as that described in section 4.1 of \citetalias{Banik_Ryan_2018}. The resulting orbital pole distribution is shown in Figure \ref{MW_SP_50} for the MW and in Figure \ref{M31_SP_50} for M31. The observed orbital poles of the MW and M31 SPs (summarized in Table \ref{SP_observed_orientations}) are obtained from section 3 of \citet{Pawlowski_2013_VPOS} and section 4 of \citet{Pawlowski_2013_LG}, respectively. These are shown as pink stars on the left-hand panels, which our \textsc{por} model matches fairly well. In Appendix \ref{Appendix_zmax}, we show that the appearance remains quite similar if we alter $z_{\text{max}}$ to 40 kpc or 60 kpc.

An important aspect of our analysis is an estimate for the dispersion in orbital pole directions. On the observational side, we use $30^\circ$ for the MW based on section 4 of \citet{Pawlowski_2013_VPOS}. For M31, we note that its SP has an aspect ratio $3\times$ smaller than for the MW \citep[table 3 of][]{Pawlowski_2013_LG}. We therefore adopt an orbital pole dispersion of $10^\circ$ for M31. These dispersions are illustrated by drawing a cone with an opening angle equal to the estimated dispersion and an axis aligned with the observed SP orbital pole direction listed in Table \ref{SP_observed_orientations}. These cones are shown on the left-hand panels of Figures \ref{MW_SP_50} and \ref{M31_SP_50} using solid pink curves.

The dashed pink curves on these figures show analogous results for the simulated tidal debris, whose preferred orbital pole $\widehat{\bm{h}}_{\text{tot}}$ is shown with a pink dot in each case. We estimate this using an iterative procedure where the initial guess is the centre of the pixel in $\left(l, b \right)$ with the most mass. We then find
\begin{eqnarray}
    \widehat{\bm{h}}_{\text{tot}} ~\propto~ \sum_i m_i \widehat{\bm{h}}_i \, ,
    \label{h_tot_SP}
\end{eqnarray}
where each particle or gas cell in the satellite region has mass $m_i$ and orbital pole direction $\widehat{\bm{h}}_i$ relative to its host galaxy. The sum is taken over only those particles whose $\widehat{\bm{h}}_i$ aligns with $\widehat{\bm{h}}_{\text{tot}}$ to better than $30^\circ$ for M31 or $90^\circ$ for the MW, i.e. $3\times$ the observed orbital pole dispersion in both cases. This restriction causes $\widehat{\bm{h}}_{\text{tot}}$ to influence which particles and gas cells contribute to the sum, so we need to repeat the process a few times until convergence is reached. We find that only a handful of iterations are required to reach convergence in $\widehat{\bm{h}}_{\text{tot}}$ to within machine precision. We then calculate the orbital pole dispersion $\theta_{\text{rms}}$ using
\begin{eqnarray}
    \theta_{\text{rms}} ~=~ \sqrt{\frac{\sum_i m_i {\theta_i}^2}{\sum_i m_i}} \, ,
    \label{theta_rms}
\end{eqnarray}
where $\theta_i \equiv \cos^{-1} \left( \widehat{\bm{h}}_i \cdot \widehat{\bm{h}}_{\text{tot}} \right)$ is the angle between $\widehat{\bm{h}}_{\text{tot}}$ and the orbital pole of particle $i$. The sum is again taken over only those particles or gas cells in the satellite region whose $\widehat{\bm{h}}_i$ is within the above-mentioned cone around $\widehat{\bm{h}}_{\text{tot}}$. The so-obtained orbital pole dispersion is $\theta_{\text{rms}} = 49^\circ$ for the MW and $19^\circ$ for M31, so our model naturally yields a lower $\theta_{\text{rms}}$ around M31. The simulated dispersions are larger than the observed ones, which we ascribe to the somewhat high 372~kK temperature floor of our \textsc{por} simulations due to resolution limitations. It is also likely that individual TDGs would form only in the densest regions, leading to a narrower spread of orbital poles than for the tidal debris considered as a whole.

The left-hand panels of Figures \ref{MW_SP_50} and \ref{M31_SP_50} reveal the expected gap in the orbital pole distribution around the disc spin vector and the opposite direction arising from our definition of the satellite region. We clarify this in the M31 case by displaying its observed disc spin vector $\left(238.65^\circ, \, -26.89^\circ \right)$ as a black star in Figure \ref{M31_SP_50} \citep[section 2.1 of][]{Banik_2018_anisotropy}. This is omitted for the MW because by definition its disc spin points towards the south Galactic pole, leading to a lack of material at very low and high Galactic latitudes in Figure \ref{MW_SP_50}.

The right-hand panels of these figures show the mass-weighted distribution of $\cos \theta_i$, which should be uniform for a completely isotropic distribution. In the MW case, we use open black bars to show the observed distribution for classical satellites at Galactocentric distances of $50-250$~kpc \citep[table 2 of][]{Pawlowski_2020}. The result is similar to what we obtain in our \textsc{por} simulation: both the model and observations show a significant clustering of orbital poles, though observationally the clustering is somewhat tighter. This could be related to the resolution and temperature floor of our model. The orbital poles are much more clustered for the tidal debris around M31 than for the MW, which as explained earlier is in line with observations as the M31 SP is much thinner than that of the MW.

The titles of Figures \ref{MW_SP_50} and \ref{M31_SP_50} indicate that the satellite region of each galaxy has a mass of $\approx 10^9 M_\odot$. Since we expect the SP material to mostly have been quite far out initially \citepalias[section 5.2.4 of][]{Banik_Ryan_2018}, our estimated SP masses are quite sensitive to the reliability of extrapolating the assumed exponential disc law to large radii. The relatively small amount of material here means that altering the initial mass distribution at large radii would hardly affect the gravitational field, leaving the simulated SP orientations unchanged. Therefore, the SP masses are not a strong test of our model. Bearing this in mind, we note that an SP mass of $10^9 M_\odot$ is reasonable for the MW because the LMC dominates the baryonic mass in the Galactic SP. The RC of the LMC has a flatline level of $\approx 70$ km/s \citep{Alves_2000, Van_der_Marel_2014, Vasiliev_2018}, which in a MOND context implies a mass of $1.5 \times 10^9 M_\odot$. While this is in reasonable agreement with our model, another important aspect of the LMC is its rather high specific angular momentum $h$ \citep{Kallivayalil_2013}. This is quite difficult to reproduce in our model, with only a small fraction of the tidal debris having a higher $h$ than that of the LMC (Section~\ref{LMC_analogues}). It is difficult to solve this problem by postulating a much larger amount of tidal debris as then the Galactic disc would need to be damaged much more significantly. Instead, the problem may lie with resolution and the temperature floor: colder gas would need more support from rotational motion, which should lead to higher $h$ initially. It is not presently clear whether this will solve the problem of the LMC, so for the time being its properties remain somewhat problematic for our flyby scenario.

In the M31 case, the simulated SP mass suggests that M32 is likely part of this structure \citepalias[see section 5.2.3 of][]{Banik_Ryan_2018}, and/or that all of the mass in the SP has not condensed into individual satellites. This is quite possible given that the orbital pole distribution is a non-uniform ring rather than a single point (Figure \ref{M31_SP_50}). It could well be that only in the densest part of this ring was the density high enough for the gas to undergo Jeans collapse into TDGs. Moreover, gas accreted onto a newly formed TDG could be subsequently expelled by feedback, with perhaps only a small fraction ending up in stars.

To address these issues more thoroughly, a higher resolution simulation would be required in which bound satellite galaxies form out of the tidal debris, which is beyond the scope of this project. A previous high-resolution hydrodynamical MOND simulation of interacting disc galaxies suggests that TDGs should form out of the tidal debris \citep{Renaud_2016}. Their work focused on the Antennae \citep{Mirabel_1992}, but our work indicates that a similar process could well have played out in the much better observed LG. We hope that the initial conditions of our best-fitting model (Table \ref{Initial_conditions}) serve as a starting point for further work on the LG in MOND.

\subsubsection{Radial distribution}
\label{Debri_r_distribution}

\begin{figure}
	\centering
	\includegraphics[width = 8.5cm] {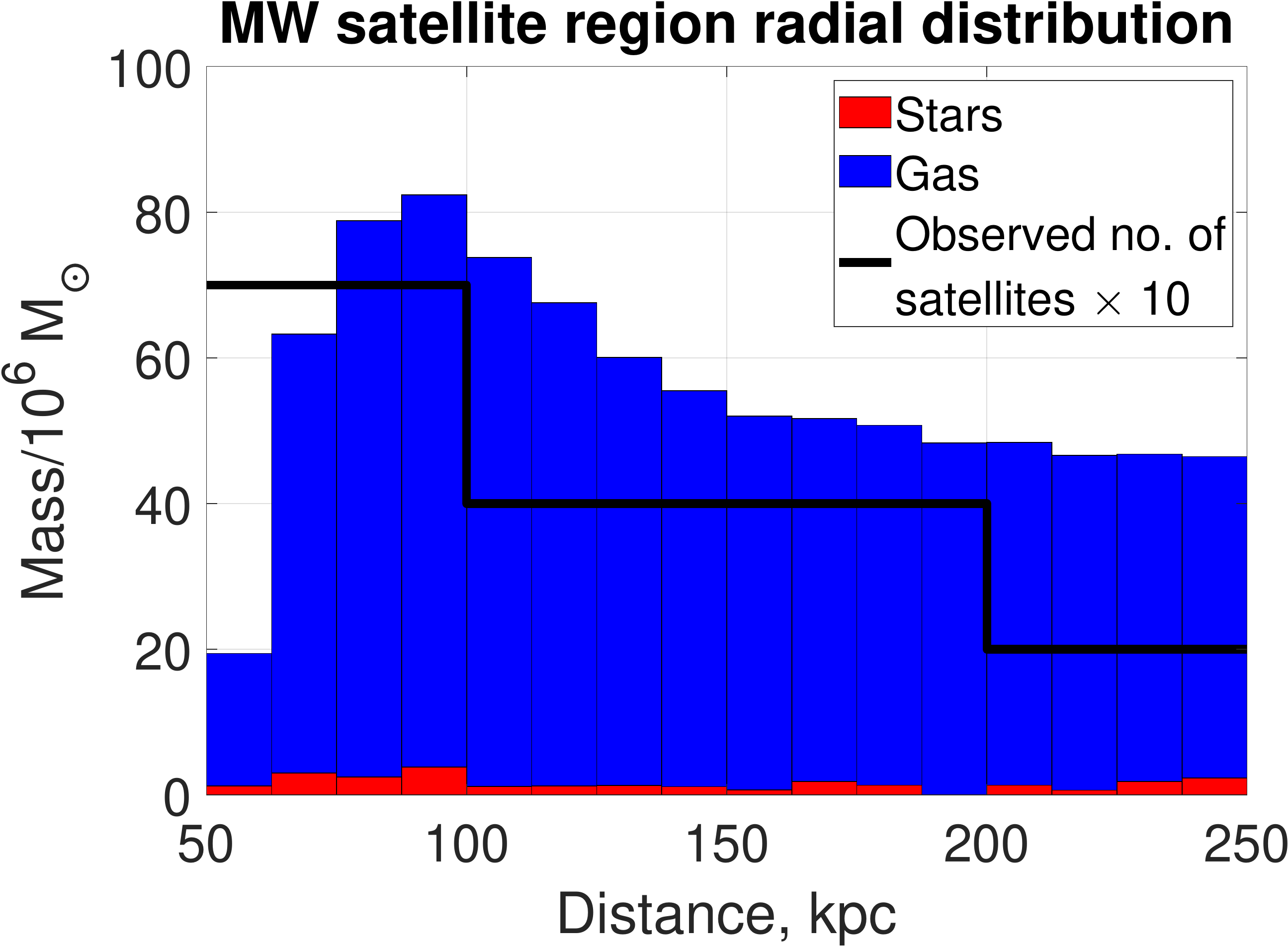}
	\includegraphics[width = 8.5cm] {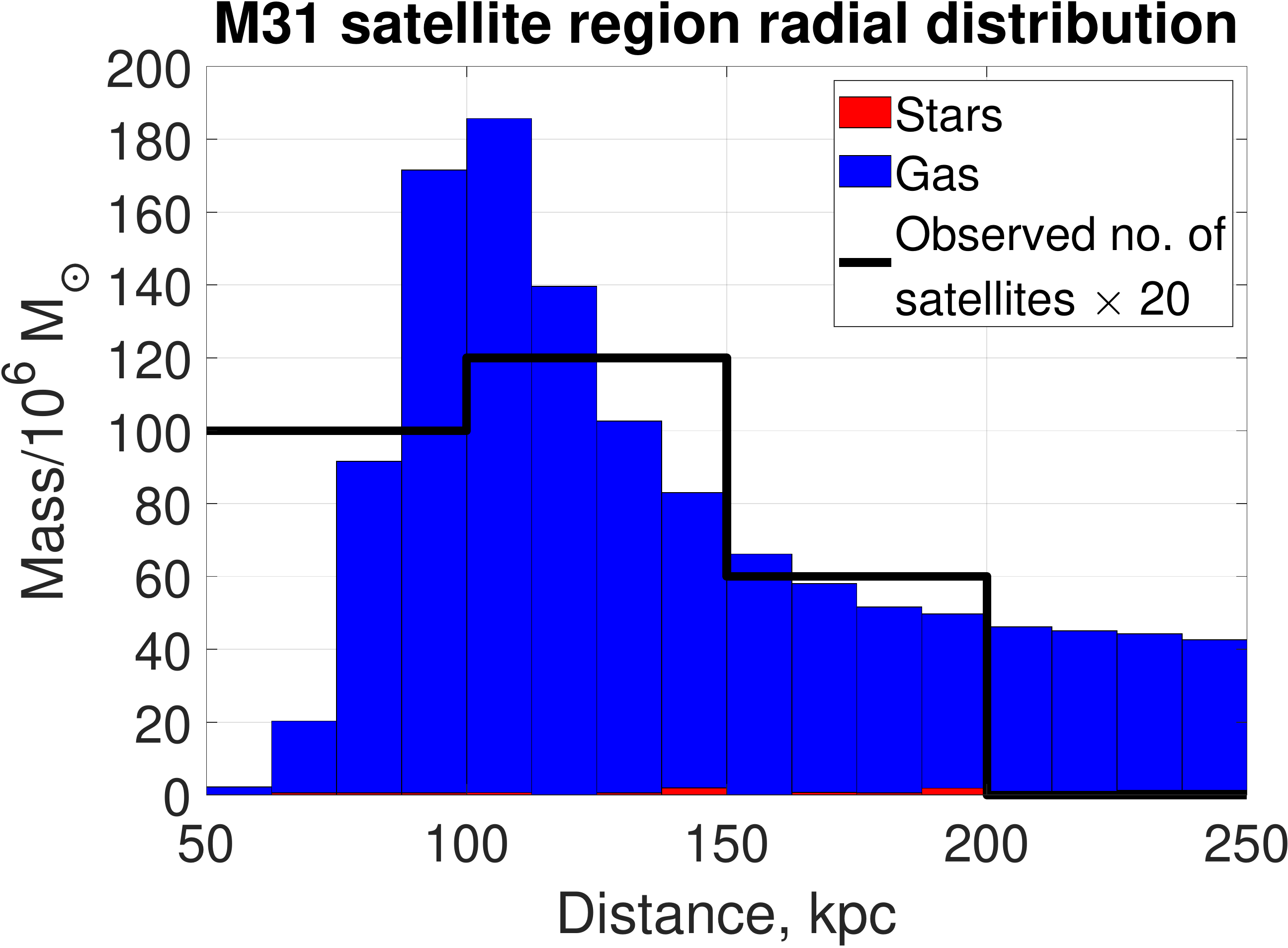}
	\caption{The radial distribution of material in the MW (\emph{top}) and M31 (\emph{bottom}) satellite regions, showing contributions from stars (red) and gas (blue). The observed satellite distribution is shown using a stepped solid black line representing the number of satellites in four equally wide bins covering $50-250$~kpc. These were obtained from the bold entries in table 4 of \citet{Pawlowski_2013_LG} and scaled for clarity as indicated in the legend. Our \textsc{por} simulation suggests that the SPs would initially have been dominated by gas, though this need not be the case today as we do not simulate star formation and feedback.}
	\label{Satellite_region_radial_distribution}
\end{figure}

We use the top and bottom panel of Figure \ref{Satellite_region_radial_distribution} to show the radial distribution of material in the satellite region of the simulated MW and M31, respectively. Each bar is divided into a red part indicating stars and a blue part indicating gas. It is clear that the satellite regions are completely dominated by gas. For the MW, this might be caused by its initial distribution of gas being more extended than that of its stars (Table \ref{Disc_parameters}). However, the stars and gas in M31 have the same initial surface density profile. The dominance of gas in the satellite region could indicate that this is more easily removed from the disc than stars due to ram pressure effects.

Our results can tentatively be compared with the observed distances of LG SP members from their host galaxy. We obtain these from the bold entries in table 4 of \citet{Pawlowski_2013_LG}, restricting further to only those satellites within $50-250$~kpc of their parent galaxy. These observational results are shown in Figure~\ref{Satellite_region_radial_distribution} using a solid black line. There is good overall agreement with our model, but observationally there is a lack of observed satellites $200-250$ kpc from their host. This could be due to the tidal debris at such large distances being too sparse to form TDGs. Selection effects could also play a role, especially around the MW where more distant satellites would be fainter. While this effect is less significant around M31, the limited size of the Pan-Andromeda Archaeological Survey \citep{Ibata_2014_PANDAS, McConnachie_2018} means that observations of M31 do not extend to galactocentric distances much beyond 150 kpc.

Our simulations do not allow star formation, so the high gas fraction in the satellite regions merely indicates that the flyby created an initially gas-rich distribution of tidal debris. In a more advanced simulation, the gas may well condense into a small number of TDGs which then form stars \citep{Renaud_2016}. While the formation of TDGs and their internal evolution is beyond the scope of this project, their overall distribution should be similar in a higher resolution simulation that allows star formation. In this respect, it is interesting that the radial distribution of simulated material in the satellite region of each galaxy broadly agrees with that of its actual SP members.

\subsubsection{Orbital eccentricity}
\label{Eccentricity}

An important aspect of the LG satellite planes is their small dispersion of orbital pole directions \citep{Pawlowski_2021}. Another aspect to consider is the motions of satellite galaxies within their preferred plane. To quantify this, we determine the position $\bm{r}$ and velocity $\bm{v}$ of each particle or gas cell relative to its parent galaxy (Section~\ref{Proper_motion_section}). We then find the unit vectors $\widehat{\bm{r}}$ and $\widehat{\bm{v}}$, where $\widehat{\bm{n}} \equiv \bm{n}/n$ for any vector $\bm{n}$. The distribution of the angle between $\widehat{\bm{r}}$ and $\widehat{\bm{v}}$ provides a measure of whether the velocity dispersion tensor is isotropic or tangentially biased, and also whether there is any net radial motion that shows up as an asymmetry between $\pm \widehat{\bm{r}} \cdot \widehat{\bm{v}}$.

\begin{figure}
	\centering
	\includegraphics[width = 8.5cm] {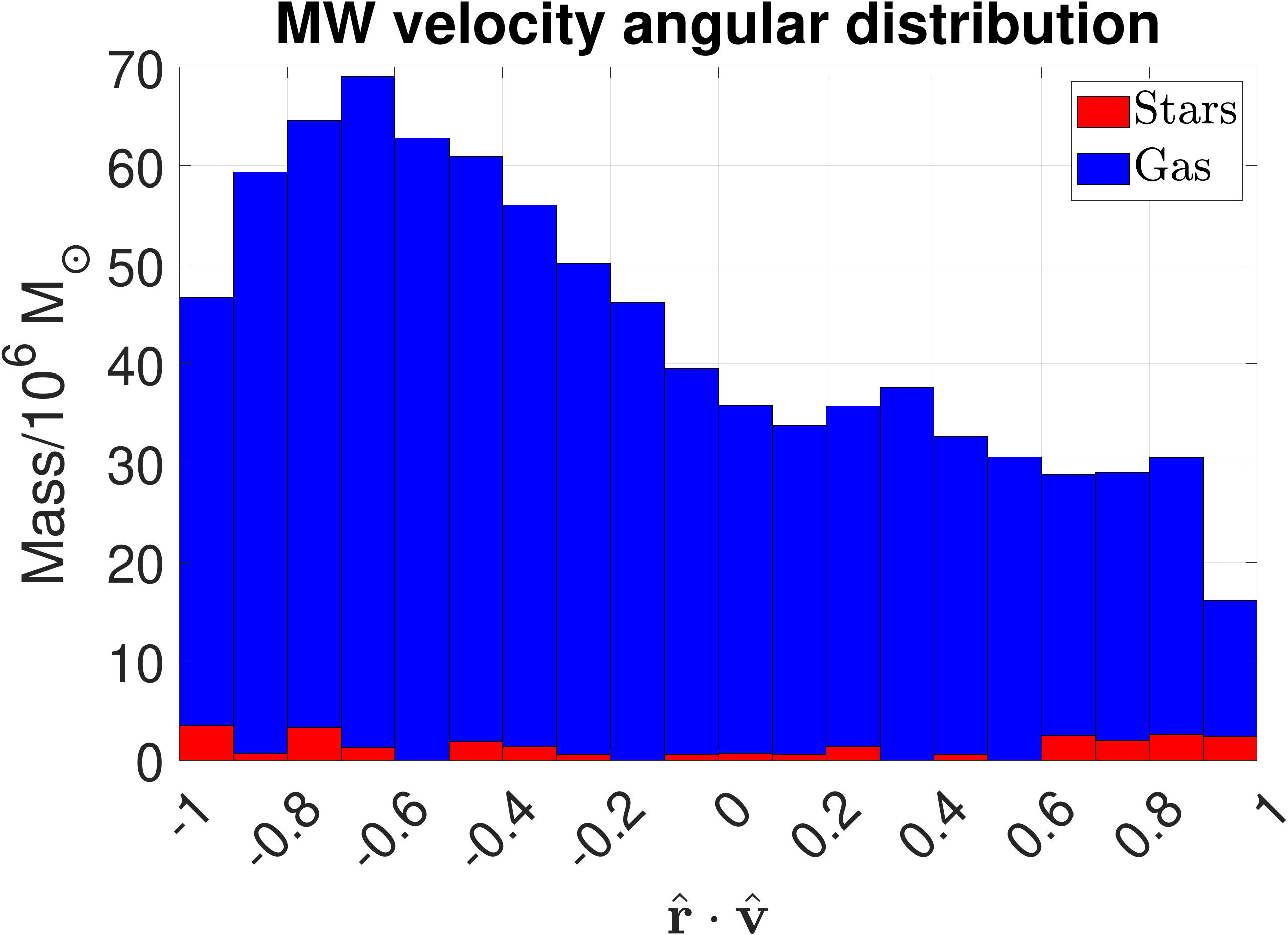}
	\includegraphics[width = 8.5cm] {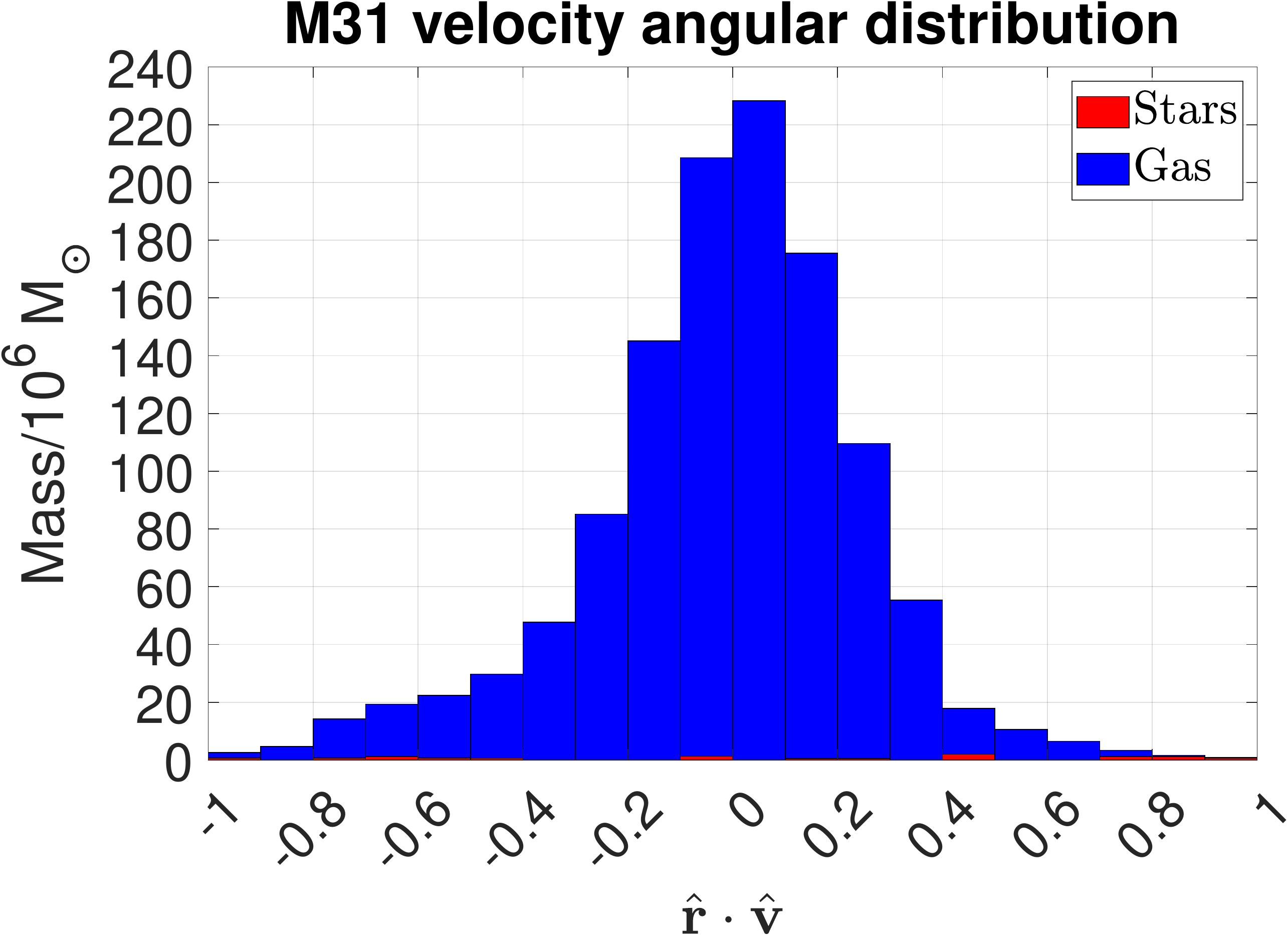}
	\caption{The cosine distribution of the angle between the position and velocity of each particle relative to its host galaxy, shown for the MW (\emph{top}) and M31 (\emph{bottom}) satellite regions. Contributions are shown from stars (red) and gas (blue). If the velocities were oriented randomly relative to the radial direction, then the distribution would be uniform. The asymmetry between $\pm \widehat{\bm{r}} \cdot \widehat{\bm{v}}$ for the MW indicates that material is still falling towards it. Notice the strong preference for tangential motion in the M31 case.}
	\label{Eccentricity_distribution}
\end{figure}

We use Figure~\ref{Eccentricity_distribution} to plot the distribution of $\widehat{\bm{r}} \cdot \widehat{\bm{v}}$. The top panel shows the result for the MW. If the $\widehat{\bm{v}}$ for each particle or gas cell were distributed randomly relative to the outwards radial direction $\widehat{\bm{r}}$, then $\widehat{\bm{r}} \cdot \widehat{\bm{v}}$ would be distributed uniformly over the range $\left( -1, 1 \right)$. This is a reasonable description of the figure, so there is not much tendency for the orbits to be tangentially biased, somewhat at odds with observations \citep{Cautun_2017, Riley_2019, Hammer_2021}. This is probably due to the high gas temperature floor of 372~kK, which is required given numerical limitations. In this sense, our result for the MW is similar to figure 8 of \citetalias{Banik_Ryan_2018}. One difference is that unlike in their work, our results show an asymmetry between $\pm \widehat{\bm{r}} \cdot \widehat{\bm{v}}$ such that radial inflow is preferred over radial outflow. This indicates that according to our simulation, some tidal debris should should still be falling back towards the MW. This could explain the properties of the Fornax dwarf spheroidal satellite \citep{Yang_2022}.

The bottom panel of Figure~\ref{Eccentricity_distribution} shows the $\widehat{\bm{r}} \cdot \widehat{\bm{v}}$ distribution for M31. There is a sharp peak near 0, indicating a strong preference for tangential motion. This is quite different to the result obtained by \citetalias{Banik_Ryan_2018}, whose figure~8 shows only a mild preference for tangential motion for the tidal debris around M31. The hydrodynamical nature of our simulations is probably the main reason for this difference. Since a higher resolution simulation with a lower temperature floor should have even more dissipation, our \textsc{por} simulation indicates that the M31 SP members should have rather low orbital eccentricities. This is in line with the observed PMs of the only two M31 SP members whose PMs are currently known \citep{Sohn_2020, Pawlowski_Sohn_2021}. There is not much tendency for radial inflow as opposed to outflow: at low values of $\left| \widehat{\bm{r}} \cdot \widehat{\bm{v}} \right| \la 0.5$, there is more material at positive values, indicating a tendency for outflow. However, this trend is reversed at high values of $\left| \widehat{\bm{r}} \cdot \widehat{\bm{v}} \right| \ga 0.5$. Thus, the M31 SP should be more nearly virialized than the MW SP. The strong tendency for tangential motion in the M31 SP also suggests that it should be dynamically colder than the MW SP, thereby having a lower aspect ratio and orbital pole dispersion. This is indeed the case in our model (Equation~\ref{theta_rms}) and observationally \citep[e.g.,][]{Pawlowski_2013_LG}. The full 6D phase space structure of the M31 SP is still unclear because this would require accurate PMs for its relatively faint member satellites, which have only recently become available in two cases \citep{Sohn_2020}. While these are indicative of corotation, a more detailed analysis will need to await further data \citep{Pawlowski_Sohn_2021}.

\subsubsection{Analogues to the LMC}
\label{LMC_analogues}

The phase space structure of the Galactic SP is much better known because of its proximity. This is clearly dominated by the LMC, so a realistic model should have material on an orbit similar to the LMC. To investigate this, we conduct a very similar analysis to section 5.4 of \citetalias{Banik_Ryan_2018}. The LMC is fairly close to pericentre at the moment \citep{Kallivayalil_2013}, which could well be something that even a quite realistic model does not reproduce simply because the considered snapshot time is not when the LMC analogue is at pericentre. Indeed, analogues to the Magellanic Clouds are not very common observationally, so it may be quite difficult to reproduce certain aspects of them even with the correct theory of nature \citep{Robotham_2012}. Therefore, we focus on quantities which should be rather more robust around the whole orbit.

\begin{figure}
	\centering
	\includegraphics[width = 8.5cm] {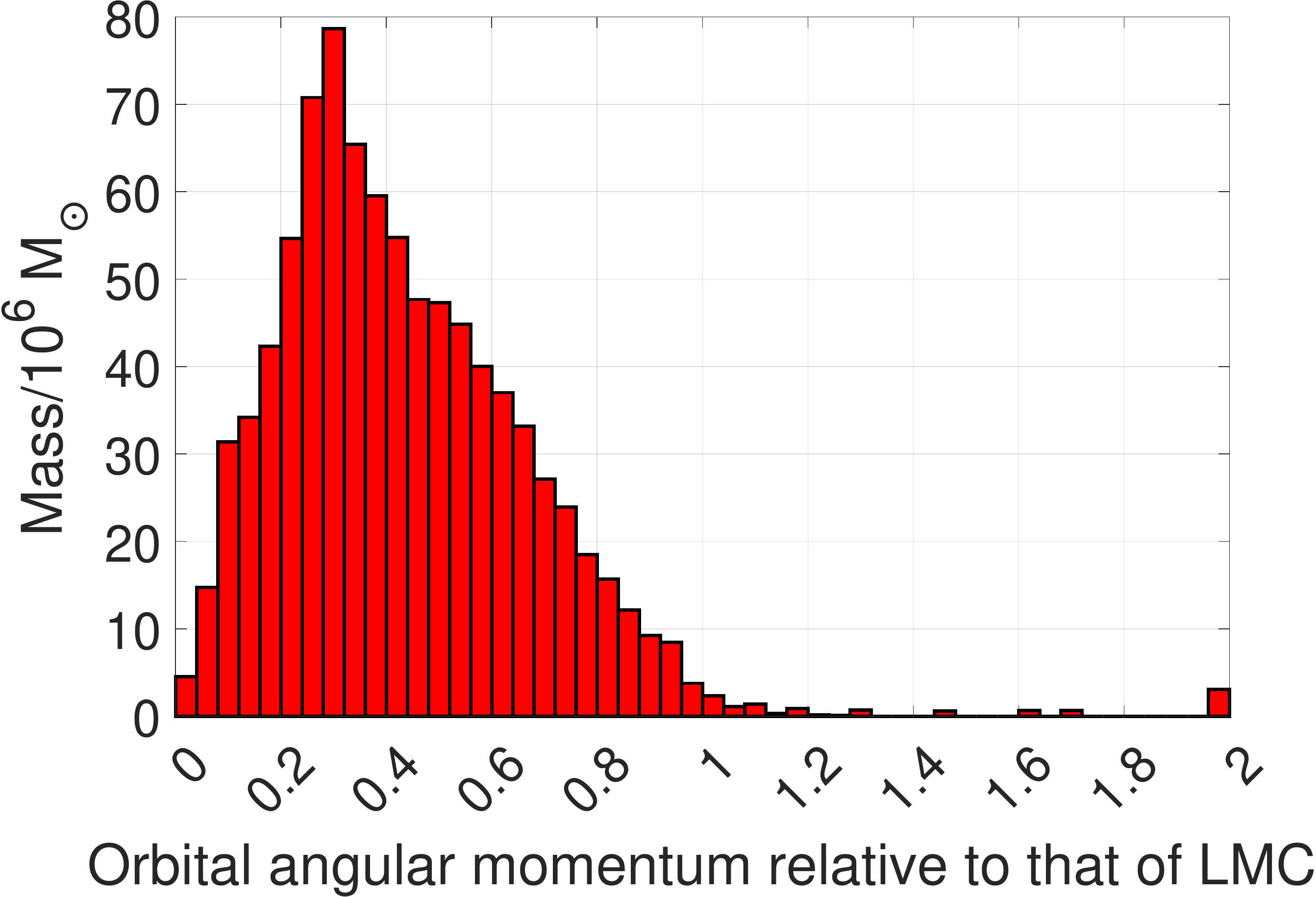}
	\caption{The specific angular momentum distribution of material in the MW satellite region, shown relative to that of the LMC. The rightmost bar includes all material with higher values.}
	\label{MW_satellite_region_h_ratio_LMC}
\end{figure}

The high tangential velocity of the LMC for its position indicates a rather high $h$ in the Galactocentric frame. To explain this, we should have material in the satellite region with a similar $h$. The distribution of $h$ is shown in Figure \ref{MW_satellite_region_h_ratio_LMC}. The LMC value is based on table 5 of \citet{Kallivayalil_2013}, which implies a Galactocentric tangential velocity of $314 \pm 24$ km/s at a distance of $r_{_{\text{LMC}}} = 49.39$ kpc \citep{Pietrzynski_2013}. More recent observations indicate a similar LMC PM \citep{Gaia_2018_Helmi}. It is clear that the high $h$ of the LMC is somewhat challenging for our model given its mass, though the simulated $h$ distribution for material in the MW satellite region does extend beyond the LMC value.

An important aspect of the LMC is its high orbital eccentricity. This is not necessarily captured merely by looking at $h$, because $h$ would be rather high even for material on a circular orbit if it is sufficiently far out. We address this by following equation 51 of \citetalias{Banik_Ryan_2018} in defining an adjusted specific energy
\begin{eqnarray}
	\widetilde{E} ~\equiv~ \Phi \left( r \right) + \frac{1}{2}v^2 - \frac{h^2}{2{r_{_{\text{LMC}}}}^2} \, ,
	\label{E_LMC}
\end{eqnarray}
where the distance $r$ and velocity $v$ are Galactocentric. $\widetilde{E}$ is the specific radial kinetic energy of a particle at $r = r_{_{\text{LMC}}}$, with $\widetilde{E} < 0$ indicating that the particle is not capable of reaching this distance. It would have little difficulty doing so if $r \approx r_{_{\text{LMC}}}$ in the simulation. But if $r \gg r_{_{\text{LMC}}}$, then the particle would need to undergo substantial radial oscillations to ever resemble the LMC at any point along its orbit. If it does so, then ideally it should match the $\widetilde{E}$ value of the LMC, i.e. have an RV of 64 km/s at $r = r_{_{\text{LMC}}}$. To calculate if this is the case, we need to assume some form for the Galactic potential $\Phi \left( r \right)$. We find this by treating the MW as an isolated point mass, neglecting M31 and the EFE. This is justified because the relevant distances are much larger than the extent of the MW disc \citep{Bovy_2013}, but much smaller than the distance to M31 \citep{McConnachie_2012} or the distance beyond which the EFE dominates \citep{Banik_2018_escape}. The form of $\Phi$ follows from integrating the simple interpolating function \citep{Famaey_Binney_2005} for a single point mass using a hyperbolic substitution.
\begin{eqnarray}
	\Phi \left( r \right) &=& \sqrt{GMa_{_0}} \left[ \ln \left(1 + \sqrt{1 + \tilde{r}^2} \right) - \frac{1}{\tilde{r}} - \sqrt{\frac{1}{\tilde{r}^2} + 1} \right] \, , \nonumber \\
	\tilde{r} &\equiv& \frac{2r}{r_{_M}} \, , \quad r_{_M} \equiv \sqrt{\frac{GM}{a_{_0}}} \, .
	\label{Point_mass_potential}
\end{eqnarray}
Here, $M = 9.15 \times 10^{10} \, M_\odot$ is the initial mass of the MW, while $r_{_M}$ is its MOND radius, the distance beyond which MOND effects become significant. Since ${r_{_M} \approx 10}$ kpc, it is not much smaller than the LMC distance, so it is not accurate to assume that the problem is in the deep-MOND limit.

\begin{figure}
	\centering
	\includegraphics[width = 8.5cm] {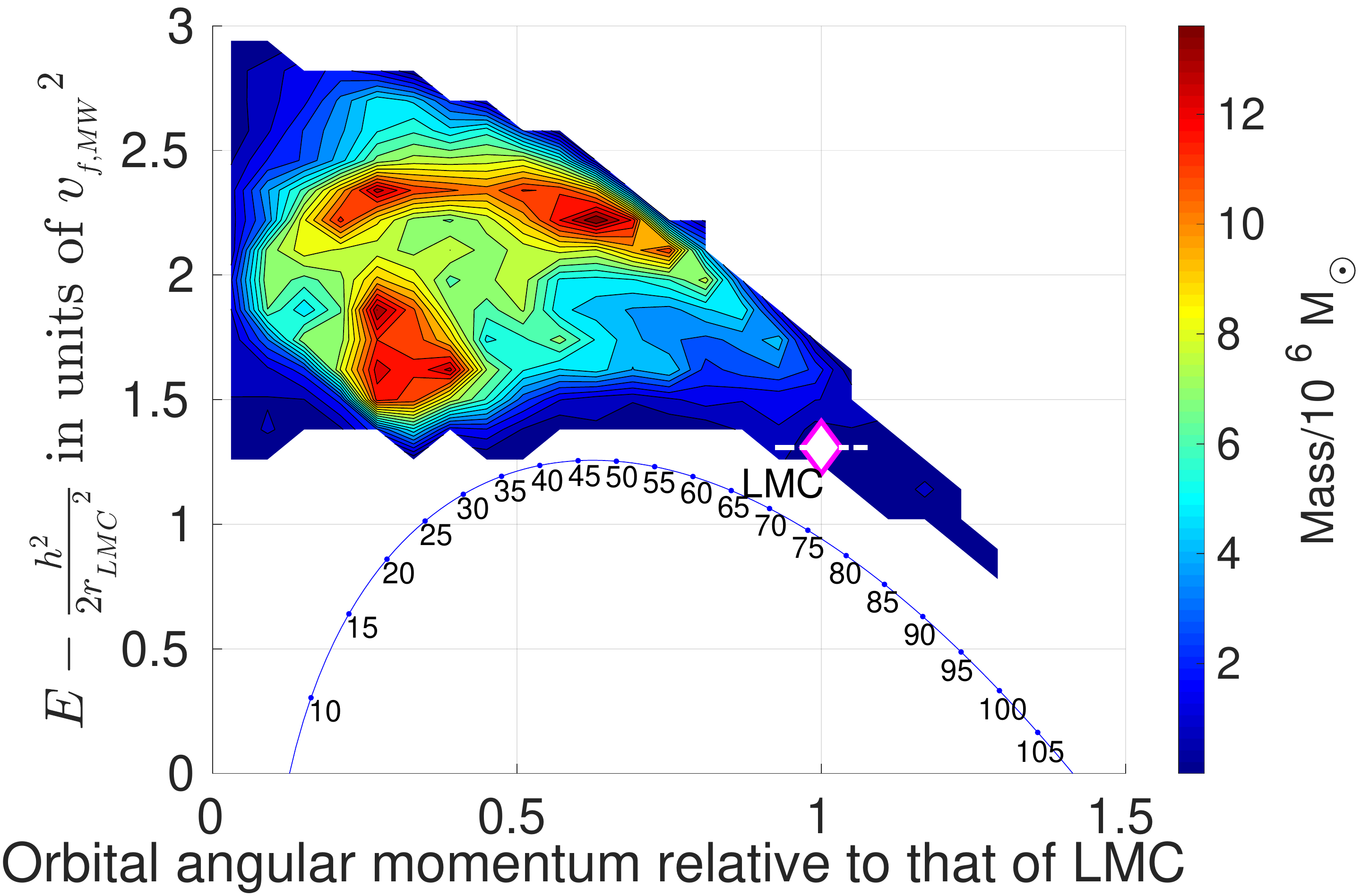}
	\caption{Distribution of the adjusted energy (Equation \ref{E_LMC}) and specific angular momentum relative to that of the LMC, shown for material in the MW satellite region assuming a point mass Galactic potential (Equation \ref{Point_mass_potential}). The LMC value is shown using a white diamond, with the horizontal dashed line through it showing a 24 km/s uncertainty in its Galactocentric tangential velocity (see the text). We neglect uncertainty in its RV because uncertainty in its PM is much larger. The solid blue curve shows values for particles on circular orbits, with the orbital radius indicated in kpc as a text label next to the corresponding filled blue dot.}
	\label{MW_satellite_region_E_h_LMC}
\end{figure}

The joint distribution of $\widetilde{E}$ and $h$ is shown in Figure \ref{MW_satellite_region_E_h_LMC}, which also has a blue curve showing values for particles on circular orbits at different $r$ (indicated in kpc). Our results show that if considering both parameters jointly, the observed LMC value is near the edge of the simulated distribution. Its $\widetilde{E}$ is somewhat on the low side, while its $h$ is on the high side. Nonetheless, the observed combination of $\left( \widetilde{E}, h \right)$ for the LMC is still within the simulated distribution for material in the MW satellite region. Therefore, the LMC does not obviously present an insurmountable difficulty for our model. Given also that material in the satellite region shows a sharp concentration of orbital poles at about the right direction (Figure \ref{MW_SP_50}), our model provides a plausible explanation for how one could get a large satellite with a rather high angular momentum on a quite eccentric nearly polar orbit aligned with the SP.

Our model strongly suggests that the LMC formed as a TDG out of the MW-M31 flyby. Its high mass could indicate that it formed at the tip of the Galactic tidal tail, where a massive TDG is more likely to form due to the weaker tidal stress and EFE combined with the tendency for material to pile up at apocentre. In this scenario, the LMC must have been orbiting the MW for ${\approx 7}$ Gyr. This is quite feasible in a MOND context as there is almost no dynamical friction 40 kpc from a purely baryonic MW. Moreover, the time dependence of the gravity from M31 could cause perigalacticon distances to vary somewhat between orbits \citep[see, e.g., figure 20 of][]{Banik_2018_Centauri}. Assuming that the SMC also formed as a TDG on a similar orbit to the LMC as part of the Galactic SP, one can imagine that the LMC and SMC subsequently underwent one or more interactions, helping to explain the observed properties of the Magellanic Stream \citep{Hammer_2015, Lucchini_2020, Lucchini_2021}. Indeed, it seems difficult to altogether avoid close interactions between a large number of TDGs confined within a 2D structure by virtue of how they formed.

The top panel of Figure \ref{Satellite_region_radial_distribution} indicates that the Galactic SP would initially have been mostly gas, though a small contribution from stars is also expected. This might explain how the LMC came to have stars older than the flyby \citep{Harris_2009, Nidever_2021}. Our scenario would struggle to explain a significant fraction of stars older than the flyby, though this depends on the efficiency with which gas in the tidal arm condenses into stars in the LMC. A lower efficiency would increase the relative importance of stars formed before the flyby. The star formation histories of MW satellites are discussed further in Section \ref{Star_formation_histories}.

A common origin during a past MW-M31 flyby provides a natural explanation for the alignment between the Galactic orbit of the LMC, the LMC-SMC orbit, and the MW SP \citep*[figure 1 of][]{Pawlowski_McGaugh_2015}. These planes also align well with that defined by the Magellanic Stream \citep[figure 1 of][]{Pawlowski_2020}. If instead the LMC and SMC were recently accreted by the MW due to dynamical friction with its CDM halo \citep{Besla_2007}, then these geometric alignments would be fortuitous. This is because the orientation of the MW SP would a priori be unrelated to the LMC-SMC binary orbit, which could itself be oriented differently to their barycentre's motion around the MW \citep[section 5.3.1 of][]{Kroupa_2015}. One exception is if the dynamical influence of the LMC is responsible for most Galactic satellites having orbital poles in a very narrow range of directions, as proposed recently in the $\Lambda$CDM context where the LMC should be more massive relative to the MW than in MOND \citep{Garavito_2021}.\footnote{\citet{Conroy_2021} recently obtained tentative evidence for the perturbation induced by a massive LMC on the Galactic stellar halo.} However, their idealized simulation only works because the test particles they consider have a very low specific angular momentum, so even a very small tidal torque can significantly change the orbital pole. This scenario is untenable for the classical MW satellites as they have quite high specific angular momenta, which is related to their strong preference to be moving tangentially rather than radially \citep{Hammer_2021}. Consequently, including the influence of the LMC in a backwards integration \citep{Correa_2022} still leaves the MW satellites with a clustered orbital pole distribution prior to the LMC infall, implying that this scenario does not explain the clustered orbital pole distribution of MW satellites in the $\Lambda$CDM context \citep{Pawlowski_2022}.

The LMC could also be related to the MW SP rather more directly if the LMC brought in its own retinue of satellites \citep{Samuel_2021}. However, their claim to explain the MW SP in $\Lambda$CDM contradicts the fact that group infall is already included in cosmological $\Lambda$CDM simulations.\footnote{An exception is if MW analogues very rarely have an LMC analogue around them, but these rare cases are also those with a satellite system similar to that of the MW. In this case, $\Lambda$CDM would not be consistent with the LMC.} Based on these, it was previously argued that ``having most satellites accreted as a single group or along a single filament is unlikely to explain the MW disc of satellites'' \citep{Shao_2018}, an important reason being that dwarf galaxy groups are typically much larger than the thickness of the LG SPs \citep[figure 1 of][]{Metz_2009_group}. It is difficult for an LMC-mass galaxy to bring in too many satellites due to its lower mass than the MW, so the Galactocentric orbit of the LMC should be aligned with a pre-existing chance alignment of Galactic satellites into a planar structure. A recently fallen in group would also span only a narrow range of angles on the sky \citep{Isabel_2021}, whereas the MW SP members go around most of the sky \citep[figure 2 of][]{Pawlowski_2020}. Group infall is already included in cosmological $\Lambda$CDM simulations like Illustris, so postulating it as a reasonably likely explanation for the MW SP is not meaningful when the low likelihood of this structure in a $\Lambda$CDM universe has already been demonstrated in a cosmological context \citep{Pawlowski_2020}, especially since high resolution hydrodynamical simulations indicate that the impact of baryonic physics is small \citep{Samuel_2021}. Moreover, the velocities of MW satellites are biased tangentially \citep{Hammer_2021}, even though a radial bias is expected from infall \citep*{Angus_2011}. Our model is quite promising in this respect because it yields a significant tangential bias to the velocities for the tidal debris around M31, though it does not achieve this for the MW (Figure~\ref{Eccentricity_distribution}). However, a dissipative origin for the Galactic SP remains a promising explanation for why its members have tangentially biased orbits, especially when considering the numerical limitations of our simulation.

A successful theory should explain not only the MW SP but also the SPs around M31 \citep{Ibata_2013, Ibata_2014} and Cen A \citep{Muller_2018, Muller_2021} without postulating a highly unusual chance alignment in all cases. Rather, it should provide a mechanism for generating the significant observed anisotropy in the only three systems where the 3D distribution of satellites is well known. Ideally, this would be done with minimal assumptions, e.g. based on the known existence of dissipative baryons and the known formation of phase space-correlated TDGs \citep{Mirabel_1992, Bournaud_2004}. In keeping with Occam's Razor, it is also preferable to use only one hypothetical galaxy interaction to explain both SPs in the LG.

\subsection{The simulated MW and M31 discs today}
\label{Disc_remnants}

Our results in Section \ref{Tidal_debris} show that only a small fraction of the initial $3.05 \times 10^{11} \, M_\odot$ in the MW and M31 discs ends up in their satellite regions. This indicates that the interaction is not very strong and should still leave recognizable disc remnants, as also suggested by Figures \ref{LG_view_part} and \ref{LG_view_gas}. The reason is that the perigalacticon distance of 81.47 kpc is much more than the disc scale length of either galaxy (Table \ref{Disc_parameters}), which also causes there to be only a small amount of dynamical friction (Figure \ref{MW_M31_trajectory}). This is encouraging for our scenario because a strong encounter that transforms the MW and M31 discs into ellipticals would not be realistic. In this section, we take a closer look at the MW and M31 disc remnants (Sections \ref{MW_disc_remnant} and \ref{M31_disc_remnant}, respectively). In each case, our analyses focus on the region within 250 kpc of the galaxy's barycentre and within 50 kpc of its disc plane. Combining our analyses of the disc and satellite regions therefore covers the entire 250 kpc sphere around each galaxy.

%Show Lagrange radii enclosing up to 90\% of the mass within the disc plane. Also need to say how much mass this is. Compare with observed distribution, which just follows an analytic scheme. Remind reader that initial distribution is the same but radially compressed by 20\% for M31 and 40\% for the MW.
%For M31: $M_{enc} \left( <r \right) \propto 1 - \left( x + 1 \right) \exp \left( -x \right)$, where $x \equiv r/r_d$.
%For MW: $M_{enc} \left( <r \right) \propto f_{in} \left( 1 - \left( x_{in} + 1 \right) \exp \left( -x_{in} \right)\right) + f_{out} \left( 1 - \left( x_{out} + 1 \right) \exp \left( -x_{out} \right) \right)$, where $x_{in} \equiv r/r_{d, in}$ (similar for $x_{out}$).

We begin by finding the total angular momentum of all material within the disc region of each galaxy relative to the galaxy's barycentre. This gives the disc spin vector of each galaxy in the analysed \textsc{por} simulation snapshot, which we list in Table \ref{Disc_orientations} in Galactic coordinates. This also shows the initial and observed disc spin vector in each case, and the amount by which the disc has precessed during the simulation. As discussed in section 4.2.5 of \citet{Banik_2020_M33}, this precession is partly due to the EFE, though we expect the flyby to also induce disc precession. The total precession is $\approx 10^\circ$ in both cases, so differential precession of different parts of the MW/M31 disc should not warp it too much. Importantly, the simulated discs end up oriented similarly to observations thanks to our iterative adjustments (Section \ref{Disc_parameters_section}), especially in the M31 case where the disc-SP misalignment is smaller (Figure \ref{M31_SP_50}). In what follows, we analyse each disc in a reference frame rotated so the $z$-axis corresponds to the observed disc orientation as listed in Table \ref{Disc_orientations}. We have checked explicitly that the results of our analyses differ very little if we instead align the $z$-axis to the spin vector of each simulated disc as given in Table \ref{Disc_orientations}. We avoid doing this for simplicity as it would require iterative adjustments to the simulated spin vector of each disc.

\begin{table}
	\centering
	\caption{Spin vectors of the MW and M31 discs at the start of our \textsc{por} simulation and in its 8.2 Gyr snapshot, comparison of which yields the angle by which each disc has precessed. We also show the observed spin vector of each disc, which for M31 follows from its observed inclination and kinematic position angle \citep[section 2.1 of][]{Banik_2018_anisotropy}.}
	\begin{tabular}{ccc}
		\hline
		Galaxy & MW & M31 \\
		\hline
		Initial & $\left(55.11^\circ, \, -84.89^\circ \right)$ & $\left(238.38^\circ, \, -33.71^\circ \right)$ \\
		Final & $\left(231.05^\circ, \, -85.63^\circ \right)$ & $\left(237.72^\circ, \, -26.15^\circ \right)$ \\
		Observed & $\equiv \left(0^\circ, \, -90^\circ \right)$ & $\left(238.65^\circ, \, -26.89^\circ \right)$ \\
		Precession & $9.47^\circ$ & $7.59^\circ$ \\
		Mismatch with & \multirow{2}{*}{$4.37^\circ$} & \multirow{2}{*}{$1.11^\circ$} \\
		observations & & \\
		\hline
	\end{tabular}
	\label{Disc_orientations}
\end{table}

\subsubsection{MW stellar disc remnant}
\label{MW_disc_remnant}

\begin{figure*}
	\centering
	\includegraphics[width = 8.5cm] {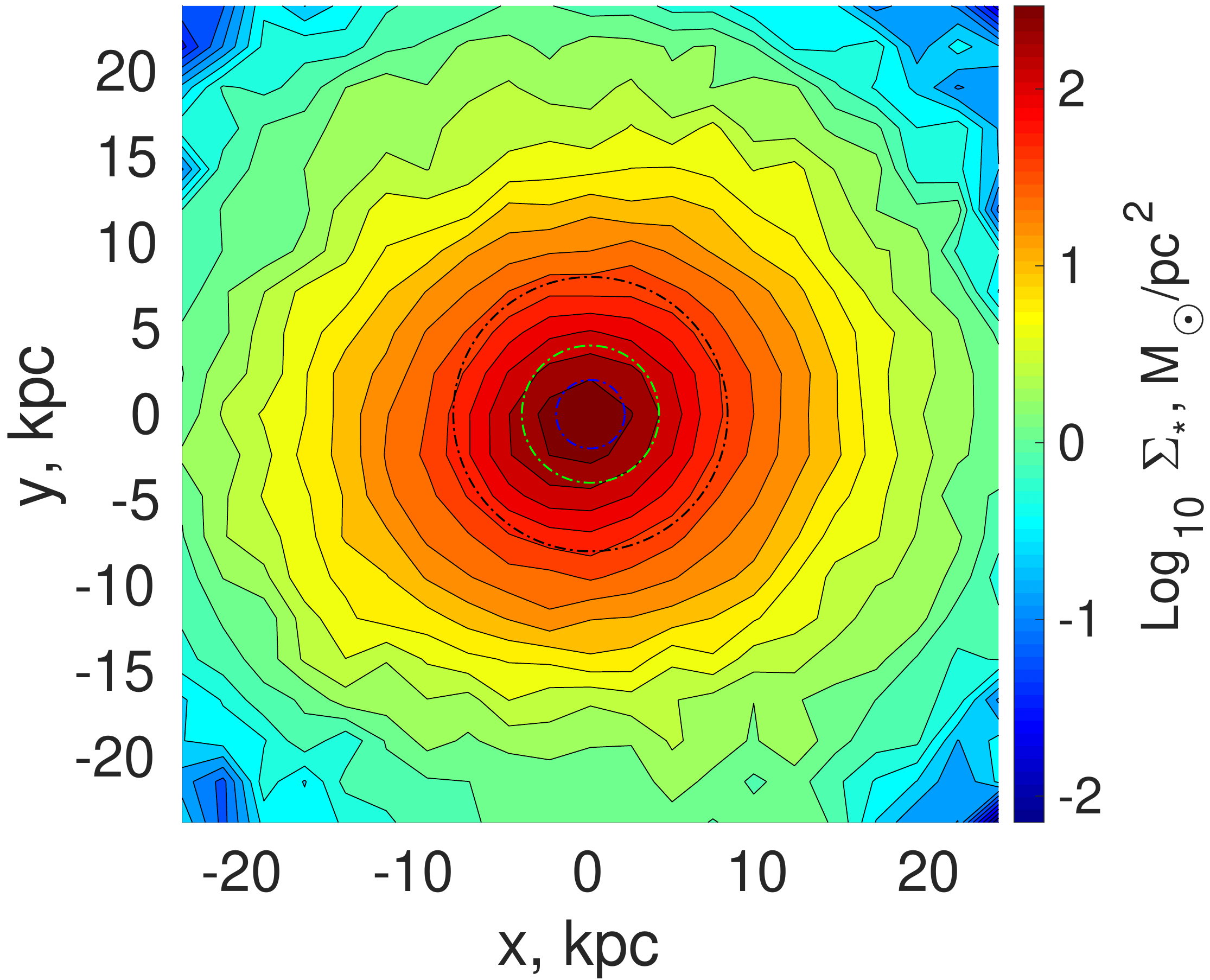}
	\includegraphics[width = 8.5cm] {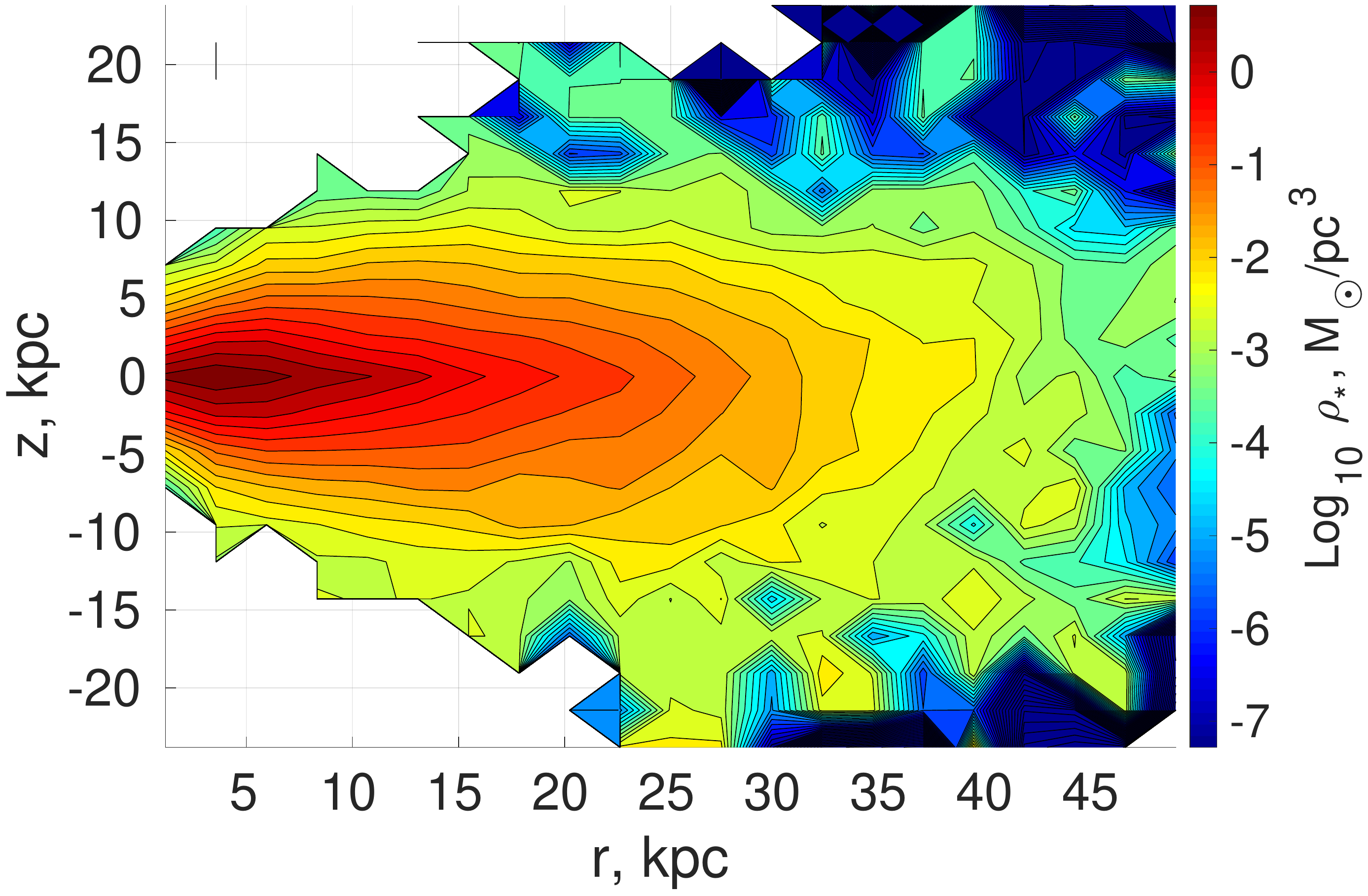}
	\hspace{0.3cm}
	\includegraphics[width = 8.5cm] {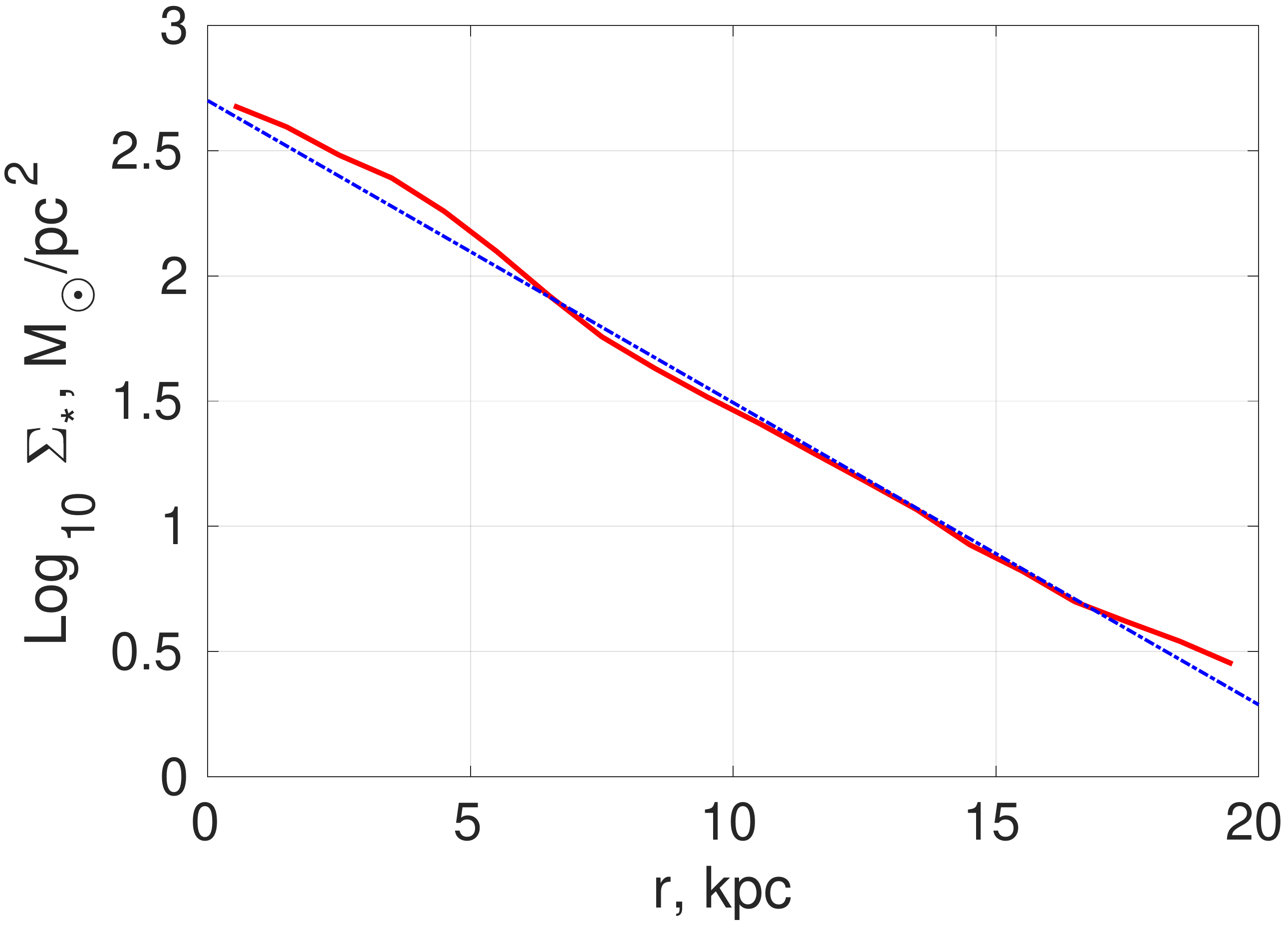}
	\includegraphics[width = 8.5cm] {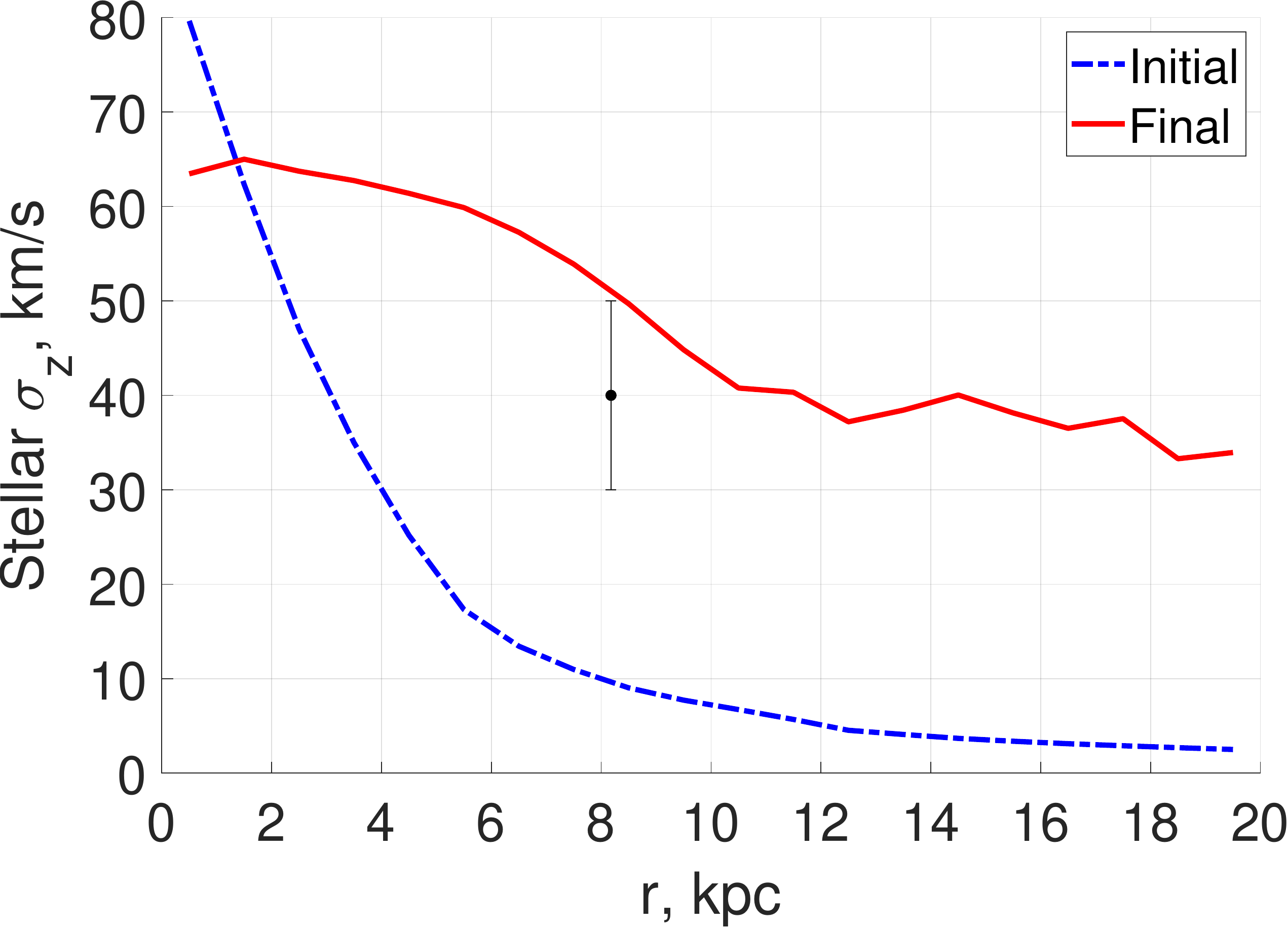}
	\caption{\emph{Top left}: The stellar surface density map of the MW as viewed from along its observed disc spin vector, which is very similar to that in our \textsc{por} simulation (Table \ref{Disc_orientations}). The positions are relative to the MW barycentre, found as described in Section \ref{Proper_motion_section}. The dot-dashed circles show radii of 2, 4, and 8 kpc. \emph{Top right}: The MW in the cylindrical $rz$ projection \citep{Banik_2020_M33}. \emph{Bottom left}: The stellar surface density profile of the MW (red curve), found by binning the stellar particles in cylindrical polar radius. The blue line shows an exponential profile whose scale length is an observationally motivated 3.6 kpc \citep{Juric_2008}. \emph{Bottom right}: The initial (dot-dashed blue) and final (solid red) stellar $\sigma_z$ of the MW disc as a function of radius. The data point at the Solar circle radius of 8.178 kpc \citep{Gravity_2019} shows $\sigma_z = 40 \pm 10$~km/s based on results for different stellar samples in \citet{Yu_2018}. Appendix \ref{Higher_resolution_simulation} shows a version of this panel for a higher resolution simulation.}
	\label{MW_disc_figures}
\end{figure*}

The top left panel of Figure \ref{MW_disc_figures} shows the face-on view of the stars in the MW 8.2 Gyr after the start of our \textsc{por} simulation. This shows a regular appearance. We also show the MW in the cylindrical $rz$ projection system described in section 3.2 of \citet{Banik_2020_M33}. The so-obtained $rz$ view of the stars in the MW is shown in the top right panel. This confirms that a disc remains despite the flyby and other simulated effects.

Our results allow us to obtain the surface density profile of the MW's stellar component. Since our simulation does not allow star formation, this should correspond to something akin to the old Galactic thick disc \citep{Gilmore_1983}, whose stars pre-date the flyby. By binning the results in cylindrical polar radius, we obtain the stellar surface density profile $\Sigma_{\star} \left( r \right)$ shown in the bottom left panel of Figure \ref{MW_disc_figures}. The red line shows the simulated result, while the blue line shows an exponential law with a scale length of 3.6 kpc \citep{Juric_2008}. This is consistent with \citet{Jayaraman_2013}, who found that the thick disc scale length is $\approx 4$ kpc (see their section 4.4). It is also in line with \citet{Li_2017} and \citet{Mateu_2018}.

Another test of our model is the vertical velocity dispersion $\sigma_z$ at different $r$, especially at the Solar circle. We find this by binning the stars in $r$ and finding their mass-weighted $\sigma_z$ using an iterative outlier rejection procedure with threshold of $3.29 \sigma$, corresponding to the 0.1\% tail of a Gaussian distribution \citep[the procedure is described further in section 3.2.1 of][]{Banik_2020_M33}. Our result is shown in the bottom right panel of Figure \ref{MW_disc_figures}. At the Solar circle of 8.2 kpc \citep{McMillan_2017, Gravity_2019}, the simulated $\sigma_z$ rises from an initial value of $\approx 10$ km/s to a present value of $\approx 50$ km/s. Note that since our model does not allow star formation, this should be compared with only those stars which existed at the time of the MW-M31 flyby. In this regard, we mention that the old metal-poor sample at large height in e.g. figure 4 of \citet{Yu_2018} has $\sigma_z \approx 40$ km/s, though results differ by $\approx 10$~km/s depending on the exact sample used. We therefore show a data point at $40 \pm 10$~km/s at the Solar circle. In a higher resolution simulation with all other parameters fixed, $\sigma_z \approx 40$ km/s at the Solar circle $-$ we present this in Appendix \ref{Higher_resolution_simulation}. While results at this level of detail may differ in a more advanced simulation, the broad agreement in the disc scale length and $\sigma_z$ is encouraging.

%MW advised initial spin axis:
%55.1106
%-84.8875

%MW spin axis:
%231.0455
%-85.6339

%Disc precession: 9.4727 degrees.
%Mismatch with observations: 4.3661 degrees.

\subsubsection{M31 stellar disc remnant}
\label{M31_disc_remnant}

%\includegraphics[width = 8.5cm] {M31_edge_on}
%\hspace{0.3cm}
%\includegraphics[width = 8.5cm] {M31_edge_on}
%\includegraphics[width = 8.5cm] {M31_edge_on}
%\hspace{0.3cm}
%\includegraphics[width = 8.5cm] {M31_edge_on}

%M31 advised initial spin axis:
%238.3848
%-33.7111

%M31 spin axis:
%237.7224
%-26.1460

%Disc precession: 7.5868 degrees.
%Mismatch with observations: 1.1146 degrees.

%The observed orientation of the M31 disc is $\left(l = 238.65^\circ, \, b = -26.89^\circ \right)$ in order to match its observed inclination and kinematic position angle \citep[section 2.1 of][]{Banik_2018_anisotropy}.

\begin{figure*}
	\centering
	\includegraphics[width = 8.5cm] {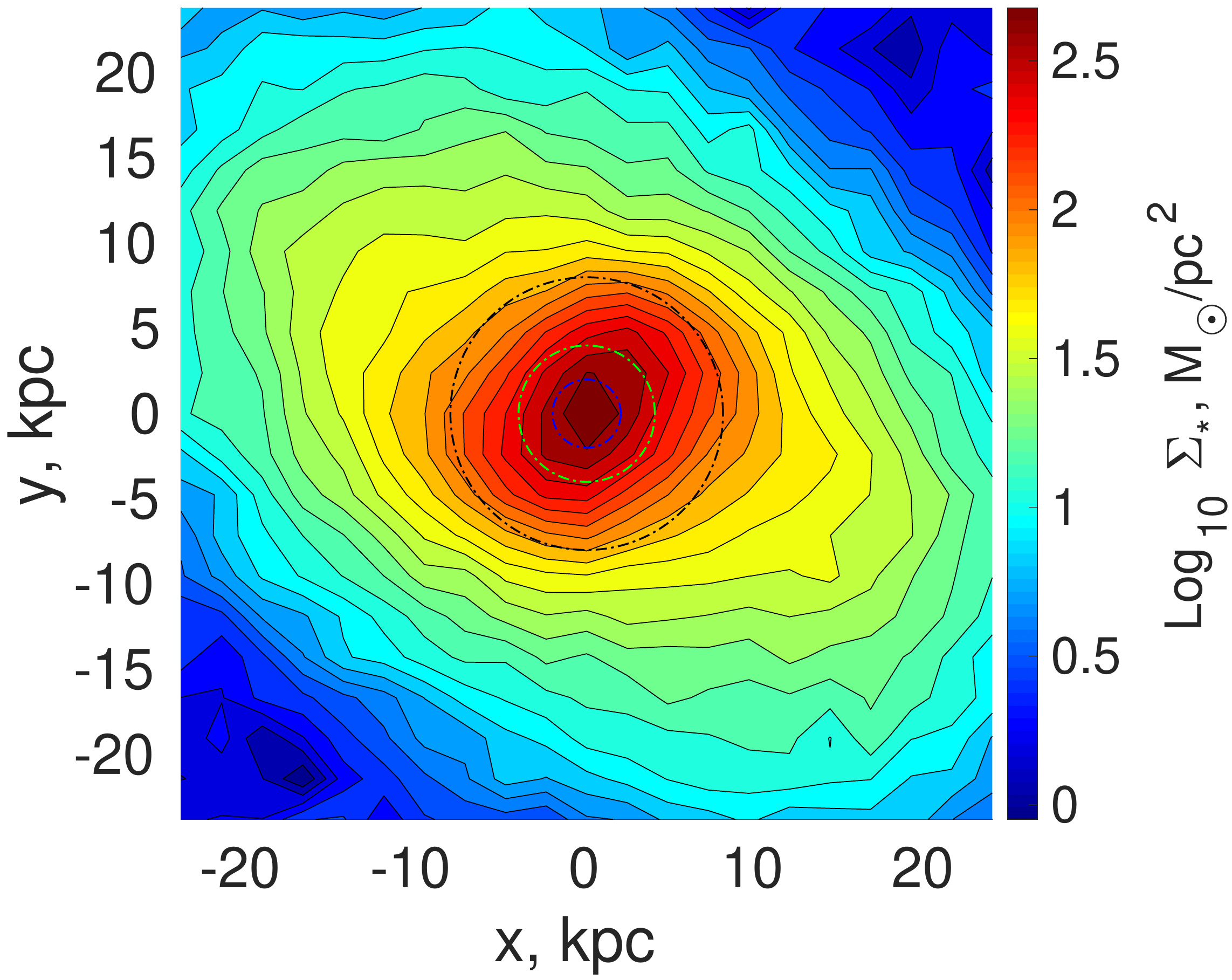}
	\includegraphics[width = 8.5cm] {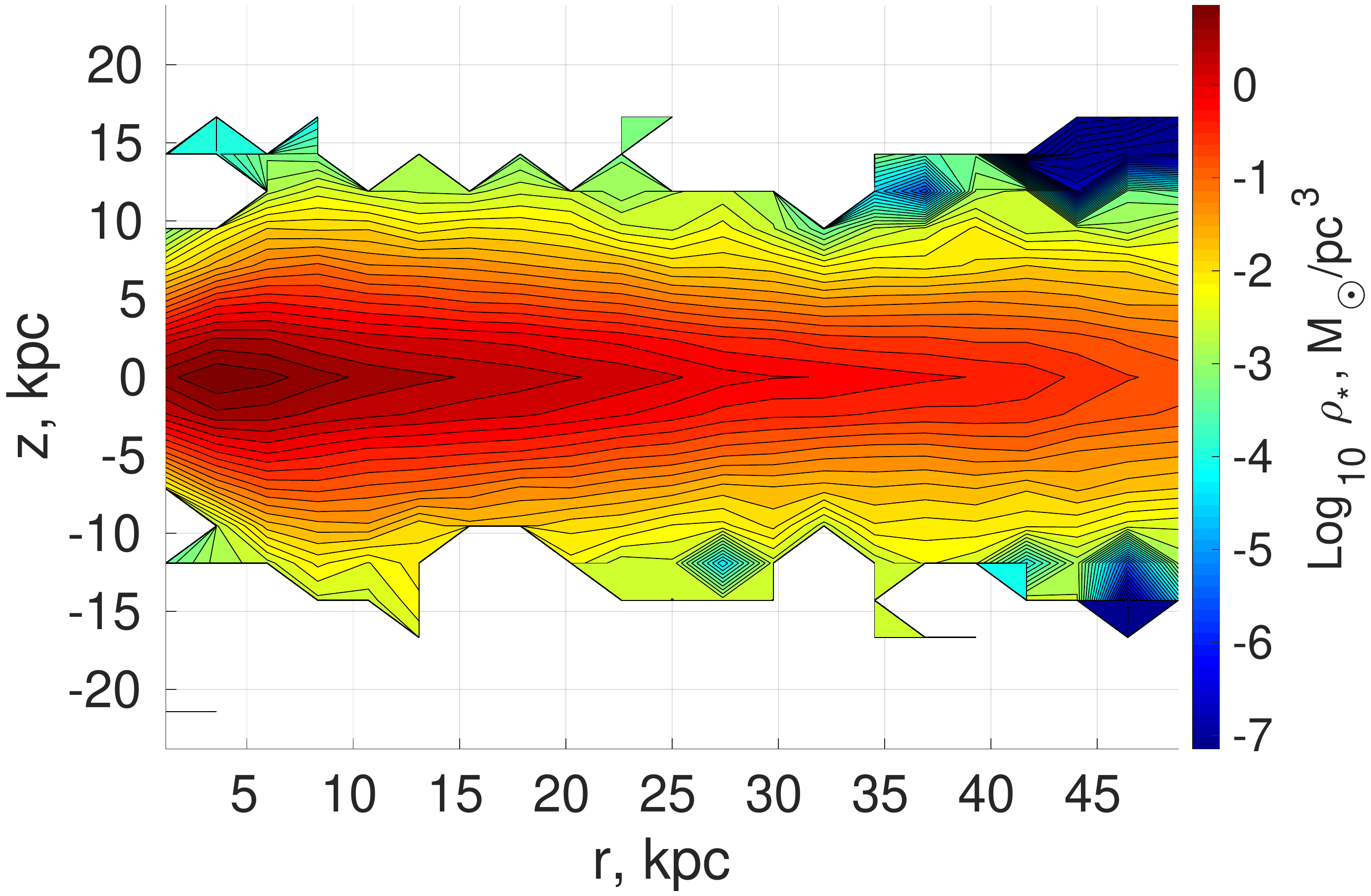}
	\hspace{0.3cm}
	\includegraphics[width = 8.5cm] {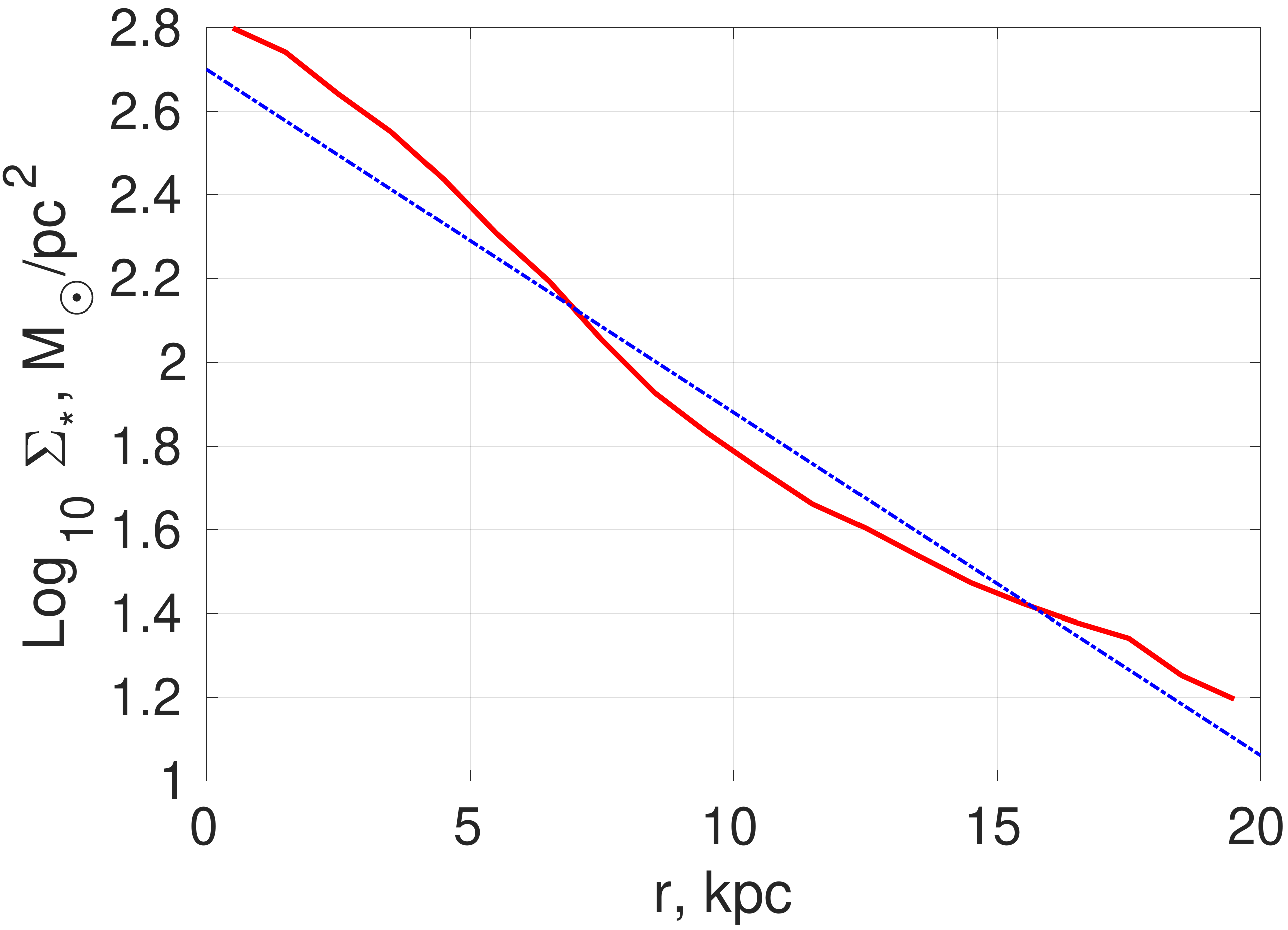}
	\includegraphics[width = 8.5cm] {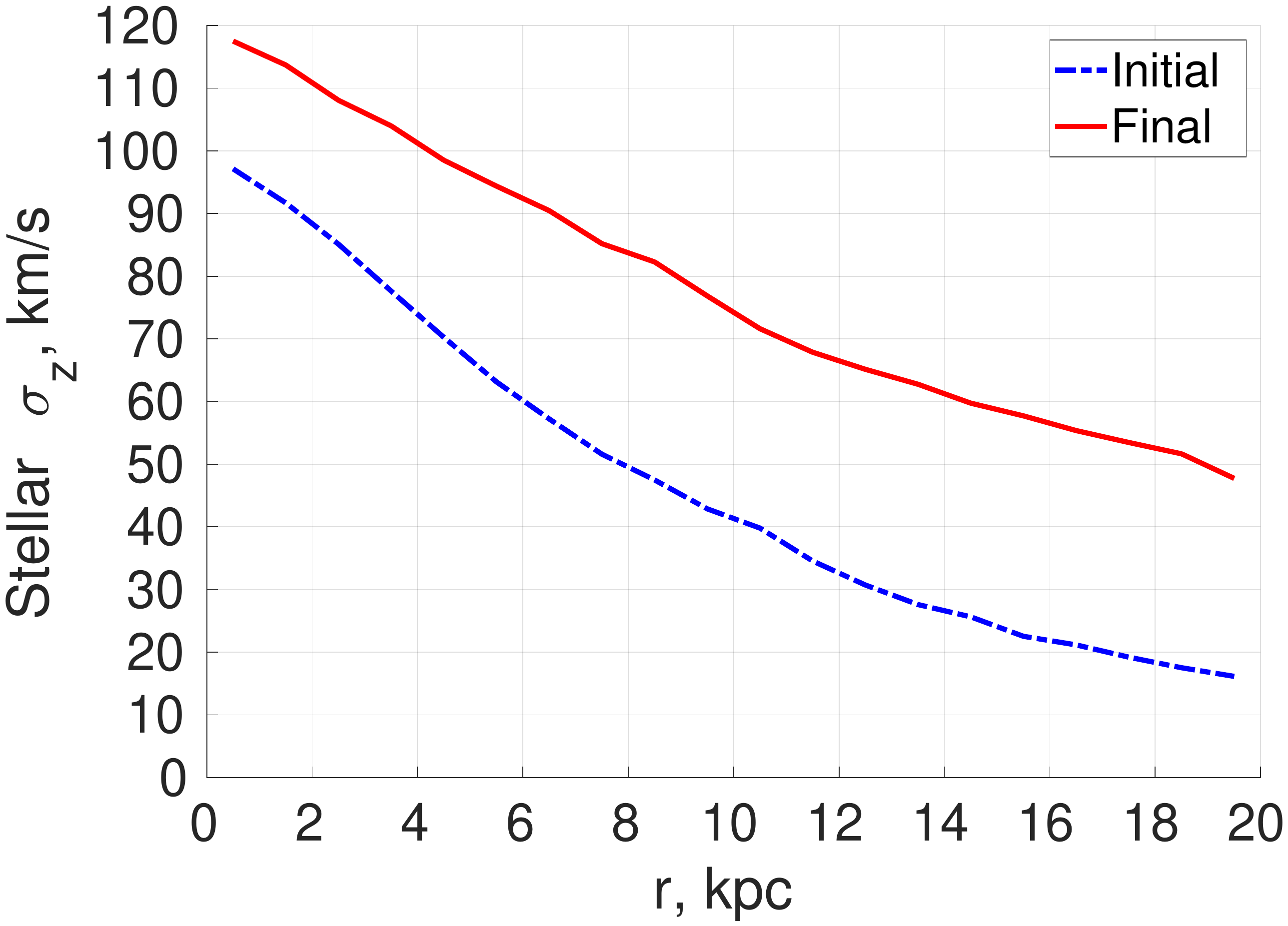}
	\caption{Similar to Figure \ref{MW_disc_figures}, but for M31. \emph{Top left}: Face-on view of M31, showing a bar (unlike the simulated MW). \emph{Top right}: Cylindrical $rz$ projection of M31. \emph{Bottom left}: The stellar surface density profile of M31 (red curve). The blue line shows an exponential law with an observationally motivated scale length of 5.3 kpc \citep{Courteau_2011}. \emph{Bottom right}: The initial (dot-dashed blue) and final (solid red) stellar $\sigma_z$ of M31 as a function of radius, showing that it is expected to be dynamically hotter than the MW (c.f. Appendix \ref{Higher_resolution_simulation}).}
	\label{M31_disc_figures}
\end{figure*}

The face-on view of M31 is shown in the top left panel of Figure \ref{M31_disc_figures}, revealing a bar. Observationally, M31 is indeed a barred spiral galaxy \citep{Beaton_2007}, with the bar having a length of $\approx 4$ kpc and a ratio of corotation radius/bar length of $\mathcal{R} = 1.6 \pm 0.2$ \citep{Blana_2018}. Detailed analysis of the bar goes beyond the scope of this contribution, though bars have been considered before in MOND \citep{Tiret_2007, Tiret_2008_gas, Combes_2014}. We refer the reader to \citet{Banik_2020_M33} for a recent study of the M33 bar in MOND, and to \citet{Sellwood_2019} for a similar study in $\Lambda$CDM. A general study of galactic bars in CDM and different theoretical frameworks is available in \citet{Roshan_2021_disc_stability}, which focused on galaxies with a central surface density similar to the MW and M31.\footnote{In an acceleration-dependent theory like MOND, there is a critical surface density. Thus, galaxies with a similar central surface density should behave similarly, albeit with dimensional quantities like lengths appropriately scaled \citep[see section 5.2 of][]{Roshan_2021_disc_stability}.} Their work showed that in isolated MOND simulations, we would typically expect such a galaxy to have a bar with $\mathcal{R} \approx 1-1.2$ (see their figure 21), so the bar of M31 is somewhat slower but consistent within uncertainties.

The $rz$ projection of M31 (top right panel of Figure \ref{M31_disc_figures}) indicates that our simulation retains a rather thin M31 disc. Its surface density profile is shown in the bottom left panel, which also shows a blue line representing an exponential law with an observationally motivated scale length of 5.3 kpc \citep{Courteau_2011}. This provides a fairly good description of our simulated M31 disc remnant, especially when bearing in mind that their estimated scale length has an uncertainty of 0.5 kpc.

Finally, the most interesting result about the M31 remnant is shown in the bottom right panel of Figure \ref{M31_disc_figures}, which plots the initial and present simulated $\sigma_z$ of the M31 stellar disc at different radii. It is dynamically much hotter than the MW disc (Figure \ref{MW_disc_figures}), which is consistent with the results of \citet{Collins_2011}. This tendency is weakened somewhat in a higher resolution model (Appendix \ref{Higher_resolution_simulation}), but even then, the M31 disc is still hotter than the MW disc at all radii. Therefore, the M31 disc in our simulation is broadly similar to the real M31 disc.

While the higher $\sigma_z$ in M31 is partly due to the initial conditions, the evolution of $\sigma_z$ differs between the MW and M31. In the M31 case, $\sigma_z$ increases by $\approx 20$ km/s towards the centre and $\approx 30$ km/s at 20 kpc, but in general, the whole $\sigma_z \left( r \right)$ curve moves upwards by a similar amount at all radii. Even in the higher resolution simulation (Appendix \ref{Higher_resolution_simulation}), the MW disc does not heat up very much at small radii ($\sigma_z$ rises by $\la 10$ km/s). The increase in $\sigma_z$ is $\approx 30$ km/s at large radii, which is similar to M31. Thus, the rise in $\sigma_z$ at any fixed radius is larger for M31 over the entire disc. However, an increase of 30 km/s is a proportionately much larger increase for the outer regions of the MW. These differences may be caused by the greater extent to which especially the outer MW disc is affected by the flyby, as evident in the larger disc-SP misalignment compared to M31 (Table \ref{MW_M31_hdir}).

\subsubsection{Gas disc remnant surface density profiles}
\label{Sigma_gas_present}

\begin{figure}
	\centering
	\includegraphics[width = 8.5cm] {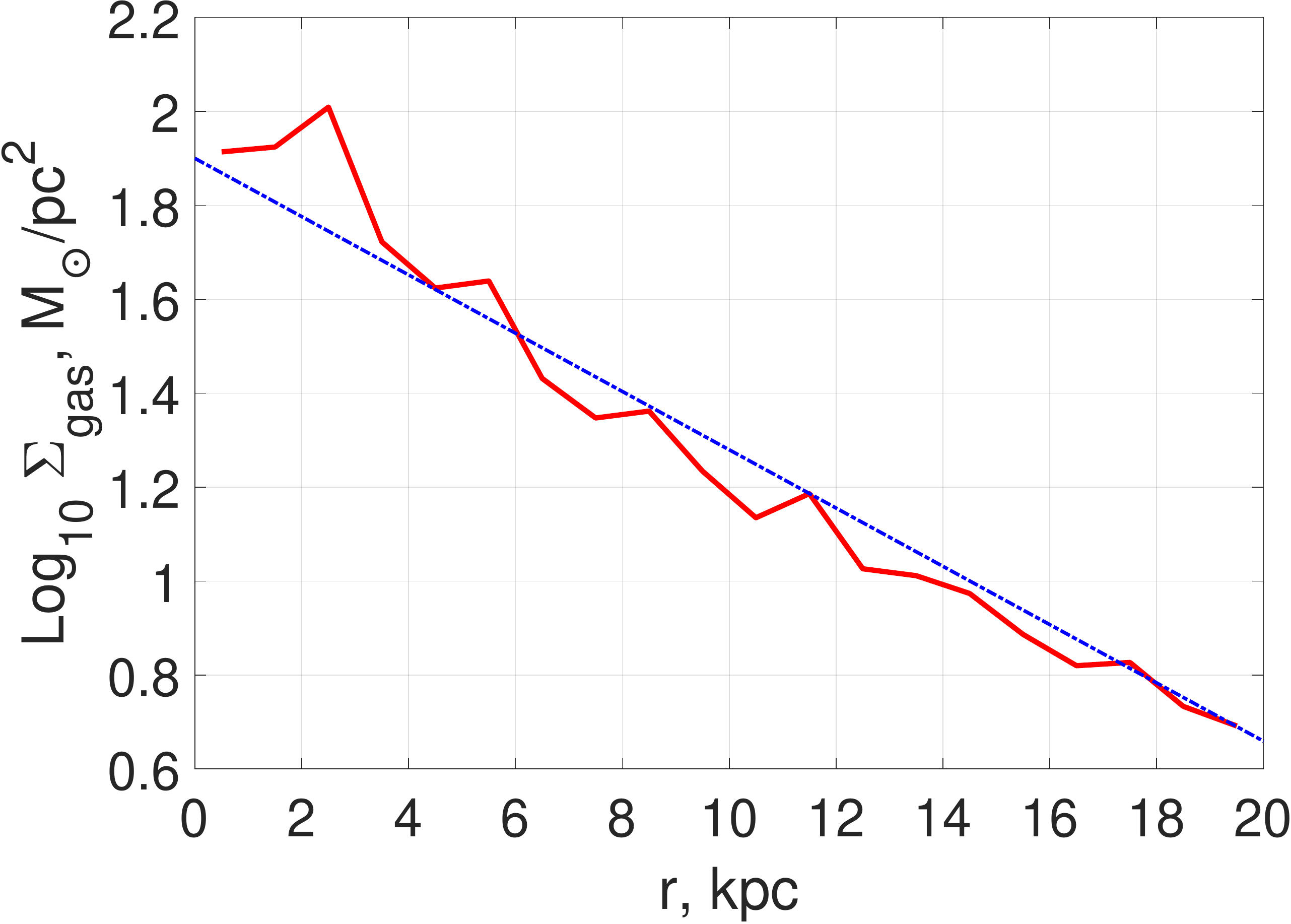}
	\includegraphics[width = 8.5cm] {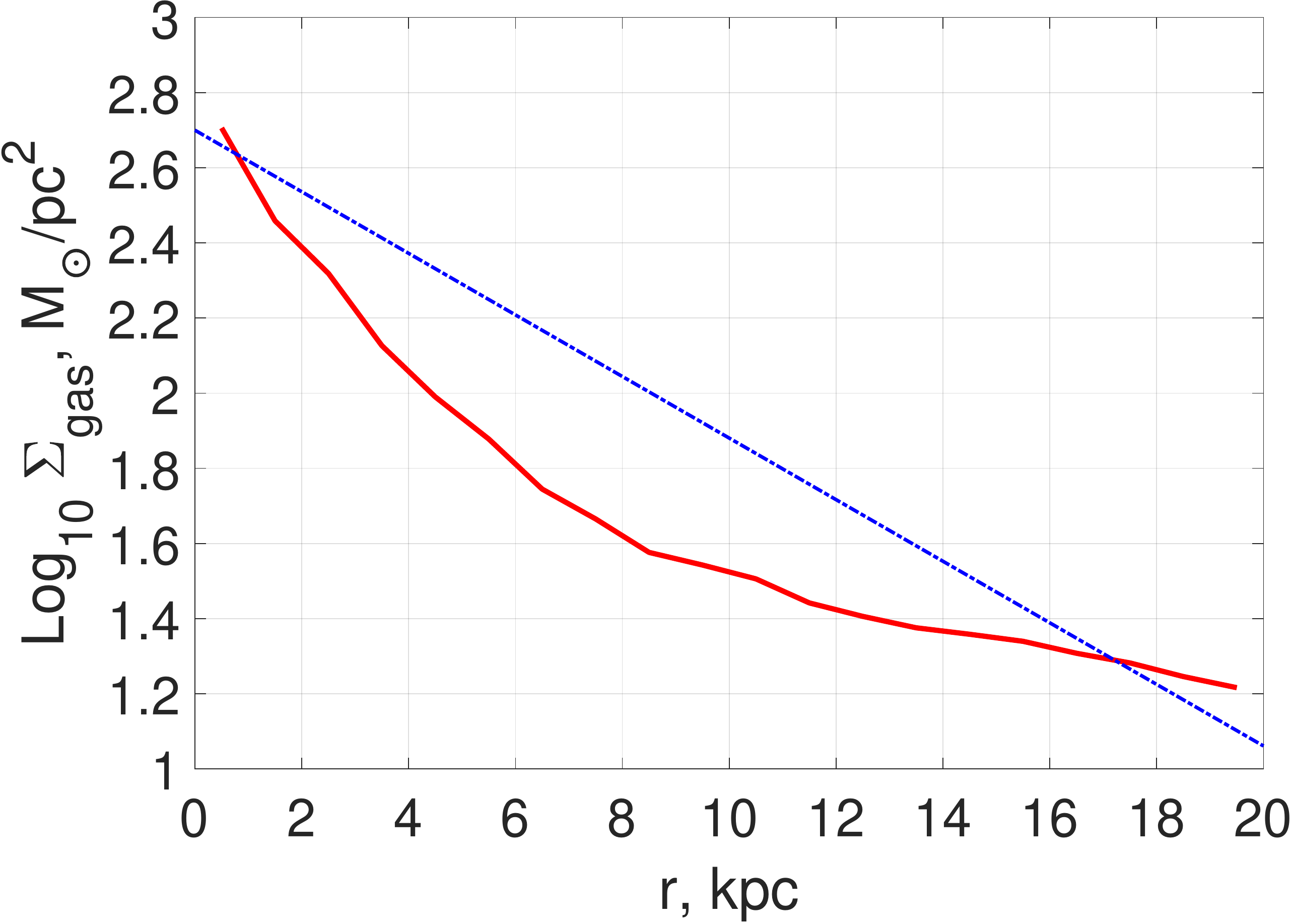}
	\caption{\emph{Top}: Surface density profile of the simulated MW gas disc remnant 8.2 Gyr into our \textsc{por} simulation (solid red curve). The dot-dashed blue line shows an exponential profile with 7 kpc scale length, which roughly describes the observed outer Galactic gas disc \citep[see table 1 of][]{McMillan_2017}. \emph{Bottom}: Similar to the top panel, but for M31. The dot-dashed blue line shows an exponential profile with a 5.3 kpc scale length. Observationally, the M31 gas disc has a rather irregular surface density profile \citep[see figure 16 of][]{Chemin_2009}.}
	\label{Gas_surface_density_profiles}
\end{figure}

Our \textsc{por} simulations do not include star formation, making it somewhat difficult to compare the gas disc remnants with observations. Nonetheless, we use Figure~\ref{Gas_surface_density_profiles} to show the surface density profiles of the MW and M31 gas disc remnants, which we find using similar procedures to those used for the stellar disc remnants.

The outer gas disc of the MW is reasonably well fit with an exponential profile of scale length 7 kpc, which works well observationally \citep[see table 1 of][]{McMillan_2017}. This is shown as a dot-dashed blue line in the top panel of Figure~\ref{Gas_surface_density_profiles} with an arbitrary normalization. It provides a reasonably good match to the simulated distribution, which is shown as a solid red line. Thus, our simulation seems to give a realistic MW gas disc surface density distribution at late times.

The results for M31 are shown in the bottom panel of Figure~\ref{Gas_surface_density_profiles}, where the dot-dashed blue line shows a 5.3 kpc scale length exponential profile with an arbitrary normalization. Our results show that this is not a good fit to the simulated gas disc, which cannot be well fit by a single exponential profile. However, we should bear in mind that a 5.3 kpc scale length is shown because it is a good fit to the observed M31 \emph{stellar} disc \citep{Courteau_2011}. The M31 gas disc has a rather irregular surface density profile \citep*[see figure 16 of][]{Chemin_2009}. Moreover, it is inevitable that the M31 gas disc is subject to significant additional processes in the MOND scenario since in addition to star formation, M31 seems to have experienced a more active interaction history \citep[e.g.,][]{Fardal_2013, Souza_2018}.

\section{Discussion}
\label{Discussion}

\subsection{The star formation histories of LG satellites}
\label{Star_formation_histories}

Our \textsc{por} simulation indicates that the satellite regions of the MW and M31 should have been dominated by gas shortly after their flyby. Since SAM puts the flyby 7.2 Gyr ago (Figure \ref{MW_M31_trajectory}), we might expect members of the MW and M31 SPs to contain very few stars formed at earlier times. This can be altered somewhat, e.g. a higher Cen A mass could push the flyby earlier by $\approx 1$ Gyr. Nonetheless, it is difficult to push the flyby much further back because this would require a much greater asymmetry between the MW-M31 orbit before and after the flyby \citepalias[section 5.1.2 of][]{Banik_Ryan_2018}.\footnote{The asymmetry could be caused by a higher past $a_{_0}$, but section 5.2.3 of \citet{Haslbauer_2020} argued against this using in particular the CMB, which would enter the MOND regime if $a_{_0}$ then exceeded its present value by $\ga 20\times$.} As a result, our flyby scenario can definitely not easily explain why a significant fraction of the stellar mass in the LMC seems to be older than 10 Gyr \citep{Harris_2009, Nidever_2021}. This is also apparent from detailed studies of Sculptor \citep{Boer_2012}, and of Galactic satellites more generally \citep{Weisz_2014}.

The tidal tails expelled from the MW and M31 during the flyby would contain both stars and gas, so some stars from before the flyby are certainly to be expected in the TDGs. A larger proportion of stars from before the flyby could arise if gas accreted onto a newly formed TDG gets expelled by feedback. This would require more mass overall in the satellite region as the observed stellar masses of the LG satellites are known fairly well. Indeed, it was argued in section 5.2.4 of \citetalias{Banik_Ryan_2018} that even a small change to the initial Galactic surface density profile could put much more mass in the satellite region with little effect on the overall potential. In this case, a lower star formation efficiency would be needed to explain the present mass in the Galactic SP members.

It is also possible that there are systematic uncertainties in the stellar ages of MW satellites. The age of the ancient globular cluster NGC 5904 has an uncertainty of 1 Gyr even though it is only 7.4 kpc away \citep*{Gontcharov_2019}. More massive systems like the LMC have a wider range of stars that allow for more constraints on their age, including from RR Lyrae. However, these could be younger than generally thought if the gas cloud from which they formed contained a higher helium abundance than the primordial value \citep{Savino_2020}. This seems likely to some extent as the flyby scenario involves TDGs forming out of gas already processed in the MW/M31, which would have been enriched in metals and helium prior to their flyby. In addition, considering binary stars generally leads to lower age estimates \citep{Stanway_2018}.

Thus, some combination of a slightly earlier flyby and reduced age estimates for the stellar populations in SP members could reduce the proportion of stars older than the flyby. The proportion might then become plausibly consistent with expectations, since after all there should be stars in a TDG that pre-date its formation from tidal debris. Nonetheless, it is clear that the very ancient nature of the Galactic satellites is by far the biggest challenge to the particular flyby scenario studied here.

\subsection{Evidence for the flyby beyond the SPs}
\label{Other_evidence_flyby}

In this section, we consider what the LG was like $\approx 9$ Gyr ago and how this might relate to an MW-M31 encounter around that time.

\subsubsection{Near the MW}
\label{Other_evidence_flyby_MW}

A past encounter with M31 could have triggered the rapid formation of the Galactic bulge \citep{Ballero_2007} by means of tidal torques driving gas into the central regions of the MW. This may be related to the formation and subsequent buckling of the Galactic bar \citep*{Grady_2020}, which could also have formed the similarly old thick disc \citep{Kilic_2017}. Its pattern of enhanced $\alpha$-element abundances \citep{Mashonkina_2019} could indicate a very early flyby or a starburst triggered by the flyby \citep[c.f.][]{Renaud_2016}.

Looking to the satellite region, the young halo globular clusters (YH GCs) in the Galactic halo likely formed out of the inner part of the tidal debris cloud around the MW because they are spatially distributed similarly to the Galactic SP \citep{Pawlowski_2012}.\footnote{The old halo globular clusters have a smaller radial extent and are more nearly isotropically distributed than the YH GCs.} Interestingly, the YH GCs have a bimodal age distribution \citep{Mackey_2004}. In addition to the ancient peak $\approx 12$~Gyr ago, their figure~9 shows a clear secondary peak $\approx 9$~Gyr ago. One advantage of their study is that the ages are relative to that of M92, which should cancel some systematic uncertainties. To obtain absolute ages, those authors assumed that M92 is 12.55 Gyr old based on averaging two earlier estimates \citep{Salaris_2002}. The more recent peak in the age distribution could be due to the flyby triggering the formation of stars and globular clusters \citep[c.f.][]{Renaud_2016}.

\subsubsection{The NGC 3109 association}
\label{Other_evidence_flyby_NGC3109}

Beyond the MW, another important line of evidence for a past encounter with M31 is provided by the kinematics of the NGC 3109 association $\approx 1.5$~Mpc away \citep{Pawlowski_McGaugh_2014}. Those authors discussed how the high RV of NGC 3109 implies that it should have been close to the MW $\approx 9$~Gyr ago based on looking at the problem backwards in time. However, their simplified dynamical analysis of the LG prevents one from drawing strong conclusions as they did not even consider gravity. Using a detailed 3D timing argument analysis of the LG and the major galaxies and galaxy groups outside it up to 8 Mpc away, \citet{Banik_Zhao_2017} showed that the RV of NGC 3109 is 110 km/s higher than in the best-fitting $\Lambda$CDM model.\footnote{Similar results were obtained by \citet{Peebles_2017} using a similar algorithm.} This conclusion was confirmed in \citet{Banik_2018_anisotropy} using a much more thorough search for trajectories that match observational constraints (see their section 4.1).

One deficiency of these few-body timing argument analyses is that they do not consider the possibility of a galaxy passing close to the MW while it was undergoing an interaction, perhaps gaining energy from the interaction in a three-body process. Such backsplash galaxies certainly exist in $\Lambda$CDM, but very rarely do they have properties resembling NGC 3109 \citep{Banik_2021}. The problem is worsened considerably when bearing in mind the filamentary nature of the NGC 3109 association, which suggests that it was a gravitationally bound galaxy group in the past with a total mass of $\approx 3.2 \times 10^{11} \, M_\odot$ \citep{Bellazzini_2013}. A close approach to a major LG galaxy would have created significant dynamical friction that precludes escape to a large distance.

The backsplash process would be much more efficient in MOND because the past high-velocity MW-M31 flyby it requires could have gravitationally slingshot galaxies out to quite large distances \citep{Banik_2018_anisotropy}. Those authors studied the backsplash process in MOND in much more detail, with their figure 6 showing that it is not necessary to very closely approach either the MW or M31. As a result, dynamical friction between the purely baryonic galaxies would be negligible and the disc of NGC 3109 might well have been preserved. However, it is likely that the NGC 3109 association as a whole would have become unbound, consistent with its unbound nature today \citep{Kourkchi_2017}. Therefore, the LG SPs are not the only anomalies faced by $\Lambda$CDM in the LG for which MOND has an explanation by means of a past MW-M31 flyby.

\subsection{Additional complications in the real LG}
\label{Additional_complications_discs}

So far, we have focused on the MW and M31. Apart from their interaction 7.2 Gyr ago and the EFE from large-scale structure, they evolve more or less in isolation. This is of course not necessarily correct in the actual LG, where we expect additional satellite and non-satellite galaxies to play some role. Although many of the LG satellites should have formed out of tidal debris from the MW and M31 in a past encounter scenario, our \textsc{por} simulation does not explicitly form individual TDGs, leaving instead a flattened gas distribution in each galaxy's halo with a preferred orbital pole aligned similarly to the observed SP (Figures \ref{MW_SP_50} and \ref{M31_SP_50}). If individual TDGs had formed, then these could perturb the disc of the parent galaxy in a way that is not captured by our simulations. Moreover, the MW and M31 have been subjected to various other perturbations that should be considered when studying the present LG.

In the M31 case, one important source of disturbance is M32, which is only 23 kpc from M31 \citep[table 4 of][]{Weisz_2014}. It is unclear whether M32 formed in the flyby, with \citetalias{Banik_Ryan_2018} arguing for a separate origin in their section 5.2.3. An origin unrelated to the MW-M31 flyby is very likely for the giant M31 southern stream as it formed in a much more recent galactic interaction \citep{Fardal_2013}. It is also possible that M31 interacted with M33 at some point \citep{Thor_2020}. M33 likely did not form as a TDG born out of the MW-M31 flyby because M33 lies outside the M31 SP \citep{Ibata_2013}. We have implicitly assumed that M33 has been orbiting M31 since at least the start of our \textsc{por} simulation and that it did not interfere with the MW-M31 flyby.

Turning to the MW, we know that the Galactic disc has been perturbed by the Sagittarius dwarf spheroidal satellite galaxy \citep*{Ibata_1994}, leaving behind the most prominent stream in the Galactic halo (see \citealt{Thomas_2017} in the MOND context). Since for every action there is an equal and opposite reaction, the Galactic disc has also been affected by Sagittarius \citep[e.g.,][]{Laporte_2018}. Its orbital motion may have imprinted oscillations on the MW's star formation history \citep{Ruiz_2020}.

The most massive MW satellite is the LMC, which in principle should have formed out of tidal debris expelled from the MW during the interaction with M31 (Section~\ref{LMC_analogues}). As our simulation does not form individual TDGs, it is possible that the formation of the LMC leads to additional subtle effects not captured in our model. The EFE from the LMC on the Galactic disc might be responsible for its warp \citep{Brada_2000}. In general, a Milgromian disc is expected to warp oppositely to the applied external field \citep[see section 4.5.1 of][]{Banik_2020_M33}. The warping of the MW disc was also explored by \citet{Bilek_2018} in $N$-body models of the MW-M31 flyby in MOND, showing that it could in principle explain the observed warp, though the orientation of the warp is likely to have changed substantially since the flyby \citep{Poggio_2020}. There might also be other causes of disc warping.

The MW seems to have undergone a minor merger with a satellite known as Gaia-Sausage-Enceladus (GSE; $M_{\star}\approx 5 \times 10^8 M_\odot$) around the time when the thick disc formed \citep{Kilic_2017}. Stars and globular clusters from GSE were deposited in the Galactic halo with high radial anisotropy \citep{Belokurov_2018, Deason_2018, Haywood_2018, Helmi_2018, Myeong_2018}. The timing of the merger is quite consistent with our estimate for the MW-M31 flyby, suggesting a connection. GSE might originally have been a satellite of M31. If instead it was a non-satellite dwarf in the LG, then the gravitational focusing effect of the combined MW$+$M31 gravity might have caused it to closely encounter the MW around the time of the flyby \citep{Banik_2018_anisotropy}. This could also be the case for other tentative halo structures potentially associated to galaxies accreted at roughly the same time, e.g. Sequoia \citep{Myeong_2019} and Thamnos \citep{Koppelman_2019}.

In any case, it is clear that the history of the real LG has been more complex than in the simulations presented here. Those nevertheless display some generic features which are encouraging, especially with regards to the LG SPs but also the disc remnants and the overall MW-M31 orbit.

%IB: references should be like \citep[e.g.][]{Laporte_2018}, without comma after (e.g.). See section 1 of MNRAS, 473, 4033.

\section{Conclusions}
\label{Conclusions}

The existence of a thin plane of satellite galaxies around the MW is highly unexpected in $\Lambda$CDM cosmology \citep{Kroupa_2005, Pawlowski_2020}, with the problem compounded by a similar SP around M31 \citep{Ibata_2013, Ibata_2014, Sohn_2020, Pawlowski_Sohn_2021} that aligns fairly well with the MW SP \citep[Table \ref{SP_observed_orientations}; see also][]{Pawlowski_2013_LG}. In this contribution, we consider whether their SPs could have formed as TDGs born in a past flyby encounter between the MW and M31 $\approx 7$ Gyr ago. Such a flyby is required in MOND \citep{Zhao_2013} due to the rather strong MW-M31 mutual gravity acting on their nearly radial orbit \citep{Van_der_Marel_2012, Van_der_Marel_2019, Salomon_2021}. For the first time, we simulate this flyby with hydrodynamic simulations by building on the earlier restricted $N$-body simulations of \citetalias{Banik_Ryan_2018} (gravity sourced by two point masses), where it was shown that the MW-M31 trajectory in MOND can plausibly be made consistent with the cosmological initial condition of little peculiar velocity at high redshift \citep[the timing argument;][]{Kahn_Woltjer_1959}. 

In this contribution, we conducted our hydrodynamical simulations of the flyby using \textsc{por}, extending the $N$-body flyby models of \citet{Bilek_2018}. The MW and M31 disc templates were initialized similarly to \citet{Banik_2020_M33}, which presented a fairly realistic MOND model of M33 that avoided some of the problems that arise when simulating it in Newtonian gravity with a live CDM halo \citep{Sellwood_2019}. The barycentric position and velocity of the MW and M31 were initialized similarly to \citetalias{Banik_Ryan_2018}. Although our \textsc{por} simulation does not allow star formation and adopts a rather high gas temperature of 465 kK, it should be enough to determine the orbital pole distribution of the tidal debris around each galaxy. Our main goal was to check if they prefer a particular orbital pole, and if so, whether this aligns with the actually observed SP of the relevant galaxy. More detailed simulations are necessary to follow the formation of individual TDGs out of the tidal debris.

We advanced the MW and M31 for 8.2 Gyr, with $\approx 1$ Gyr of this before the flyby. The timing argument mass of the LG in this model is $3.457 \times 10^{11} \, M_\odot$ as estimated by SAM, which is slightly on the high side but still reasonable (Section \ref{Section_MW_M31_orbit}). The galaxies separate to a large distance and retain rather thin discs, with very little material ending up outside both discs (Figures \ref{LG_view_part} and \ref{LG_view_gas}). This is because the relatively distant encounter compared to the disc scale lengths (Figure \ref{MW_M31_trajectory}) limits dynamical friction and tidal disruption during the flyby. The MW-M31 orbital geometry in our best-fitting model is consistent with the observed M31 PM (Section \ref{Proper_motion_section}). This is a non-trivial success because it was not considered a priori when exploring the parameter space and because the latest constraints \citep{Salomon_2021} are tight enough that not all orbital geometries are allowed for fixed $h$ (Figure \ref{Proper_motion_loop}). 

The tidal debris around the MW prefer a particular orbital pole, with the preferred direction aligning fairly well with the observed orbital pole of its SP (Figure \ref{MW_SP_50}). The same is true for M31 (Figure \ref{M31_SP_50}). Our model naturally yields a more concentrated distribution of orbital poles around M31, in line with observations \citep{Pawlowski_2013_LG}. These results are not much affected by the precise definition of the satellite region (Appendix \ref{Appendix_zmax}).

The encounter does not destroy the discs of the MW and M31, both of which retain a flattened disc by the end of the simulation (top right panels of Figures \ref{MW_disc_figures} and \ref{M31_disc_figures}, respectively). The spin vectors of their discs are quite similar to observations (Table \ref{Disc_orientations}). In the flyby scenario, we expect that the Galactic thick disc formed by dynamical heating of a pre-existing stellar disc \citep[c.f. the thickening of both discs in the $N$-body simulations of][]{Bilek_2018}. Bearing in mind that the simulated Galactic stellar disc today corresponds to its observed thick disc due to the lack of star formation in our models, the simulated $\Sigma_{\star}$ profiles of the disc remnants are broadly consistent with an exponential law with a reasonable scale length (Section \ref{Disc_remnants}). This also applies to the Galactic gas disc remnant. Most importantly, our model indicates that the M31 disc should be dynamically hotter than the MW disc, as observed \citep[e.g.,][]{Collins_2011}.

While our model is promising in many respects, a higher resolution simulation would be required to form individual TDGs and compare their properties with observations of LG satellites. For instance, it would be interesting to determine the mass-metallicity relation, which is expected to be rather similar to primordial satellites because the MW-M31 flyby was so long ago \citep{Collins_2015, Recchi_2015} and because the tidal debris would mostly come from the outskirts of the MW and M31 discs \citepalias{Banik_Ryan_2018}. Any TDGs formed out of the MW-M31 flyby could have undergone significant enrichment at later times, especially given their strong self-gravity in MOND. This is related to the star formation histories of the TDGs, which are potentially problematic for our model \citep{Weisz_2014} because we would in general expect only a small fraction of their stars to be substantially older than the flyby. A significant fraction of the stellar mass in the LMC seems to be older than plausible estimates for when the flyby occurred \citep{Harris_2009, Nidever_2021}. The solution might lie in a slightly earlier flyby combined with a low star formation efficiency in the tidal tails, increasing the relative importance of stars drawn from the parent galaxy (Section~\ref{Star_formation_histories}).

%It is also possible that a significant stellar overdensity like a globular cluster was expelled from the MW disc by tides from M31, later forming the core of a TDG that today is the LMC. BF: NOT CONVINCING, NOT ONLY THE LMC NEEDS TO BE EXPLAINED.

Although our study focused on the LG SPs, a similar scenario could be responsible for the Cen A SP \citep{Muller_2018, Muller_2021}. This might have formed due to a past M83 flyby, or it could be related to the major merger that Cen A likely experienced $\approx 2$ Gyr ago \citep{Wang_2020}. In the latter case, the SP members should be more metal-rich \citep{Duc_2014}, though this need not be the case in the LG due to the much more ancient interaction \citep{Recchi_2015}. Forming an SP out of a merger in principle allows a much wider range of initial orbital configurations because the merger geometry is not constrained independently of the SP. There is also only one SP to match rather than the two in the LG. As a result, similar modelling attempts around Cen A are likely under-determined unless additional observational constraints become available. Alternatively, the Cen A SP might be related to a past interaction with M83, whose satellite system also appears to be flattened \citep*{Muller_2018_M83}. One difficulty is that distance uncertainties make it challenging to identify a distant plane of satellites viewed close to face-on. Further observations are needed to clarify the situation for M83 and to better characterize the SP around Cen A, whose properties are less well known than for the SPs in the LG.

The LG SPs \citep[reviewed in][]{Pawlowski_2018, Pawlowski_2021} pose a severe challenge to $\Lambda$CDM due to the combination of their extreme flatness and coherent rotation (suggestive of dissipation), their mutual alignment within $37^\circ$ (suggestive of a common origin), and the high $\sigma_{\text{int}}$ of their member satellites (ruling out that they are TDGs obeying Newtonian gravity). The issue is not much affected by baryonic physics \citep{Pawlowski_2020, Samuel_2021}, suggesting a fundamental problem. We therefore considered a MOND model of the LG \citep[for a discussion of larger scale issues, see e.g. section 3.1 of][]{Haslbauer_2020}. We showed here that the past close MW-M31 flyby required by this framework \citep{Zhao_2013} leads to the formation of tidal debris around each galaxy, with a preferred orbital pole in each case. This direction aligns with that of the observed SP for both the MW and M31. The two galaxies reach a realistic post-flyby separation with a reasonable relative PM in accordance with the latest constraints. The MW and M31 also retain thin discs with realistic properties by the end of the simulation. Therefore, it might be possible to explain their SPs in a Milgromian framework while remaining consistent with other constraints, the only truly significant challenge for the particular flyby scenario studied here being the ages of the stellar populations within LG satellite galaxies.

\section*{Data availability}

The data underlying this article are available in the article. The algorithms used to prepare and run \textsc{por} simulations of disc galaxies and to extract their results into human-readable form are discussed further in \citet{Banik_2020_M33} and are publicly available.\footnote{\url{https://bitbucket.org/SrikanthTN/bonnpor/src/master/} \newline This includes a version of \textsc{rdramses} rated to work in parallel.} A user manual describing the operation of these algorithms is also available \citep{Nagesh_2021}.

\section*{Acknowledgements}

IB is supported by Science and Technology Facilities Council grant ST/V000861/1. He acknowledges support from a `Pathways to Research' fellowship from the University of Bonn in 2021 after an Alexander von Humboldt Foundation postdoctoral research fellowship (2018-2020). IT acknowledges support through the Stellar Populations and Dynamics research group at the University of Bonn. GC acknowledges support from Chile's National Fund for Scientific and Technological Development (FONDECYT) Regular No. 1181708. BF and RI acknowledge funding from the Agence Nationale de la Recherche (ANR project ANR-18-CE31-0006 and ANR-19-CE31-0017) and from the European Research Council (ERC) under the European Union's Horizon 2020 Framework Programme (grant agreement number 834148). MSP was supported by the Leibniz-Junior Research Group grant J94/2020 via the Leibniz Competition, and a Klaus Tschira Boost Fund provided by the Klaus Tschira Stiftung and the German Scholars Organization. IB is grateful to Hongsheng Zhao for supervising a masters project in which RT adapted \textsc{dice} to prepare a MOND disc template algorithm that forms the basis for this and other work. The authors are grateful to the anonymous referee for comments which helped to improve this publication.

\bibliographystyle{mnras}
\DeclareRobustCommand{\VAN}[3]{#3}
\DeclareRobustCommand{\DE}[3]{#3}
\bibliography{LSP_bbl}

\begin{thebibliography}{}
\makeatletter
\relax
\def\mn@urlcharsother{\let\do\@makeother \do\$\do\&\do\#\do\^\do\_\do\%\do\~}
\def\mn@doi{\begingroup\mn@urlcharsother \@ifnextchar [ {\mn@doi@}
  {\mn@doi@[]}}
\def\mn@doi@[#1]#2{\def\@tempa{#1}\ifx\@tempa\@empty \href
  {http://dx.doi.org/#2} {doi:#2}\else \href {http://dx.doi.org/#2} {#1}\fi
  \endgroup}
\def\mn@eprint#1#2{\mn@eprint@#1:#2::\@nil}
\def\mn@eprint@arXiv#1{\href {http://arxiv.org/abs/#1} {{\tt arXiv:#1}}}
\def\mn@eprint@dblp#1{\href {http://dblp.uni-trier.de/rec/bibtex/#1.xml}
  {dblp:#1}}
\def\mn@eprint@#1:#2:#3:#4\@nil{\def\@tempa {#1}\def\@tempb {#2}\def\@tempc
  {#3}\ifx \@tempc \@empty \let \@tempc \@tempb \let \@tempb \@tempa \fi \ifx
  \@tempb \@empty \def\@tempb {arXiv}\fi \@ifundefined
  {mn@eprint@\@tempb}{\@tempb:\@tempc}{\expandafter \expandafter \csname
  mn@eprint@\@tempb\endcsname \expandafter{\@tempc}}}

\bibitem[\protect\citeauthoryear{{Ahmed}, {Brooks}  \& {Christensen}}{{Ahmed}
  et~al.}{2017}]{Ahmed_2017}
{Ahmed} S.~H.,  {Brooks} A.~M.,   {Christensen} C.~R.,  2017, \mn@doi [MNRAS]
  {10.1093/mnras/stw3271}, \href
  {http://adsabs.harvard.edu/abs/2017MNRAS.466.3119A} {466, 3119}

\bibitem[\protect\citeauthoryear{{Aiola} et~al.,}{{Aiola}
  et~al.}{2020}]{Aiola_2020}
{Aiola} S.,  et~al., 2020, \mn@doi [JCAP] {10.1088/1475-7516/2020/12/047},
  \href {https://ui.adsabs.harvard.edu/abs/2020JCAP...12..047A} {2020, 047}

\bibitem[\protect\citeauthoryear{{Algorry} et~al.,}{{Algorry}
  et~al.}{2017}]{Algorry_2017}
{Algorry} D.~G.,  et~al., 2017, \mn@doi [MNRAS] {10.1093/mnras/stx1008}, \href
  {https://ui.adsabs.harvard.edu/abs/2017MNRAS.469.1054A} {469, 1054}

\bibitem[\protect\citeauthoryear{{Alves} \& {Nelson}}{{Alves} \&
  {Nelson}}{2000}]{Alves_2000}
{Alves} D.~R.,  {Nelson} C.~A.,  2000, \mn@doi [ApJ] {10.1086/317023}, \href
  {http://adsabs.harvard.edu/abs/2000ApJ...542..789A} {542, 789}

\bibitem[\protect\citeauthoryear{{Angus}}{{Angus}}{2009}]{Angus_2009}
{Angus} G.~W.,  2009, \mn@doi [MNRAS] {10.1111/j.1365-2966.2008.14341.x}, \href
  {http://adsabs.harvard.edu/abs/2009MNRAS.394..527A} {394, 527}

\bibitem[\protect\citeauthoryear{{Angus}}{{Angus}}{2010}]{Angus_2010_mass}
{Angus} G.~W.,  2010, \mn@doi [JCAP] {10.1088/1475-7516/2010/03/026}, \href
  {https://ui.adsabs.harvard.edu/abs/2010JCAP...03..026A} {2010, 026}

\bibitem[\protect\citeauthoryear{{Angus}, {Famaey}  \& {Diaferio}}{{Angus}
  et~al.}{2010}]{Angus_2010}
{Angus} G.~W.,  {Famaey} B.,   {Diaferio} A.,  2010, \mn@doi [MNRAS]
  {10.1111/j.1365-2966.2009.15895.x}, \href
  {http://adsabs.harvard.edu/abs/2010MNRAS.402..395A} {402, 395}

\bibitem[\protect\citeauthoryear{{Angus}, {Diaferio}  \& {Kroupa}}{{Angus}
  et~al.}{2011}]{Angus_2011}
{Angus} G.~W.,  {Diaferio} A.,   {Kroupa} P.,  2011, \mn@doi [MNRAS]
  {10.1111/j.1365-2966.2011.19138.x}, \href
  {http://adsabs.harvard.edu/abs/2011MNRAS.416.1401A} {416, 1401}

\bibitem[\protect\citeauthoryear{{Angus}, {Diaferio}, {Famaey}  \&
  {\VAN{Van}{Van}{van}}~der Heyden}{{Angus} et~al.}{2013}]{Angus_2013}
{Angus} G.~W.,  {Diaferio} A.,  {Famaey} B.,   {\VAN{Van}{Van}{van}}~der Heyden
  K.~J.,  2013, \mn@doi [MNRAS] {10.1093/mnras/stt1564}, \href
  {https://ui.adsabs.harvard.edu/abs/2013MNRAS.436..202A/abstract} {436, 202}

\bibitem[\protect\citeauthoryear{{Asencio}, {Banik}  \& {Kroupa}}{{Asencio}
  et~al.}{2021}]{Asencio_2021}
{Asencio} E.,  {Banik} I.,   {Kroupa} P.,  2021, \mn@doi [MNRAS]
  {10.1093/mnras/staa3441}, \href
  {https://ui.adsabs.harvard.edu/abs/2021MNRAS.500.5249A} {500, 5249}

\bibitem[\protect\citeauthoryear{{Athanassoula}}{{Athanassoula}}{2002}]{Athanassoula_2002}
{Athanassoula} E.,  2002, \mn@doi [ApJL] {10.1086/340784}, \href
  {https://ui.adsabs.harvard.edu/abs/2002ApJ...569L..83A} {569, L83}

\bibitem[\protect\citeauthoryear{{Ballero}, {Matteucci}, {Origlia}  \&
  {Rich}}{{Ballero} et~al.}{2007}]{Ballero_2007}
{Ballero} S.~K.,  {Matteucci} F.,  {Origlia} L.,   {Rich} R.~M.,  2007, \mn@doi
  [A\&A] {10.1051/0004-6361:20066596}, \href
  {https://ui.adsabs.harvard.edu/abs/2007A\&A...467..123B} {467, 123}

\bibitem[\protect\citeauthoryear{{Banik} \& {Zhao}}{{Banik} \&
  {Zhao}}{2016}]{Banik_Zhao_2016}
{Banik} I.,  {Zhao} H.,  2016, \mn@doi [MNRAS] {10.1093/mnras/stw787}, \href
  {http://adsabs.harvard.edu/abs/2016MNRAS.459.2237B} {459, 2237}

\bibitem[\protect\citeauthoryear{{Banik} \& {Zhao}}{{Banik} \&
  {Zhao}}{2017}]{Banik_Zhao_2017}
{Banik} I.,  {Zhao} H.,  2017, \mn@doi [MNRAS] {10.1093/mnras/stx151}, \href
  {http://adsabs.harvard.edu/abs/2017MNRAS.467.2180B} {467, 2180}

\bibitem[\protect\citeauthoryear{{Banik} \& {Zhao}}{{Banik} \&
  {Zhao}}{2018a}]{Banik_2015}
{Banik} I.,  {Zhao} H.,  2018a, SciFed Journal of Astrophysics, \href
  {https://ui.adsabs.harvard.edu/abs/2015arXiv150908457B} {1, 1000008}

\bibitem[\protect\citeauthoryear{{Banik} \& {Zhao}}{{Banik} \&
  {Zhao}}{2018b}]{Banik_2018_escape}
{Banik} I.,  {Zhao} H.,  2018b, \mn@doi [MNRAS] {10.1093/mnras/stx2350}, \href
  {http://adsabs.harvard.edu/abs/2018MNRAS.473..419B} {473, 419}

\bibitem[\protect\citeauthoryear{{Banik} \& {Zhao}}{{Banik} \&
  {Zhao}}{2018c}]{Banik_2018_anisotropy}
{Banik} I.,  {Zhao} H.,  2018c, \mn@doi [MNRAS] {10.1093/mnras/stx2596}, \href
  {http://adsabs.harvard.edu/abs/2018MNRAS.473.4033B} {473, 4033}

\bibitem[\protect\citeauthoryear{{Banik} \& {Zhao}}{{Banik} \&
  {Zhao}}{2018d}]{Banik_2018_Centauri}
{Banik} I.,  {Zhao} H.,  2018d, \mn@doi [MNRAS] {10.1093/mnras/sty2007}, \href
  {https://ui.adsabs.harvard.edu/#abs/2018MNRAS.480.2660B} {480, 2660}

\bibitem[\protect\citeauthoryear{{Banik} \& {Zhao}}{{Banik} \&
  {Zhao}}{2022}]{Banik_2022}
{Banik} I.,  {Zhao} H.,  2022, preprint, \href
  {https://ui.adsabs.harvard.edu/abs/2021arXiv211006936B} {Arxiv} (\mn@eprint
  {arXiv} {2110.06936})

\bibitem[\protect\citeauthoryear{{Banik}, {O'Ryan}  \& {Zhao}}{{Banik}
  et~al.}{2018}]{Banik_Ryan_2018}
{Banik} I.,  {O'Ryan} D.,   {Zhao} H.,  2018, \mn@doi [MNRAS]
  {10.1093/mnras/sty919}, \href
  {http://adsabs.harvard.edu/abs/2018MNRAS.477.4768B} {477, 4768}

\bibitem[\protect\citeauthoryear{{Banik}, {Thies}, {Candlish}, {Famaey},
  {Ibata}  \& {Kroupa}}{{Banik} et~al.}{2020}]{Banik_2020_M33}
{Banik} I.,  {Thies} I.,  {Candlish} G.,  {Famaey} B.,  {Ibata} R.,   {Kroupa}
  P.,  2020, \mn@doi [ApJ] {10.3847/1538-4357/abc623}, \href
  {https://ui.adsabs.harvard.edu/abs/2020ApJ...905..135B} {905, 135}

\bibitem[\protect\citeauthoryear{{Banik}, {Haslbauer}, {Pawlowski}, {Famaey}
  \& {Kroupa}}{{Banik} et~al.}{2021}]{Banik_2021}
{Banik} I.,  {Haslbauer} M.,  {Pawlowski} M.~S.,  {Famaey} B.,   {Kroupa} P.,
  2021, \mn@doi [MNRAS] {10.1093/mnras/stab751}, \href
  {https://ui.adsabs.harvard.edu/abs/2021MNRAS.503.6170B} {503, 6170}

\bibitem[\protect\citeauthoryear{{Barnes} \& {Hernquist}}{{Barnes} \&
  {Hernquist}}{1992}]{Barnes_1992}
{Barnes} J.~E.,  {Hernquist} L.,  1992, \mn@doi [Nature] {10.1038/360715a0},
  \href {http://adsabs.harvard.edu/abs/1992Natur.360..715B} {360, 715}

\bibitem[\protect\citeauthoryear{{Beaton} et~al.,}{{Beaton}
  et~al.}{2007}]{Beaton_2007}
{Beaton} R.~L.,  et~al., 2007, \mn@doi [ApJL] {10.1086/514333}, \href
  {https://ui.adsabs.harvard.edu/abs/2007ApJ...658L..91B} {658, L91}

\bibitem[\protect\citeauthoryear{{Begeman}, {Broeils}  \& {Sanders}}{{Begeman}
  et~al.}{1991}]{Begeman_1991}
{Begeman} K.~G.,  {Broeils} A.~H.,   {Sanders} R.~H.,  1991, \mn@doi [MNRAS]
  {10.1093/mnras/249.3.523}, \href
  {http://adsabs.harvard.edu/abs/1991MNRAS.249..523B} {249, 523}

\bibitem[\protect\citeauthoryear{{Bekenstein} \& {Milgrom}}{{Bekenstein} \&
  {Milgrom}}{1984}]{Bekenstein_Milgrom_1984}
{Bekenstein} J.,  {Milgrom} M.,  1984, \mn@doi [ApJ] {10.1086/162570}, \href
  {http://adsabs.harvard.edu/abs/1984ApJ...286....7B} {286, 7}

\bibitem[\protect\citeauthoryear{{Bellazzini}, {Oosterloo}, {Fraternali}  \&
  {Beccari}}{{Bellazzini} et~al.}{2013}]{Bellazzini_2013}
{Bellazzini} M.,  {Oosterloo} T.,  {Fraternali} F.,   {Beccari} G.,  2013,
  \mn@doi [A\&A] {10.1051/0004-6361/201322744}, \href
  {http://adsabs.harvard.edu/abs/2013A\%26A...559L..11B} {559, L11}

\bibitem[\protect\citeauthoryear{{Belokurov}, {Erkal}, {Evans}, {Koposov}  \&
  {Deason}}{{Belokurov} et~al.}{2018}]{Belokurov_2018}
{Belokurov} V.,  {Erkal} D.,  {Evans} N.~W.,  {Koposov} S.~E.,   {Deason}
  A.~J.,  2018, \mn@doi [MNRAS] {10.1093/mnras/sty982}, \href
  {http://adsabs.harvard.edu/abs/2018arXiv180203414B} {478, 611}

\bibitem[\protect\citeauthoryear{{Besla}, {Kallivayalil}, {Hernquist},
  {Robertson}, {Cox}, {\VAN{Van}{Van}{van}}~der Marel  \& {Alcock}}{{Besla}
  et~al.}{2007}]{Besla_2007}
{Besla} G.,  {Kallivayalil} N.,  {Hernquist} L.,  {Robertson} B.,  {Cox} T.~J.,
   {\VAN{Van}{Van}{van}}~der Marel R.~P.,   {Alcock} C.,  2007, \mn@doi [ApJ]
  {10.1086/521385}, \href {http://adsabs.harvard.edu/abs/2007ApJ...668..949B}
  {668, 949}

\bibitem[\protect\citeauthoryear{{B{\'{\i}}lek}, {Thies}, {Kroupa}  \&
  {Famaey}}{{B{\'{\i}}lek} et~al.}{2018}]{Bilek_2018}
{B{\'{\i}}lek} M.,  {Thies} I.,  {Kroupa} P.,   {Famaey} B.,  2018, \mn@doi
  [A\&A] {10.1051/0004-6361/201731939}, \href
  {http://adsabs.harvard.edu/abs/2017arXiv171204938B} {614, A59}

\bibitem[\protect\citeauthoryear{{B{\'\i}lek}, {M{\"u}ller}  \&
  {Famaey}}{{B{\'\i}lek} et~al.}{2019}]{Bilek_2019}
{B{\'\i}lek} M.,  {M{\"u}ller} O.,   {Famaey} B.,  2019, \mn@doi [A\&A]
  {10.1051/0004-6361/201935840}, \href
  {https://ui.adsabs.harvard.edu/abs/2019A\&A...627L...1B} {627, L1}

\bibitem[\protect\citeauthoryear{{B{\'\i}lek}, {Thies}, {Kroupa}  \&
  {Famaey}}{{B{\'\i}lek} et~al.}{2021}]{Bilek_2021}
{B{\'\i}lek} M.,  {Thies} I.,  {Kroupa} P.,   {Famaey} B.,  2021, \mn@doi
  [Galaxies] {10.3390/galaxies9040100}, \href
  {https://ui.adsabs.harvard.edu/abs/2021Galax...9..100B} {9, 100}

\bibitem[\protect\citeauthoryear{{Bla{\~n}a D{\'\i}az} et~al.,}{{Bla{\~n}a
  D{\'\i}az} et~al.}{2018}]{Blana_2018}
{Bla{\~n}a D{\'\i}az} M.,  et~al., 2018, \mn@doi [MNRAS]
  {10.1093/mnras/sty2311}, \href
  {https://ui.adsabs.harvard.edu/abs/2018MNRAS.481.3210B} {481, 3210}

\bibitem[\protect\citeauthoryear{{Bournaud}, {Duc}, {Amram}, {Combes}  \&
  {Gach}}{{Bournaud} et~al.}{2004}]{Bournaud_2004}
{Bournaud} F.,  {Duc} P.~A.,  {Amram} P.,  {Combes} F.,   {Gach} J.~L.,  2004,
  \mn@doi [A\&A] {10.1051/0004-6361:20040394}, \href
  {https://ui.adsabs.harvard.edu/abs/2004A\&A...425..813B} {425, 813}

\bibitem[\protect\citeauthoryear{{Bovy} \& {Rix}}{{Bovy} \&
  {Rix}}{2013}]{Bovy_2013}
{Bovy} J.,  {Rix} H.-W.,  2013, \mn@doi [ApJ] {10.1088/0004-637X/779/2/115},
  \href {http://adsabs.harvard.edu/abs/2013ApJ...779..115B} {779, 115}

\bibitem[\protect\citeauthoryear{{Brada} \& {Milgrom}}{{Brada} \&
  {Milgrom}}{2000}]{Brada_2000}
{Brada} R.,  {Milgrom} M.,  2000, \mn@doi [ApJL] {10.1086/312510}, \href
  {https://ui.adsabs.harvard.edu/abs/2000ApJ...531L..21B} {531, L21}

\bibitem[\protect\citeauthoryear{{Brimioulle}, {Seitz}, {Lerchster}, {Bender}
  \& {Snigula}}{{Brimioulle} et~al.}{2013}]{Brimioulle_2013}
{Brimioulle} F.,  {Seitz} S.,  {Lerchster} M.,  {Bender} R.,   {Snigula} J.,
  2013, \mn@doi [MNRAS] {10.1093/mnras/stt525}, \href
  {http://adsabs.harvard.edu/abs/2013MNRAS.432.1046B} {432, 1046}

\bibitem[\protect\citeauthoryear{{Brouwer} et~al.,}{{Brouwer}
  et~al.}{2017}]{Brouwer_2017}
{Brouwer} M.~M.,  et~al., 2017, \mn@doi [MNRAS] {10.1093/mnras/stw3192}, \href
  {https://ui.adsabs.harvard.edu/abs/2017MNRAS.466.2547B} {466, 2547}

\bibitem[\protect\citeauthoryear{{Brouwer} et~al.,}{{Brouwer}
  et~al.}{2021}]{Brouwer_2021}
{Brouwer} M.~M.,  et~al., 2021, \mn@doi [A\&A] {10.1051/0004-6361/202040108},
  \href {https://ui.adsabs.harvard.edu/abs/2021A\&A...650A.113B} {650, A113}

\bibitem[\protect\citeauthoryear{{Buch}, {Leung}  \& {Fan}}{{Buch}
  et~al.}{2019}]{Buch_2019}
{Buch} J.,  {Leung} J. S.~C.,   {Fan} J.,  2019, \mn@doi [JCAP]
  {10.1088/1475-7516/2019/04/026}, \href
  {https://ui.adsabs.harvard.edu/abs/2019JCAP...04..026B} {2019, 026}

\bibitem[\protect\citeauthoryear{{Candlish}}{{Candlish}}{2016}]{Candlish_2016}
{Candlish} G.~N.,  2016, \mn@doi [MNRAS] {10.1093/mnras/stw1130}, \href
  {http://adsabs.harvard.edu/abs/2016MNRAS.460.2571C} {460, 2571}

\bibitem[\protect\citeauthoryear{{Casas}, {Arias}, {Pe{\~n}a Ram{\'\i}rez}  \&
  {Kroupa}}{{Casas} et~al.}{2012}]{Casas_2012}
{Casas} R.~A.,  {Arias} V.,  {Pe{\~n}a Ram{\'\i}rez} K.,   {Kroupa} P.,  2012,
  \mn@doi [MNRAS] {10.1111/j.1365-2966.2012.21319.x}, \href
  {https://ui.adsabs.harvard.edu/abs/2012MNRAS.424.1941C} {424, 1941}

\bibitem[\protect\citeauthoryear{{Cautun} \& {Frenk}}{{Cautun} \&
  {Frenk}}{2017}]{Cautun_2017}
{Cautun} M.,  {Frenk} C.~S.,  2017, \mn@doi [MNRAS] {10.1093/mnrasl/slx025},
  \href {https://ui.adsabs.harvard.edu/abs/2017MNRAS.468L..41C} {468, L41}

\bibitem[\protect\citeauthoryear{{Chae}, {Bernardi}, {Dom{\'\i}nguez
  S{\'a}nchez}  \& {Sheth}}{{Chae} et~al.}{2020a}]{Chae_2020_elliptical}
{Chae} K.-H.,  {Bernardi} M.,  {Dom{\'\i}nguez S{\'a}nchez} H.,   {Sheth}
  R.~K.,  2020a, \mn@doi [ApJL] {10.3847/2041-8213/abc2d3}, \href
  {https://ui.adsabs.harvard.edu/abs/2020ApJ...903L..31C} {903, L31}

\bibitem[\protect\citeauthoryear{{Chae}, {Lelli}, {Desmond}, {McGaugh}, {Li}
  \& {Schombert}}{{Chae} et~al.}{2020b}]{Chae_2020_EFE}
{Chae} K.-H.,  {Lelli} F.,  {Desmond} H.,  {McGaugh} S.~S.,  {Li} P.,
  {Schombert} J.~M.,  2020b, \mn@doi [ApJ] {10.3847/1538-4357/abbb96}, \href
  {http://adsabs.harvard.edu/abs/2020ApJ...904...51C} {904, 51}

\bibitem[\protect\citeauthoryear{{Chae}, {Desmond}, {Lelli}, {McGaugh}  \&
  {Schombert}}{{Chae} et~al.}{2021}]{Chae_2021}
{Chae} K.-H.,  {Desmond} H.,  {Lelli} F.,  {McGaugh} S.~S.,   {Schombert}
  J.~M.,  2021, \mn@doi [ApJ] {10.3847/1538-4357/ac1bba}, \href
  {https://ui.adsabs.harvard.edu/abs/2021ApJ...921..104C} {921, 104}

\bibitem[\protect\citeauthoryear{{Chemin}, {Carignan}  \& {Foster}}{{Chemin}
  et~al.}{2009}]{Chemin_2009}
{Chemin} L.,  {Carignan} C.,   {Foster} T.,  2009, \mn@doi [ApJ]
  {10.1088/0004-637X/705/2/1395}, \href
  {http://adsabs.harvard.edu/abs/2009ApJ...705.1395C} {705, 1395}

\bibitem[\protect\citeauthoryear{{Collins} et~al.,}{{Collins}
  et~al.}{2011}]{Collins_2011}
{Collins} M.~L.~M.,  et~al., 2011, \mn@doi [MNRAS]
  {10.1111/j.1365-2966.2011.18238.x}, \href
  {https://ui.adsabs.harvard.edu/abs/2011MNRAS.413.1548C} {413, 1548}

\bibitem[\protect\citeauthoryear{{Collins} et~al.,}{{Collins}
  et~al.}{2015}]{Collins_2015}
{Collins} M.~L.~M.,  et~al., 2015, \mn@doi [ApJL]
  {10.1088/2041-8205/799/1/L13}, \href
  {http://adsabs.harvard.edu/abs/2015ApJ...799L..13C} {799, L13}

\bibitem[\protect\citeauthoryear{{Combes}}{{Combes}}{2014}]{Combes_2014}
{Combes} F.,  2014, \mn@doi [A\&A] {10.1051/0004-6361/201424990}, \href
  {https://ui.adsabs.harvard.edu/abs/2014A\&A...571A..82C} {571, A82}

\bibitem[\protect\citeauthoryear{{Conn} et~al.,}{{Conn}
  et~al.}{2013}]{Conn_2013}
{Conn} A.~R.,  et~al., 2013, \mn@doi [ApJ] {10.1088/0004-637X/766/2/120}, \href
  {https://ui.adsabs.harvard.edu/abs/2013ApJ...766..120C} {766, 120}

\bibitem[\protect\citeauthoryear{{Conroy}, {Naidu}, {Garavito-Camargo},
  {Besla}, {Zaritsky}, {Bonaca}  \& {Johnson}}{{Conroy}
  et~al.}{2021}]{Conroy_2021}
{Conroy} C.,  {Naidu} R.~P.,  {Garavito-Camargo} N.,  {Besla} G.,  {Zaritsky}
  D.,  {Bonaca} A.,   {Johnson} B.~D.,  2021, \mn@doi [Nature]
  {10.1038/s41586-021-03385-7}, \href
  {https://ui.adsabs.harvard.edu/abs/2021Natur.592..534C} {592, 534}

\bibitem[\protect\citeauthoryear{{Corbelli}, {Lorenzoni}, {Walterbos}, {Braun}
  \& {Thilker}}{{Corbelli} et~al.}{2010}]{Corbelli_2010}
{Corbelli} E.,  {Lorenzoni} S.,  {Walterbos} R.,  {Braun} R.,   {Thilker} D.,
  2010, \mn@doi [A\&A] {10.1051/0004-6361/200913297}, \href
  {https://ui.adsabs.harvard.edu/abs/2010A\&A...511A..89C} {511, A89}

\bibitem[\protect\citeauthoryear{{Correa Magnus} \& {Vasiliev}}{{Correa Magnus}
  \& {Vasiliev}}{2022}]{Correa_2022}
{Correa Magnus} L.,  {Vasiliev} E.,  2022, \mn@doi [MNRAS]
  {10.1093/mnras/stab3726}, \href
  {https://ui.adsabs.harvard.edu/abs/2022MNRAS.511.2610C} {511, 2610}

\bibitem[\protect\citeauthoryear{{Courteau}, {Widrow}, {McDonald},
  {Guhathakurta}, {Gilbert}, {Zhu}, {Beaton}  \& {Majewski}}{{Courteau}
  et~al.}{2011}]{Courteau_2011}
{Courteau} S.,  {Widrow} L.~M.,  {McDonald} M.,  {Guhathakurta} P.,  {Gilbert}
  K.~M.,  {Zhu} Y.,  {Beaton} R.~L.,   {Majewski} S.~R.,  2011, \mn@doi [ApJ]
  {10.1088/0004-637X/739/1/20}, \href
  {http://adsabs.harvard.edu/abs/2011ApJ...739...20C} {739, 20}

\bibitem[\protect\citeauthoryear{{D'Souza} \& {Bell}}{{D'Souza} \&
  {Bell}}{2018}]{Souza_2018}
{D'Souza} R.,  {Bell} E.~F.,  2018, \mn@doi [Nature Astronomy]
  {10.1038/s41550-018-0533-x}, \href
  {https://ui.adsabs.harvard.edu/abs/2018NatAs...2..737D} {2, 737}

\bibitem[\protect\citeauthoryear{{\DE{De}{De}{de}}~Boer
  et~al.,}{{\DE{De}{De}{de}}~Boer et~al.}{2012}]{Boer_2012}
{\DE{De}{De}{de}}~Boer T.~J.~L.,  et~al., 2012, \mn@doi [A\&A]
  {10.1051/0004-6361/201118378}, \href
  {https://ui.adsabs.harvard.edu/abs/2012A\&A...539A.103D} {539, A103}

\bibitem[\protect\citeauthoryear{{Deason}, {Belokurov}, {Koposov}  \&
  {Lancaster}}{{Deason} et~al.}{2018}]{Deason_2018}
{Deason} A.~J.,  {Belokurov} V.,  {Koposov} S.~E.,   {Lancaster} L.,  2018,
  \mn@doi [ApJL] {10.3847/2041-8213/aad0ee}, \href
  {https://ui.adsabs.harvard.edu/abs/2018ApJ...862L...1D} {862, L1}

\bibitem[\protect\citeauthoryear{{Debattista} \& {Sellwood}}{{Debattista} \&
  {Sellwood}}{2000}]{Debattista_2000}
{Debattista} V.~P.,  {Sellwood} J.~A.,  2000, \mn@doi [ApJ] {10.1086/317148},
  \href {https://ui.adsabs.harvard.edu/abs/2000ApJ...543..704D} {543, 704}

\bibitem[\protect\citeauthoryear{{Di Valentino}}{{Di
  Valentino}}{2021}]{Valentino_2021}
{Di Valentino} E.,  2021, \mn@doi [MNRAS] {10.1093/mnras/stab187}, \href
  {https://ui.adsabs.harvard.edu/abs/2021MNRAS.502.2065D} {502, 2065}

\bibitem[\protect\citeauthoryear{{Duc}, {Paudel}, {McDermid}, {Cuillandre},
  {Serra}, {Bournaud}, {Cappellari}  \& {Emsellem}}{{Duc}
  et~al.}{2014}]{Duc_2014}
{Duc} P.-A.,  {Paudel} S.,  {McDermid} R.~M.,  {Cuillandre} J.-C.,  {Serra} P.,
   {Bournaud} F.,  {Cappellari} M.,   {Emsellem} E.,  2014, \mn@doi [MNRAS]
  {10.1093/mnras/stu330}, \href
  {https://ui.adsabs.harvard.edu/abs/2014MNRAS.440.1458D} {440, 1458}

\bibitem[\protect\citeauthoryear{{Efstathiou}, {Sutherland}  \&
  {Maddox}}{{Efstathiou} et~al.}{1990}]{Efstathiou_1990}
{Efstathiou} G.,  {Sutherland} W.~J.,   {Maddox} S.~J.,  1990, \mn@doi [Nature]
  {10.1038/348705a0}, \href
  {https://ui.adsabs.harvard.edu/abs/1990Natur.348..705E} {348, 705}

\bibitem[\protect\citeauthoryear{{Famaey} \& {Binney}}{{Famaey} \&
  {Binney}}{2005}]{Famaey_Binney_2005}
{Famaey} B.,  {Binney} J.,  2005, \mn@doi [MNRAS]
  {10.1111/j.1365-2966.2005.09474.x}, \href
  {http://adsabs.harvard.edu/abs/2005MNRAS.363..603F} {363, 603}

\bibitem[\protect\citeauthoryear{{Famaey} \& {McGaugh}}{{Famaey} \&
  {McGaugh}}{2012}]{Famaey_McGaugh_2012}
{Famaey} B.,  {McGaugh} S.~S.,  2012, \mn@doi [Living Reviews in Relativity]
  {10.12942/lrr-2012-10}, \href
  {http://adsabs.harvard.edu/abs/2012LRR....15...10F} {15, 10}

\bibitem[\protect\citeauthoryear{{Famaey}, {McGaugh}  \& {Milgrom}}{{Famaey}
  et~al.}{2018}]{Famaey_2018}
{Famaey} B.,  {McGaugh} S.,   {Milgrom} M.,  2018, \mn@doi [MNRAS]
  {10.1093/mnras/sty1884}, \href
  {http://adsabs.harvard.edu/abs/2018arXiv180404167F} {480, 473}

\bibitem[\protect\citeauthoryear{{Fardal} et~al.,}{{Fardal}
  et~al.}{2013}]{Fardal_2013}
{Fardal} M.~A.,  et~al., 2013, \mn@doi [MNRAS] {10.1093/mnras/stt1121}, \href
  {http://adsabs.harvard.edu/abs/2013MNRAS.434.2779F} {434, 2779}

\bibitem[\protect\citeauthoryear{{Fattahi}, {Navarro}, {Frenk}, {Oman},
  {Sawala}  \& {Schaller}}{{Fattahi} et~al.}{2018}]{Fattahi_2018}
{Fattahi} A.,  {Navarro} J.~F.,  {Frenk} C.~S.,  {Oman} K.~A.,  {Sawala} T.,
  {Schaller} M.,  2018, \mn@doi [MNRAS] {10.1093/mnras/sty408}, \href
  {https://ui.adsabs.harvard.edu/abs/2018MNRAS.476.3816F} {476, 3816}

\bibitem[\protect\citeauthoryear{Fletcher \& Powell}{Fletcher \&
  Powell}{1963}]{Fletcher_1963}
Fletcher R.,  Powell M. J.~D.,  1963, \mn@doi [The Computer Journal]
  {10.1093/comjnl/6.2.163}, \href {http://dx.doi.org/10.1093/comjnl/6.2.163}
  {6, 163}

\bibitem[\protect\citeauthoryear{{Fragkoudi}, {Grand}, {Pakmor}, {Springel},
  {White}, {Marinacci}, {Gomez}  \& {Navarro}}{{Fragkoudi}
  et~al.}{2021}]{Fragkoudi_2021}
{Fragkoudi} F.,  {Grand} R.~J.~J.,  {Pakmor} R.,  {Springel} V.,  {White}
  S.~D.~M.,  {Marinacci} F.,  {Gomez} F.~A.,   {Navarro} J.~F.,  2021, \mn@doi
  [A\&A] {10.1051/0004-6361/202140320}, \href
  {https://ui.adsabs.harvard.edu/abs/2021A\&A...650L..16F} {650, L16}

\bibitem[\protect\citeauthoryear{{Francis} \& {Anderson}}{{Francis} \&
  {Anderson}}{2014}]{Francis_2014}
{Francis} C.,  {Anderson} E.,  2014, \mn@doi [Celestial Mechanics and Dynamical
  Astronomy] {10.1007/s10569-014-9541-z}, \href
  {http://adsabs.harvard.edu/abs/2014CeMDA.118..399F} {118, 399}

\bibitem[\protect\citeauthoryear{{Freeman}}{{Freeman}}{1970}]{Freeman_1970}
{Freeman} K.~C.,  1970, \mn@doi [ApJ] {10.1086/150474}, \href
  {http://adsabs.harvard.edu/abs/1970ApJ...160..811F} {160, 811}

\bibitem[\protect\citeauthoryear{{Freundlich}, {Famaey}, {Oria}, {B{\'\i}lek},
  {M{\"u}ller}  \& {Ibata}}{{Freundlich} et~al.}{2022}]{Freundlich_2022}
{Freundlich} J.,  {Famaey} B.,  {Oria} P.-A.,  {B{\'\i}lek} M.,  {M{\"u}ller}
  O.,   {Ibata} R.,  2022, \mn@doi [A\&A] {10.1051/0004-6361/202142060}, \href
  {https://ui.adsabs.harvard.edu/abs/2022A\&A...658A..26F} {658, A26}

\bibitem[\protect\citeauthoryear{{Gaia Collaboration}}{{Gaia
  Collaboration}}{2018}]{Gaia_2018_Helmi}
{Gaia Collaboration} 2018, \mn@doi [A\&A] {10.1051/0004-6361/201832698}, \href
  {https://ui.adsabs.harvard.edu/abs/2018A\&A...616A..12G} {616, A12}

\bibitem[\protect\citeauthoryear{{Garavito-Camargo}, {Patel}, {Besla},
  {Price-Whelan}, {G{\'o}mez}, {Laporte}  \& {Johnston}}{{Garavito-Camargo}
  et~al.}{2021}]{Garavito_2021}
{Garavito-Camargo} N.,  {Patel} E.,  {Besla} G.,  {Price-Whelan} A.~M.,
  {G{\'o}mez} F.~A.,  {Laporte} C. F.~P.,   {Johnston} K.~V.,  2021, \mn@doi
  [ApJ] {10.3847/1538-4357/ac2c05}, \href
  {https://ui.adsabs.harvard.edu/abs/2021ApJ...923..140G} {923, 140}

\bibitem[\protect\citeauthoryear{{Gentile}, {Famaey}  \&
  {\DE{De}{De}{de}}~Blok}{{Gentile} et~al.}{2011}]{Gentile_2011}
{Gentile} G.,  {Famaey} B.,   {\DE{De}{De}{de}}~Blok W.~J.~G.,  2011, \mn@doi
  [A\&A] {10.1051/0004-6361/201015283}, \href
  {http://adsabs.harvard.edu/abs/2011A\%26A...527A..76G} {527, A76}

\bibitem[\protect\citeauthoryear{{Genzel} et~al.,}{{Genzel}
  et~al.}{2017}]{Genzel_2017}
{Genzel} R.,  et~al., 2017, \mn@doi [Nature] {10.1038/nature21685}, \href
  {https://ui.adsabs.harvard.edu/abs/2017Natur.543..397G} {543, 397}

\bibitem[\protect\citeauthoryear{{Genzel} et~al.,}{{Genzel}
  et~al.}{2020}]{Genzel_2020}
{Genzel} R.,  et~al., 2020, \mn@doi [ApJ] {10.3847/1538-4357/abb0ea}, \href
  {https://ui.adsabs.harvard.edu/abs/2020ApJ...902...98G} {902, 98}

\bibitem[\protect\citeauthoryear{{Gilmore} \& {Reid}}{{Gilmore} \&
  {Reid}}{1983}]{Gilmore_1983}
{Gilmore} G.,  {Reid} N.,  1983, \mn@doi [MNRAS] {10.1093/mnras/202.4.1025},
  \href {http://adsabs.harvard.edu/abs/1983MNRAS.202.1025G} {202, 1025}

\bibitem[\protect\citeauthoryear{{Gontcharov}, {Mosenkov}  \&
  {Khovritchev}}{{Gontcharov} et~al.}{2019}]{Gontcharov_2019}
{Gontcharov} G.~A.,  {Mosenkov} A.~V.,   {Khovritchev} M.~Y.,  2019, \mn@doi
  [MNRAS] {10.1093/mnras/sty3439}, \href
  {https://ui.adsabs.harvard.edu/abs/2019MNRAS.483.4949G} {483, 4949}

\bibitem[\protect\citeauthoryear{{Grady}, {Belokurov}  \& {Evans}}{{Grady}
  et~al.}{2020}]{Grady_2020}
{Grady} J.,  {Belokurov} V.,   {Evans} N.~W.,  2020, \mn@doi [MNRAS]
  {10.1093/mnras/stz3617}, \href
  {https://ui.adsabs.harvard.edu/abs/2020MNRAS.492.3128G} {492, 3128}

\bibitem[\protect\citeauthoryear{{Gravity Collaboration}}{{Gravity
  Collaboration}}{2019}]{Gravity_2019}
{Gravity Collaboration} 2019, \mn@doi [A\&A] {10.1051/0004-6361/201935656},
  \href {https://ui.adsabs.harvard.edu/abs/2019A\&A...625L..10G} {625, L10}

\bibitem[\protect\citeauthoryear{{Haghi}, {Bazkiaei}, {Zonoozi}  \&
  {Kroupa}}{{Haghi} et~al.}{2016}]{Haghi_2016}
{Haghi} H.,  {Bazkiaei} A.~E.,  {Zonoozi} A.~H.,   {Kroupa} P.,  2016, \mn@doi
  [MNRAS] {10.1093/mnras/stw573}, \href
  {http://adsabs.harvard.edu/abs/2016MNRAS.458.4172H} {458, 4172}

\bibitem[\protect\citeauthoryear{{Haghi} et~al.,}{{Haghi}
  et~al.}{2019a}]{Haghi_2019}
{Haghi} H.,  et~al., 2019a, \mn@doi [MNRAS] {10.1093/mnras/stz1465}, \href
  {https://academic.oup.com/mnras/advance-article-abstract/doi/10.1093/mnras/stz1465/5505850}
  {487, 2441}

\bibitem[\protect\citeauthoryear{{Haghi}, {Amiri}, {Hasani Zonoozi}, {Banik},
  {Kroupa}  \& {Haslbauer}}{{Haghi} et~al.}{2019b}]{Haghi_2019_DF44}
{Haghi} H.,  {Amiri} V.,  {Hasani Zonoozi} A.,  {Banik} I.,  {Kroupa} P.,
  {Haslbauer} M.,  2019b, ApJL, \href
  {https://ui.adsabs.harvard.edu/abs/2019arXiv190907978H} {884, L25}

\bibitem[\protect\citeauthoryear{{Hammer}, {Yang}, {Wang}, {Puech}, {Flores}
  \& {Fouquet}}{{Hammer} et~al.}{2010}]{Hammer_2010}
{Hammer} F.,  {Yang} Y.~B.,  {Wang} J.~L.,  {Puech} M.,  {Flores} H.,
  {Fouquet} S.,  2010, \mn@doi [ApJ] {10.1088/0004-637X/725/1/542}, \href
  {https://ui.adsabs.harvard.edu/abs/2010ApJ...725..542H} {725, 542}

\bibitem[\protect\citeauthoryear{{Hammer}, {Yang}, {Fouquet}, {Pawlowski},
  {Kroupa}, {Puech}, {Flores}  \& {Wang}}{{Hammer} et~al.}{2013}]{Hammer_2013}
{Hammer} F.,  {Yang} Y.,  {Fouquet} S.,  {Pawlowski} M.~S.,  {Kroupa} P.,
  {Puech} M.,  {Flores} H.,   {Wang} J.,  2013, \mn@doi [MNRAS]
  {10.1093/mnras/stt435}, \href
  {http://adsabs.harvard.edu/abs/2013MNRAS.431.3543H} {431, 3543}

\bibitem[\protect\citeauthoryear{{Hammer}, {Yang}, {Flores}, {Puech}  \&
  {Fouquet}}{{Hammer} et~al.}{2015}]{Hammer_2015}
{Hammer} F.,  {Yang} Y.~B.,  {Flores} H.,  {Puech} M.,   {Fouquet} S.,  2015,
  \mn@doi [ApJ] {10.1088/0004-637X/813/2/110}, \href
  {http://adsabs.harvard.edu/abs/2015ApJ...813..110H} {813, 110}

\bibitem[\protect\citeauthoryear{{Hammer}, {Wang}, {Pawlowski}, {Yang},
  {Bonifacio}, {Li}, {Babusiaux}  \& {Arenou}}{{Hammer}
  et~al.}{2021}]{Hammer_2021}
{Hammer} F.,  {Wang} J.,  {Pawlowski} M.~S.,  {Yang} Y.,  {Bonifacio} P.,  {Li}
  H.,  {Babusiaux} C.,   {Arenou} F.,  2021, \mn@doi [ApJ]
  {10.3847/1538-4357/ac27a8}, \href
  {https://ui.adsabs.harvard.edu/abs/2021ApJ...922...93H} {922, 93}

\bibitem[\protect\citeauthoryear{{Harris} \& {Zaritsky}}{{Harris} \&
  {Zaritsky}}{2009}]{Harris_2009}
{Harris} J.,  {Zaritsky} D.,  2009, \mn@doi [AJ]
  {10.1088/0004-6256/138/5/1243}, \href
  {https://ui.adsabs.harvard.edu/abs/2009AJ....138.1243H} {138, 1243}

\bibitem[\protect\citeauthoryear{{Harrison} et~al.,}{{Harrison}
  et~al.}{2017}]{Harrison_2017}
{Harrison} C.~M.,  et~al., 2017, \mn@doi [MNRAS] {10.1093/mnras/stx217}, \href
  {https://ui.adsabs.harvard.edu/abs/2017MNRAS.467.1965H} {467, 1965}

\bibitem[\protect\citeauthoryear{{Haslbauer}, {Dabringhausen}, {Kroupa},
  {Javanmardi}  \& {Banik}}{{Haslbauer} et~al.}{2019}]{Haslbauer_2019}
{Haslbauer} M.,  {Dabringhausen} J.,  {Kroupa} P.,  {Javanmardi} B.,   {Banik}
  I.,  2019, \mn@doi [A\&A] {10.1051/0004-6361/201833771}, \href
  {https://ui.adsabs.harvard.edu/abs/2019A\&A...626A..47H} {626, A47}

\bibitem[\protect\citeauthoryear{{Haslbauer}, {Banik}  \& {Kroupa}}{{Haslbauer}
  et~al.}{2020}]{Haslbauer_2020}
{Haslbauer} M.,  {Banik} I.,   {Kroupa} P.,  2020, \mn@doi [MNRAS]
  {10.1093/mnras/staa2348}, \href
  {https://ui.adsabs.harvard.edu/abs/2020MNRAS.499.2845H} {499, 2845}

\bibitem[\protect\citeauthoryear{{Haslbauer}, {Banik}, {Kroupa}, {Wittenburg}
  \& {Javanmardi}}{{Haslbauer} et~al.}{2022}]{Haslbauer_2022}
{Haslbauer} M.,  {Banik} I.,  {Kroupa} P.,  {Wittenburg} N.,   {Javanmardi} B.,
   2022, \mn@doi [ApJ] {10.3847/1538-4357/ac46ac}, \href
  {https://ui.adsabs.harvard.edu/abs/2022arXiv220201221H} {925, 183}

\bibitem[\protect\citeauthoryear{{Haywood}, {Di Matteo}, {Lehnert}, {Snaith},
  {Khoperskov}  \& {G{\'o}mez}}{{Haywood} et~al.}{2018}]{Haywood_2018}
{Haywood} M.,  {Di Matteo} P.,  {Lehnert} M.~D.,  {Snaith} O.,  {Khoperskov}
  S.,   {G{\'o}mez} A.,  2018, \mn@doi [ApJ] {10.3847/1538-4357/aad235}, \href
  {https://ui.adsabs.harvard.edu/abs/2018ApJ...863..113H} {863, 113}

\bibitem[\protect\citeauthoryear{{Hees}, {Folkner}, {Jacobson}  \&
  {Park}}{{Hees} et~al.}{2014}]{Hees_2014}
{Hees} A.,  {Folkner} W.~M.,  {Jacobson} R.~A.,   {Park} R.~S.,  2014, \mn@doi
  [Physical Review D] {10.1103/PhysRevD.89.102002}, \href
  {http://adsabs.harvard.edu/abs/2014PhRvD..89j2002H} {89, 102002}

\bibitem[\protect\citeauthoryear{{Hees}, {Famaey}, {Angus}  \&
  {Gentile}}{{Hees} et~al.}{2016}]{Hees_2016}
{Hees} A.,  {Famaey} B.,  {Angus} G.~W.,   {Gentile} G.,  2016, \mn@doi [MNRAS]
  {10.1093/mnras/stv2330}, \href
  {http://adsabs.harvard.edu/abs/2016MNRAS.455..449H} {455, 449}

\bibitem[\protect\citeauthoryear{{Helmi}, {Babusiaux}, {Koppelman}, {Massari},
  {Veljanoski}  \& {Brown}}{{Helmi} et~al.}{2018}]{Helmi_2018}
{Helmi} A.,  {Babusiaux} C.,  {Koppelman} H.~H.,  {Massari} D.,  {Veljanoski}
  J.,   {Brown} A. G.~A.,  2018, \mn@doi [Nature] {10.1038/s41586-018-0625-x},
  \href {https://ui.adsabs.harvard.edu/abs/2018Natur.563...85H} {563, 85}

\bibitem[\protect\citeauthoryear{{Ibata}, {Gilmore}  \& {Irwin}}{{Ibata}
  et~al.}{1994}]{Ibata_1994}
{Ibata} R.~A.,  {Gilmore} G.,   {Irwin} M.~J.,  1994, \mn@doi [Nature]
  {10.1038/370194a0}, \href
  {https://ui.adsabs.harvard.edu/abs/1994Natur.370..194I} {370, 194}

\bibitem[\protect\citeauthoryear{{Ibata} et~al.,}{{Ibata}
  et~al.}{2013}]{Ibata_2013}
{Ibata} R.~A.,  et~al., 2013, \mn@doi [Nature] {10.1038/nature11717}, \href
  {http://adsabs.harvard.edu/abs/2013Natur.493...62I} {493, 62}

\bibitem[\protect\citeauthoryear{{Ibata} et~al.,}{{Ibata}
  et~al.}{2014a}]{Ibata_2014_PANDAS}
{Ibata} R.~A.,  et~al., 2014a, \mn@doi [ApJ] {10.1088/0004-637X/780/2/128},
  \href {https://ui.adsabs.harvard.edu/abs/2014ApJ...780..128I} {780, 128}

\bibitem[\protect\citeauthoryear{{Ibata}, {Ibata}, {Lewis}, {Martin}, {Conn},
  {Elahi}, {Arias}  \& {Fernando}}{{Ibata} et~al.}{2014b}]{Ibata_2014}
{Ibata} R.~A.,  {Ibata} N.~G.,  {Lewis} G.~F.,  {Martin} N.~F.,  {Conn} A.,
  {Elahi} P.,  {Arias} V.,   {Fernando} N.,  2014b, \mn@doi [ApJL]
  {10.1088/2041-8205/784/1/L6}, \href
  {http://adsabs.harvard.edu/abs/2014ApJ...784L...6I} {784, L6}

\bibitem[\protect\citeauthoryear{{Jayaraman}, {Gilmore}, {Wyse}, {Norris}  \&
  {Belokurov}}{{Jayaraman} et~al.}{2013}]{Jayaraman_2013}
{Jayaraman} A.,  {Gilmore} G.,  {Wyse} R.~F.~G.,  {Norris} J.~E.,   {Belokurov}
  V.,  2013, \mn@doi [MNRAS] {10.1093/mnras/stt221}, \href
  {http://adsabs.harvard.edu/abs/2013MNRAS.431..930J} {431, 930}

\bibitem[\protect\citeauthoryear{{Juri{\'c}} et~al.,}{{Juri{\'c}}
  et~al.}{2008}]{Juric_2008}
{Juri{\'c}} M.,  et~al., 2008, \mn@doi [ApJ] {10.1086/523619}, \href
  {http://adsabs.harvard.edu/abs/2008ApJ...673..864J} {673, 864}

\bibitem[\protect\citeauthoryear{{Kahn} \& {Woltjer}}{{Kahn} \&
  {Woltjer}}{1959}]{Kahn_Woltjer_1959}
{Kahn} F.~D.,  {Woltjer} L.,  1959, \mn@doi [ApJ] {10.1086/146762}, \href
  {http://adsabs.harvard.edu/abs/1959ApJ...130..705K} {130, 705}

\bibitem[\protect\citeauthoryear{{Kallivayalil}, {\VAN{Van}{Van}{van}}~der
  Marel, {Besla}, {Anderson}  \& {Alcock}}{{Kallivayalil}
  et~al.}{2013}]{Kallivayalil_2013}
{Kallivayalil} N.,  {\VAN{Van}{Van}{van}}~der Marel R.~P.,  {Besla} G.,
  {Anderson} J.,   {Alcock} C.,  2013, \mn@doi [ApJ]
  {10.1088/0004-637X/764/2/161}, \href
  {http://adsabs.harvard.edu/abs/2013ApJ...764..161K} {764, 161}

\bibitem[\protect\citeauthoryear{{Katz}, {McGaugh}, {Teuben}  \&
  {Angus}}{{Katz} et~al.}{2013}]{Katz_2013}
{Katz} H.,  {McGaugh} S.,  {Teuben} P.,   {Angus} G.~W.,  2013, \mn@doi [ApJ]
  {10.1088/0004-637x/772/1/10}, \href
  {https://ui.adsabs.harvard.edu/abs/2013ApJ...772...10K/abstract} {772, 10}

\bibitem[\protect\citeauthoryear{{Keenan}, {Barger}  \& {Cowie}}{{Keenan}
  et~al.}{2013}]{Keenan_2013}
{Keenan} R.~C.,  {Barger} A.~J.,   {Cowie} L.~L.,  2013, \mn@doi [ApJ]
  {10.1088/0004-637X/775/1/62}, \href
  {https://ui.adsabs.harvard.edu/abs/2013ApJ...775...62K} {775, 62}

\bibitem[\protect\citeauthoryear{{Kilic}, {Munn}, {Harris}, {von Hippel},
  {Liebert}, {Williams}, {Jeffery}  \& {DeGennaro}}{{Kilic}
  et~al.}{2017}]{Kilic_2017}
{Kilic} M.,  {Munn} J.~A.,  {Harris} H.~C.,  {von Hippel} T.,  {Liebert} J.~W.,
   {Williams} K.~A.,  {Jeffery} E.,   {DeGennaro} S.,  2017, \mn@doi [ApJ]
  {10.3847/1538-4357/aa62a5}, \href
  {https://ui.adsabs.harvard.edu/abs/2017ApJ...837..162K} {837, 162}

\bibitem[\protect\citeauthoryear{{Klessen} \& {Kroupa}}{{Klessen} \&
  {Kroupa}}{1998}]{Klessen_1998}
{Klessen} R.~S.,  {Kroupa} P.,  1998, \mn@doi [ApJ] {10.1086/305540}, \href
  {https://ui.adsabs.harvard.edu/abs/1998ApJ...498..143K} {498, 143}

\bibitem[\protect\citeauthoryear{{Kogut} et~al.,}{{Kogut}
  et~al.}{1993}]{Kogut_1993}
{Kogut} A.,  et~al., 1993, \mn@doi [ApJ] {10.1086/173453}, \href
  {http://cdsads.u-strasbg.fr/abs/1993ApJ...419....1K} {419, 1}

\bibitem[\protect\citeauthoryear{{Koppelman}, {Helmi}, {Massari},
  {Price-Whelan}  \& {Starkenburg}}{{Koppelman} et~al.}{2019}]{Koppelman_2019}
{Koppelman} H.~H.,  {Helmi} A.,  {Massari} D.,  {Price-Whelan} A.~M.,
  {Starkenburg} T.~K.,  2019, \mn@doi [A\&A] {10.1051/0004-6361/201936738},
  \href {https://ui.adsabs.harvard.edu/abs/2019A\&A...631L...9K} {631, L9}

\bibitem[\protect\citeauthoryear{{Kourkchi} \& {Tully}}{{Kourkchi} \&
  {Tully}}{2017}]{Kourkchi_2017}
{Kourkchi} E.,  {Tully} R.~B.,  2017, \mn@doi [ApJ] {10.3847/1538-4357/aa76db},
  \href {http://adsabs.harvard.edu/abs/2017ApJ...843...16K} {843, 16}

\bibitem[\protect\citeauthoryear{{Kroupa}}{{Kroupa}}{1997}]{Kroupa_1997}
{Kroupa} P.,  1997, \mn@doi [New Astronomy] {10.1016/S1384-1076(97)00012-2},
  \href {https://ui.adsabs.harvard.edu/abs/1997NewA....2..139K} {2, 139}

\bibitem[\protect\citeauthoryear{{Kroupa}}{{Kroupa}}{2015}]{Kroupa_2015}
{Kroupa} P.,  2015, \mn@doi [Canadian Journal of Physics]
  {10.1139/cjp-2014-0179}, \href
  {http://adsabs.harvard.edu/abs/2015CaJPh..93..169K} {93, 169}

\bibitem[\protect\citeauthoryear{{Kroupa}, {Theis}  \& {Boily}}{{Kroupa}
  et~al.}{2005}]{Kroupa_2005}
{Kroupa} P.,  {Theis} C.,   {Boily} C.~M.,  2005, \mn@doi [A\&A]
  {10.1051/0004-6361:20041122}, \href
  {http://adsabs.harvard.edu/abs/2005A\%26A...431..517K} {431, 517}

\bibitem[\protect\citeauthoryear{{Kroupa} et~al.,}{{Kroupa}
  et~al.}{2018a}]{Kroupa_2018}
{Kroupa} P.,  et~al., 2018a, \mn@doi [Nature Astronomy]
  {10.1038/s41550-018-0622-x}, \href
  {http://adsabs.harvard.edu/abs/2018NatAs...2..925K} {2, 925}

\bibitem[\protect\citeauthoryear{{Kroupa} et~al.,}{{Kroupa}
  et~al.}{2018b}]{Kroupa_2018_Nature}
{Kroupa} P.,  et~al., 2018b, \mn@doi [Nature] {10.1038/s41586-018-0429-z},
  \href {http://adsabs.harvard.edu/abs/2018Natur.561E....4} {561, E4}

\bibitem[\protect\citeauthoryear{{Kunkel} \& {Demers}}{{Kunkel} \&
  {Demers}}{1976}]{Kunkel_1976}
{Kunkel} W.~E.,  {Demers} S.,  1976, in The Galaxy and the Local Group. Royal
  Greenwich Observatory Bulletins, p.~241

\bibitem[\protect\citeauthoryear{{Laporte}, {Johnston}, {G{\'o}mez},
  {Garavito-Camargo}  \& {Besla}}{{Laporte} et~al.}{2018}]{Laporte_2018}
{Laporte} C. F.~P.,  {Johnston} K.~V.,  {G{\'o}mez} F.~A.,  {Garavito-Camargo}
  N.,   {Besla} G.,  2018, \mn@doi [MNRAS] {10.1093/mnras/sty1574}, \href
  {https://ui.adsabs.harvard.edu/abs/2018MNRAS.481..286L} {481, 286}

\bibitem[\protect\citeauthoryear{{Lelli}, {McGaugh}, {Schombert}  \&
  {Pawlowski}}{{Lelli} et~al.}{2017}]{Lelli_2017}
{Lelli} F.,  {McGaugh} S.~S.,  {Schombert} J.~M.,   {Pawlowski} M.~S.,  2017,
  \mn@doi [ApJ] {10.3847/1538-4357/836/2/152}, \href
  {http://adsabs.harvard.edu/abs/2017ApJ...836..152L} {836, 152}

\bibitem[\protect\citeauthoryear{{Lelli}, {\DE{De}{De}{de}}~Breuck,
  {Falkendal}, {Fraternali}, {Man}, {Nesvadba}  \& {Lehnert}}{{Lelli}
  et~al.}{2018}]{Lelli_2018}
{Lelli} F.,  {\DE{De}{De}{de}}~Breuck C.,  {Falkendal} T.,  {Fraternali} F.,
  {Man} A. W.~S.,  {Nesvadba} N. P.~H.,   {Lehnert} M.~D.,  2018, \mn@doi
  [MNRAS] {10.1093/mnras/sty1795}, \href
  {https://ui.adsabs.harvard.edu/abs/2018MNRAS.479.5440L} {479, 5440}

\bibitem[\protect\citeauthoryear{{Lelli}, {McGaugh}, {Schombert}, {Desmond}  \&
  {Katz}}{{Lelli} et~al.}{2019}]{Lelli_2019}
{Lelli} F.,  {McGaugh} S.~S.,  {Schombert} J.~M.,  {Desmond} H.,   {Katz} H.,
  2019, \mn@doi [MNRAS] {10.1093/mnras/stz205}, \href
  {https://ui.adsabs.harvard.edu/abs/2019MNRAS.484.3267L} {484, 3267}

\bibitem[\protect\citeauthoryear{{Lelli}, {Di Teodoro}, {Fraternali}, {Man},
  {Zhang}, {\DE{De}{De}{de}}~Breuck, {Davis}  \& {Maiolino}}{{Lelli}
  et~al.}{2021}]{Lelli_2021}
{Lelli} F.,  {Di Teodoro} E.~M.,  {Fraternali} F.,  {Man} A. W.~S.,  {Zhang}
  Z.-Y.,  {\DE{De}{De}{de}}~Breuck C.,  {Davis} T.~A.,   {Maiolino} R.,  2021,
  \mn@doi [Science] {10.1126/science.abc1893}, \href
  {https://ui.adsabs.harvard.edu/abs/2021Sci...371..713L} {371, 713}

\bibitem[\protect\citeauthoryear{{Li} \& {Zhao}}{{Li} \&
  {Zhao}}{2017}]{Li_2017}
{Li} C.,  {Zhao} G.,  2017, \mn@doi [ApJ] {10.3847/1538-4357/aa93f4}, \href
  {https://ui.adsabs.harvard.edu/abs/2017ApJ...850...25L} {850, 25}

\bibitem[\protect\citeauthoryear{{Li}, {Lelli}, {McGaugh}  \& {Schombert}}{{Li}
  et~al.}{2018}]{Li_2018}
{Li} P.,  {Lelli} F.,  {McGaugh} S.,   {Schombert} J.,  2018, \mn@doi [A\&A]
  {10.1051/0004-6361/201732547}, \href
  {http://adsabs.harvard.edu/abs/2018arXiv180300022L} {615, A3}

\bibitem[\protect\citeauthoryear{{Li}, {Hammer}, {Babusiaux}, {Pawlowski},
  {Yang}, {Arenou}, {Du}  \& {Wang}}{{Li} et~al.}{2021}]{Li_2021}
{Li} H.,  {Hammer} F.,  {Babusiaux} C.,  {Pawlowski} M.~S.,  {Yang} Y.,
  {Arenou} F.,  {Du} C.,   {Wang} J.,  2021, \mn@doi [ApJ]
  {10.3847/1538-4357/ac0436}, \href
  {https://ui.adsabs.harvard.edu/abs/2021ApJ...916....8L} {916, 8}

\bibitem[\protect\citeauthoryear{{Lucchini}, {D'Onghia}, {Fox}, {Bustard},
  {Bland-Hawthorn}  \& {Zweibel}}{{Lucchini} et~al.}{2020}]{Lucchini_2020}
{Lucchini} S.,  {D'Onghia} E.,  {Fox} A.~J.,  {Bustard} C.,  {Bland-Hawthorn}
  J.,   {Zweibel} E.,  2020, Nature, \href
  {https://ui.adsabs.harvard.edu/abs/2020Natur.585..203L} {585, 203}

\bibitem[\protect\citeauthoryear{{Lucchini}, {D'Onghia}  \& {Fox}}{{Lucchini}
  et~al.}{2021}]{Lucchini_2021}
{Lucchini} S.,  {D'Onghia} E.,   {Fox} A.~J.,  2021, \mn@doi [ApJL]
  {10.3847/2041-8213/ac3338}, \href
  {https://ui.adsabs.harvard.edu/abs/2021ApJ...921L..36L} {921, L36}

\bibitem[\protect\citeauthoryear{{L{\"u}ghausen}, {Famaey}  \&
  {Kroupa}}{{L{\"u}ghausen} et~al.}{2015}]{Lughausen_2015}
{L{\"u}ghausen} F.,  {Famaey} B.,   {Kroupa} P.,  2015, \mn@doi [Canadian
  Journal of Physics] {10.1139/cjp-2014-0168}, \href
  {http://adsabs.harvard.edu/abs/2015CaJPh..93..232L} {93, 232}

\bibitem[\protect\citeauthoryear{{Lynden-Bell}}{{Lynden-Bell}}{1976}]{Lynden_Bell_1976}
{Lynden-Bell} D.,  1976, \mn@doi [MNRAS] {10.1093/mnras/174.3.695}, \href
  {http://adsabs.harvard.edu/abs/1976MNRAS.174..695L} {174, 695}

\bibitem[\protect\citeauthoryear{{Lynden-Bell}}{{Lynden-Bell}}{1982}]{Lynden_Bell_1982}
{Lynden-Bell} D.,  1982, The Observatory, \href
  {http://adsabs.harvard.edu/abs/1982Obs...102..202L} {102, 202}

\bibitem[\protect\citeauthoryear{{Ma} et~al.,}{{Ma} et~al.}{1998}]{Ma_1998}
{Ma} C.,  et~al., 1998, \mn@doi [AJ] {10.1086/300408}, \href
  {http://adsabs.harvard.edu/abs/1998AJ....116..516M} {116, 516}

\bibitem[\protect\citeauthoryear{{Mackey} \& {Gilmore}}{{Mackey} \&
  {Gilmore}}{2004}]{Mackey_2004}
{Mackey} A.~D.,  {Gilmore} G.~F.,  2004, \mn@doi [MNRAS]
  {10.1111/j.1365-2966.2004.08343.x}, \href
  {https://ui.adsabs.harvard.edu/abs/2004MNRAS.355..504M} {355, 504}

\bibitem[\protect\citeauthoryear{{Mashonkina}, {Neretina}, {Sitnova}  \&
  {Pakhomov}}{{Mashonkina} et~al.}{2019}]{Mashonkina_2019}
{Mashonkina} L.~I.,  {Neretina} M.~D.,  {Sitnova} T.~M.,   {Pakhomov} Y.~V.,
  2019, \mn@doi [Astronomy Reports] {10.1134/S1063772919090063}, \href
  {https://ui.adsabs.harvard.edu/abs/2019ARep...63..726M} {63, 726}

\bibitem[\protect\citeauthoryear{{Mateu} \& {Vivas}}{{Mateu} \&
  {Vivas}}{2018}]{Mateu_2018}
{Mateu} C.,  {Vivas} A.~K.,  2018, \mn@doi [MNRAS] {10.1093/mnras/sty1373},
  \href {https://ui.adsabs.harvard.edu/abs/2018MNRAS.479..211M} {479, 211}

\bibitem[\protect\citeauthoryear{{McConnachie}}{{McConnachie}}{2012}]{McConnachie_2012}
{McConnachie} A.~W.,  2012, \mn@doi [AJ] {10.1088/0004-6256/144/1/4}, \href
  {http://adsabs.harvard.edu/abs/2012AJ....144....4M} {144, 4}

\bibitem[\protect\citeauthoryear{{McConnachie} et~al.,}{{McConnachie}
  et~al.}{2018}]{McConnachie_2018}
{McConnachie} A.~W.,  et~al., 2018, \mn@doi [ApJ] {10.3847/1538-4357/aae8e7},
  \href {https://ui.adsabs.harvard.edu/abs/2018ApJ...868...55M} {868, 55}

\bibitem[\protect\citeauthoryear{{McGaugh}}{{McGaugh}}{2012}]{McGaugh_2012}
{McGaugh} S.~S.,  2012, \mn@doi [AJ] {10.1088/0004-6256/143/2/40}, \href
  {http://adsabs.harvard.edu/abs/2012AJ....143...40M} {143, 40}

\bibitem[\protect\citeauthoryear{{McGaugh}}{{McGaugh}}{2021}]{McGaugh_2020}
{McGaugh} S.~S.,  2021, \mn@doi [Studies in History and Philosophy of Science]
  {10.1016/j.shpsa.2021.05.008}, \href
  {https://ui.adsabs.harvard.edu/abs/2021SHPSA..88..220M} {88, 220}

\bibitem[\protect\citeauthoryear{{McGaugh} \& {Milgrom}}{{McGaugh} \&
  {Milgrom}}{2013a}]{McGaugh_2013a}
{McGaugh} S.,  {Milgrom} M.,  2013a, \mn@doi [ApJ]
  {10.1088/0004-637X/766/1/22}, \href
  {https://ui.adsabs.harvard.edu/abs/2013ApJ...766...22M} {766, 22}

\bibitem[\protect\citeauthoryear{{McGaugh} \& {Milgrom}}{{McGaugh} \&
  {Milgrom}}{2013b}]{McGaugh_2013b}
{McGaugh} S.,  {Milgrom} M.,  2013b, \mn@doi [ApJ]
  {10.1088/0004-637X/775/2/139}, \href
  {http://adsabs.harvard.edu/abs/2013ApJ...775..139M} {775, 139}

\bibitem[\protect\citeauthoryear{{McGaugh} \& {Wolf}}{{McGaugh} \&
  {Wolf}}{2010}]{McGaugh_2010}
{McGaugh} S.~S.,  {Wolf} J.,  2010, \mn@doi [ApJ]
  {10.1088/0004-637X/722/1/248}, \href
  {http://adsabs.harvard.edu/abs/2010ApJ...722..248M} {722, 248}

\bibitem[\protect\citeauthoryear{{McGaugh}, {Lelli}  \& {Schombert}}{{McGaugh}
  et~al.}{2016}]{McGaugh_Lelli_2016}
{McGaugh} S.,  {Lelli} F.,   {Schombert} J.,  2016, \mn@doi [Phys. Rev. Lett.]
  {10.1103/PhysRevLett.117.201101}, \href
  {http://adsabs.harvard.edu/abs/2016arXiv160905917M} {117, 201101}

\bibitem[\protect\citeauthoryear{{McGaugh}, {Lelli}, {Schombert}, {Li},
  {Visgaitis}, {Parker}  \& {Pawlowski}}{{McGaugh} et~al.}{2021}]{McGaugh_2021}
{McGaugh} S.~S.,  {Lelli} F.,  {Schombert} J.~M.,  {Li} P.,  {Visgaitis} T.,
  {Parker} K.~S.,   {Pawlowski} M.~S.,  2021, \mn@doi [AJ]
  {10.3847/1538-3881/ac2502}, \href
  {https://ui.adsabs.harvard.edu/abs/2021AJ....162..202M} {162, 202}

\bibitem[\protect\citeauthoryear{{McMillan}}{{McMillan}}{2011}]{McMillan_2011}
{McMillan} P.~J.,  2011, \mn@doi [MNRAS] {10.1111/j.1365-2966.2011.18564.x},
  \href {http://adsabs.harvard.edu/abs/2011MNRAS.414.2446M} {414, 2446}

\bibitem[\protect\citeauthoryear{{McMillan}}{{McMillan}}{2017}]{McMillan_2017}
{McMillan} P.~J.,  2017, \mn@doi [MNRAS] {10.1093/mnras/stw2759}, \href
  {http://adsabs.harvard.edu/abs/2017MNRAS.465...76M} {465, 76}

\bibitem[\protect\citeauthoryear{{Metz}, {Kroupa}  \& {Jerjen}}{{Metz}
  et~al.}{2007}]{Metz_2007}
{Metz} M.,  {Kroupa} P.,   {Jerjen} H.,  2007, \mn@doi [MNRAS]
  {10.1111/j.1365-2966.2006.11228.x}, \href
  {http://adsabs.harvard.edu/abs/2007MNRAS.374.1125M} {374, 1125}

\bibitem[\protect\citeauthoryear{{Metz}, {Kroupa}  \& {Libeskind}}{{Metz}
  et~al.}{2008}]{Metz_2008}
{Metz} M.,  {Kroupa} P.,   {Libeskind} N.~I.,  2008, \mn@doi [ApJ]
  {10.1086/587833}, \href {http://adsabs.harvard.edu/abs/2008ApJ...680..287M}
  {680, 287}

\bibitem[\protect\citeauthoryear{{Metz}, {Kroupa}  \& {Jerjen}}{{Metz}
  et~al.}{2009a}]{Metz_2009}
{Metz} M.,  {Kroupa} P.,   {Jerjen} H.,  2009a, \mn@doi [MNRAS]
  {10.1111/j.1365-2966.2009.14489.x}, \href
  {http://adsabs.harvard.edu/abs/2009MNRAS.394.2223M} {394, 2223}

\bibitem[\protect\citeauthoryear{{Metz}, {Kroupa}, {Theis}, {Hensler}  \&
  {Jerjen}}{{Metz} et~al.}{2009b}]{Metz_2009_group}
{Metz} M.,  {Kroupa} P.,  {Theis} C.,  {Hensler} G.,   {Jerjen} H.,  2009b,
  \mn@doi [ApJ] {10.1088/0004-637X/697/1/269}, \href
  {https://ui.adsabs.harvard.edu/abs/2009ApJ...697..269M} {697, 269}

\bibitem[\protect\citeauthoryear{{Milgrom}}{{Milgrom}}{1983}]{Milgrom_1983}
{Milgrom} M.,  1983, \mn@doi [ApJ] {10.1086/161130}, \href
  {http://adsabs.harvard.edu/abs/1983ApJ...270..365M} {270, 365}

\bibitem[\protect\citeauthoryear{{Milgrom}}{{Milgrom}}{1986}]{Milgrom_1986}
{Milgrom} M.,  1986, \mn@doi [ApJ] {10.1086/164021}, \href
  {http://adsabs.harvard.edu/abs/1986ApJ...302..617M} {302, 617}

\bibitem[\protect\citeauthoryear{{Milgrom}}{{Milgrom}}{2010}]{QUMOND}
{Milgrom} M.,  2010, \mn@doi [MNRAS] {10.1111/j.1365-2966.2009.16184.x}, \href
  {http://adsabs.harvard.edu/abs/2010MNRAS.403..886M} {403, 886}

\bibitem[\protect\citeauthoryear{{Milgrom}}{{Milgrom}}{2012}]{Milgrom_2012}
{Milgrom} M.,  2012, \mn@doi [Phys. Rev. Lett.]
  {10.1103/PhysRevLett.109.131101}, \href
  {https://ui.adsabs.harvard.edu/abs/2012PhRvL.109m1101M} {109, 131101}

\bibitem[\protect\citeauthoryear{{Milgrom}}{{Milgrom}}{2013}]{Milgrom_2013}
{Milgrom} M.,  2013, \mn@doi [Physical Review Letters]
  {10.1103/PhysRevLett.111.041105}, \href
  {http://adsabs.harvard.edu/abs/2013PhRvL.111d1105M} {111, 041105}

\bibitem[\protect\citeauthoryear{{Milgrom}}{{Milgrom}}{2015}]{Milgrom_2015_a0_variation}
{Milgrom} M.,  2015, \mn@doi [Physical Review D] {10.1103/PhysRevD.91.044009},
  \href {https://ui.adsabs.harvard.edu/abs/2015PhRvD..91d4009M} {91, 044009}

\bibitem[\protect\citeauthoryear{{Milgrom}}{{Milgrom}}{2017}]{Milgrom_2017}
{Milgrom} M.,  2017, preprint, \href
  {https://ui.adsabs.harvard.edu/abs/2017arXiv170306110M} {Arxiv} (\mn@eprint
  {arXiv} {1703.06110})

\bibitem[\protect\citeauthoryear{{Mirabel}, {Dottori}  \& {Lutz}}{{Mirabel}
  et~al.}{1992}]{Mirabel_1992}
{Mirabel} I.~F.,  {Dottori} H.,   {Lutz} D.,  1992, A\&A, \href
  {http://adsabs.harvard.edu/abs/1992A\%26A...256L..19M} {256, L19}

\bibitem[\protect\citeauthoryear{{M{\"u}ller}, {Jerjen}, {Pawlowski}  \&
  {Binggeli}}{{M{\"u}ller} et~al.}{2016}]{Muller_2016}
{M{\"u}ller} O.,  {Jerjen} H.,  {Pawlowski} M.~S.,   {Binggeli} B.,  2016,
  \mn@doi [A\&A] {10.1051/0004-6361/201629298}, \href
  {http://adsabs.harvard.edu/abs/2016A\%26A...595A.119M} {595, A119}

\bibitem[\protect\citeauthoryear{{M{\"u}ller}, {Pawlowski}, {Jerjen}  \&
  {Lelli}}{{M{\"u}ller} et~al.}{2018a}]{Muller_2018}
{M{\"u}ller} O.,  {Pawlowski} M.~S.,  {Jerjen} H.,   {Lelli} F.,  2018a,
  \mn@doi [Science] {10.1126/science.aao1858}, \href
  {http://adsabs.harvard.edu/abs/2018arXiv180200081M} {359, 534}

\bibitem[\protect\citeauthoryear{{M{\"u}ller}, {Rejkuba}  \&
  {Jerjen}}{{M{\"u}ller} et~al.}{2018b}]{Muller_2018_M83}
{M{\"u}ller} O.,  {Rejkuba} M.,   {Jerjen} H.,  2018b, \mn@doi [A\&A]
  {10.1051/0004-6361/201732455}, \href
  {https://ui.adsabs.harvard.edu/abs/2018A\&A...615A..96M} {615, A96}

\bibitem[\protect\citeauthoryear{{M{\"u}ller} et~al.,}{{M{\"u}ller}
  et~al.}{2021}]{Muller_2021}
{M{\"u}ller} O.,  et~al., 2021, \mn@doi [A\&A] {10.1051/0004-6361/202039973},
  \href {https://ui.adsabs.harvard.edu/abs/2021A\%26A...645L...5M} {645, L5}

\bibitem[\protect\citeauthoryear{{Myeong}, {Evans}, {Belokurov}, {Sanders}  \&
  {Koposov}}{{Myeong} et~al.}{2018}]{Myeong_2018}
{Myeong} G.~C.,  {Evans} N.~W.,  {Belokurov} V.,  {Sanders} J.~L.,   {Koposov}
  S.~E.,  2018, \mn@doi [ApJL] {10.3847/2041-8213/aad7f7}, \href
  {https://ui.adsabs.harvard.edu/abs/2018ApJ...863L..28M} {863, L28}

\bibitem[\protect\citeauthoryear{{Myeong}, {Vasiliev}, {Iorio}, {Evans}  \&
  {Belokurov}}{{Myeong} et~al.}{2019}]{Myeong_2019}
{Myeong} G.~C.,  {Vasiliev} E.,  {Iorio} G.,  {Evans} N.~W.,   {Belokurov} V.,
  2019, \mn@doi [MNRAS] {10.1093/mnras/stz1770}, \href
  {https://ui.adsabs.harvard.edu/abs/2019MNRAS.488.1235M} {488, 1235}

\bibitem[\protect\citeauthoryear{{Nagesh}, {Banik}, {Thies}, {Kroupa},
  {Famaey}, {Wittenburg}, {Parziale}  \& {Haslbauer}}{{Nagesh}
  et~al.}{2021}]{Nagesh_2021}
{Nagesh} S.~T.,  {Banik} I.,  {Thies} I.,  {Kroupa} P.,  {Famaey} B.,
  {Wittenburg} N.,  {Parziale} R.,   {Haslbauer} M.,  2021, \mn@doi [Canadian
  Journal of Physics] {10.1139/cjp-2020-0624}, \href
  {https://ui.adsabs.harvard.edu/abs/2021CaJPh..99..607N} {99, 607}

\bibitem[\protect\citeauthoryear{{Neeleman}, {Prochaska}, {Kanekar}  \&
  {Rafelski}}{{Neeleman} et~al.}{2020}]{Neeleman_2020}
{Neeleman} M.,  {Prochaska} J.~X.,  {Kanekar} N.,   {Rafelski} M.,  2020,
  \mn@doi [Nature] {10.1038/s41586-020-2276-y}, \href
  {https://ui.adsabs.harvard.edu/abs/2020Natur.581..269N} {581, 269}

\bibitem[\protect\citeauthoryear{{Nidever} et~al.,}{{Nidever}
  et~al.}{2021}]{Nidever_2021}
{Nidever} D.~L.,  et~al., 2021, \mn@doi [AJ] {10.3847/1538-3881/abceb7}, \href
  {https://ui.adsabs.harvard.edu/abs/2021AJ....161...74N} {161, 74}

\bibitem[\protect\citeauthoryear{{Oria} et~al.,}{{Oria}
  et~al.}{2021}]{Oria_2021}
{Oria} P.~A.,  et~al., 2021, \mn@doi [\apj] {10.3847/1538-4357/ac273d}, \href
  {https://ui.adsabs.harvard.edu/abs/2021ApJ...923...68O} {923, 68}

\bibitem[\protect\citeauthoryear{{Ostriker} \& {Steinhardt}}{{Ostriker} \&
  {Steinhardt}}{1995}]{Ostriker_Steinhardt_1995}
{Ostriker} J.~P.,  {Steinhardt} P.~J.,  1995, \mn@doi [Nature]
  {10.1038/377600a0}, \href {http://adsabs.harvard.edu/abs/1995Natur.377..600O}
  {377, 600}

\bibitem[\protect\citeauthoryear{{Pawlowski}}{{Pawlowski}}{2018}]{Pawlowski_2018}
{Pawlowski} M.~S.,  2018, \mn@doi [Modern Physics Letters A]
  {10.1142/S0217732318300045}, \href
  {http://adsabs.harvard.edu/abs/2018arXiv180202579P} {33, 1830004}

\bibitem[\protect\citeauthoryear{{Pawlowski}}{{Pawlowski}}{2021}]{Pawlowski_2021}
{Pawlowski} M.~S.,  2021, \mn@doi [Galaxies] {10.3390/galaxies9030066}, \href
  {https://ui.adsabs.harvard.edu/abs/2021Galax...9...66P} {9, 66}

\bibitem[\protect\citeauthoryear{{Pawlowski} \& {Kroupa}}{{Pawlowski} \&
  {Kroupa}}{2013}]{Pawlowski_2013_VPOS}
{Pawlowski} M.~S.,  {Kroupa} P.,  2013, \mn@doi [MNRAS]
  {10.1093/mnras/stt1429}, \href
  {http://adsabs.harvard.edu/abs/2013MNRAS.435.2116P} {435, 2116}

\bibitem[\protect\citeauthoryear{{Pawlowski} \& {Kroupa}}{{Pawlowski} \&
  {Kroupa}}{2020}]{Pawlowski_2020}
{Pawlowski} M.~S.,  {Kroupa} P.,  2020, \mn@doi [MNRAS]
  {10.1093/mnras/stz3163}, \href
  {https://ui.adsabs.harvard.edu/abs/2020MNRAS.491.3042P} {491, 3042}

\bibitem[\protect\citeauthoryear{{Pawlowski} \& {McGaugh}}{{Pawlowski} \&
  {McGaugh}}{2014}]{Pawlowski_McGaugh_2014}
{Pawlowski} M.~S.,  {McGaugh} S.~S.,  2014, \mn@doi [MNRAS]
  {10.1093/mnras/stu321}, \href
  {http://adsabs.harvard.edu/abs/2014MNRAS.440..908P} {440, 908}

\bibitem[\protect\citeauthoryear{{Pawlowski} \& {Tony Sohn}}{{Pawlowski} \&
  {Tony Sohn}}{2021}]{Pawlowski_Sohn_2021}
{Pawlowski} M.~S.,  {Tony Sohn} S.,  2021, \mn@doi [ApJ]
  {10.3847/1538-4357/ac2aa9}, \href
  {https://ui.adsabs.harvard.edu/abs/2021ApJ...923...42P} {923, 42}

\bibitem[\protect\citeauthoryear{{Pawlowski}, {Kroupa}  \&
  {\DE{De}{De}{de}}~Boer}{{Pawlowski} et~al.}{2011}]{Pawlowski_2011}
{Pawlowski} M.~S.,  {Kroupa} P.,   {\DE{De}{De}{de}}~Boer K.~S.,  2011, \mn@doi
  [A\&A] {10.1051/0004-6361/201015021}, \href
  {http://adsabs.harvard.edu/abs/2011A\%26A...532A.118P} {532, A118}

\bibitem[\protect\citeauthoryear{{Pawlowski}, {Pflamm-Altenburg}  \&
  {Kroupa}}{{Pawlowski} et~al.}{2012}]{Pawlowski_2012}
{Pawlowski} M.~S.,  {Pflamm-Altenburg} J.,   {Kroupa} P.,  2012, \mn@doi
  [MNRAS] {10.1111/j.1365-2966.2012.20937.x}, \href
  {http://adsabs.harvard.edu/abs/2012MNRAS.423.1109P} {423, 1109}

\bibitem[\protect\citeauthoryear{{Pawlowski}, {Kroupa}  \&
  {Jerjen}}{{Pawlowski} et~al.}{2013}]{Pawlowski_2013_LG}
{Pawlowski} M.~S.,  {Kroupa} P.,   {Jerjen} H.,  2013, \mn@doi [MNRAS]
  {10.1093/mnras/stt1384}, \href
  {http://adsabs.harvard.edu/abs/2013MNRAS.435.1928P} {435, 1928}

\bibitem[\protect\citeauthoryear{{Pawlowski} et~al.,}{{Pawlowski}
  et~al.}{2014}]{Pawlowski_2014}
{Pawlowski} M.~S.,  et~al., 2014, \mn@doi [MNRAS] {10.1093/mnras/stu1005},
  \href {http://adsabs.harvard.edu/abs/2014MNRAS.442.2362P} {442, 2362}

\bibitem[\protect\citeauthoryear{{Pawlowski}, {McGaugh}  \&
  {Jerjen}}{{Pawlowski} et~al.}{2015}]{Pawlowski_McGaugh_2015}
{Pawlowski} M.~S.,  {McGaugh} S.~S.,   {Jerjen} H.,  2015, \mn@doi [MNRAS]
  {10.1093/mnras/stv1588}, \href
  {http://adsabs.harvard.edu/abs/2015MNRAS.453.1047P} {453, 1047}

\bibitem[\protect\citeauthoryear{{Pawlowski} et~al.,}{{Pawlowski}
  et~al.}{2017}]{Pawlowski_2017}
{Pawlowski} M.~S.,  et~al., 2017, \mn@doi [Astronomische Nachrichten]
  {10.1002/asna.201713366}, \href
  {http://adsabs.harvard.edu/abs/2017AN....338..854P} {338, 854}

\bibitem[\protect\citeauthoryear{{Pawlowski}, {Oria}, {Taibi}, {Famaey}  \&
  {Ibata}}{{Pawlowski} et~al.}{2021}]{Pawlowski_2022}
{Pawlowski} M.~S.,  {Oria} P.-A.,  {Taibi} S.,  {Famaey} B.,   {Ibata} R.,
  2021, preprint, \href {https://ui.adsabs.harvard.edu/abs/2021arXiv211105358P}
  {Arxiv} (\mn@eprint {arXiv} {2111.05358})

\bibitem[\protect\citeauthoryear{{Peebles}}{{Peebles}}{2017}]{Peebles_2017}
{Peebles} P.~J.~E.,  2017, preprint, \href
  {http://adsabs.harvard.edu/abs/2017arXiv170510683P} {Arxiv} (\mn@eprint
  {arXiv} {1705.10683})

\bibitem[\protect\citeauthoryear{{Peebles}}{{Peebles}}{2020}]{Peebles_2020}
{Peebles} P.~J.~E.,  2020, \mn@doi [MNRAS] {10.1093/mnras/staa2649}, \href
  {https://ui.adsabs.harvard.edu/abs/2020MNRAS.498.4386P} {498, 4386}

\bibitem[\protect\citeauthoryear{{Peebles} \& {Nusser}}{{Peebles} \&
  {Nusser}}{2010}]{Peebles_2010}
{Peebles} P.~J.~E.,  {Nusser} A.,  2010, \mn@doi [Nature]
  {10.1038/nature09101}, \href
  {http://adsabs.harvard.edu/abs/2010Natur.465..565P} {465, 565}

\bibitem[\protect\citeauthoryear{{Perret}, {Renaud}, {Epinat}, {Amram},
  {Bournaud}, {Contini}, {Teyssier}  \& {Lambert}}{{Perret}
  et~al.}{2014}]{Perret_2014}
{Perret} V.,  {Renaud} F.,  {Epinat} B.,  {Amram} P.,  {Bournaud} F.,
  {Contini} T.,  {Teyssier} R.,   {Lambert} J.-C.,  2014, \mn@doi [A\&A]
  {10.1051/0004-6361/201322395}, \href
  {https://ui.adsabs.harvard.edu/abs/2014A\%26A...562A...1P} {562, A1}

\bibitem[\protect\citeauthoryear{{Peschken} \& {{\L}okas}}{{Peschken} \&
  {{\L}okas}}{2019}]{Peschken_2019}
{Peschken} N.,  {{\L}okas} E.~L.,  2019, \mn@doi [MNRAS]
  {10.1093/mnras/sty3277}, \href
  {https://ui.adsabs.harvard.edu/abs/2019MNRAS.483.2721P} {483, 2721}

\bibitem[\protect\citeauthoryear{{Pietrzy{\'n}ski} et~al.,}{{Pietrzy{\'n}ski}
  et~al.}{2013}]{Pietrzynski_2013}
{Pietrzy{\'n}ski} G.,  et~al., 2013, \mn@doi [Nature] {10.1038/nature11878},
  \href {http://adsabs.harvard.edu/abs/2013Natur.495...76P} {495, 76}

\bibitem[\protect\citeauthoryear{{Planck Collaboration VI}}{{Planck
  Collaboration VI}}{2020}]{Planck_2020}
{Planck Collaboration VI} 2020, \mn@doi [A\&A] {10.1051/0004-6361/201833910},
  \href {https://ui.adsabs.harvard.edu/abs/2020A\&A...641A...6P} {641, A6}

\bibitem[\protect\citeauthoryear{{Planck Collaboration XVI}}{{Planck
  Collaboration XVI}}{2014}]{Planck_2014_cosmology}
{Planck Collaboration XVI} 2014, \mn@doi [A\&A] {10.1051/0004-6361/201321591},
  \href {https://ui.adsabs.harvard.edu/abs/2014A\&A...571A..16P} {571, A16}

\bibitem[\protect\citeauthoryear{{Poggio}, {Drimmel}, {Andrae}, {Bailer-Jones},
  {Fouesneau}, {Lattanzi}, {Smart}  \& {Spagna}}{{Poggio}
  et~al.}{2020}]{Poggio_2020}
{Poggio} E.,  {Drimmel} R.,  {Andrae} R.,  {Bailer-Jones} C.~A.~L.,
  {Fouesneau} M.,  {Lattanzi} M.~G.,  {Smart} R.~L.,   {Spagna} A.,  2020,
  \mn@doi [Nature Astronomy] {10.1038/s41550-020-1017-3}, \href
  {https://ui.adsabs.harvard.edu/abs/2020NatAs...4..590P} {4, 590}

\bibitem[\protect\citeauthoryear{{Recchi}, {Kroupa}  \& {Ploeckinger}}{{Recchi}
  et~al.}{2015}]{Recchi_2015}
{Recchi} S.,  {Kroupa} P.,   {Ploeckinger} S.,  2015, \mn@doi [MNRAS]
  {10.1093/mnras/stv798}, \href
  {http://adsabs.harvard.edu/abs/2015MNRAS.450.2367R} {450, 2367}

\bibitem[\protect\citeauthoryear{{Renaud}, {Famaey}  \& {Kroupa}}{{Renaud}
  et~al.}{2016}]{Renaud_2016}
{Renaud} F.,  {Famaey} B.,   {Kroupa} P.,  2016, \mn@doi [MNRAS]
  {10.1093/mnras/stw2331}, \href
  {http://adsabs.harvard.edu/abs/2016MNRAS.463.3637R} {463, 3637}

\bibitem[\protect\citeauthoryear{{Riess}}{{Riess}}{2020}]{Riess_2020}
{Riess} A.~G.,  2020, \mn@doi [Nat. Rev. Phys.] {10.1038/s42254-019-0137-0},
  \href {https://ui.adsabs.harvard.edu/abs/2019NatRP...2...10R} {2, 10}

\bibitem[\protect\citeauthoryear{{Riess}, {Casertano}, {Yuan}, {Bowers},
  {Macri}, {Zinn}  \& {Scolnic}}{{Riess} et~al.}{2021a}]{Riess_2021}
{Riess} A.~G.,  {Casertano} S.,  {Yuan} W.,  {Bowers} J.~B.,  {Macri} L.,
  {Zinn} J.~C.,   {Scolnic} D.,  2021a, \mn@doi [ApJL]
  {10.3847/2041-8213/abdbaf}, \href
  {https://ui.adsabs.harvard.edu/abs/2021ApJ...908L...6R} {908, L6}

\bibitem[\protect\citeauthoryear{{Riess} et~al.,}{{Riess}
  et~al.}{2021b}]{Riess_2022}
{Riess} A.~G.,  et~al., 2021b, preprint, \href
  {https://ui.adsabs.harvard.edu/abs/2021arXiv211204510R} {Arxiv} (\mn@eprint
  {arXiv} {2112.04510})

\bibitem[\protect\citeauthoryear{{Riley} et~al.,}{{Riley}
  et~al.}{2019}]{Riley_2019}
{Riley} A.~H.,  et~al., 2019, \mn@doi [MNRAS] {10.1093/mnras/stz973}, \href
  {https://ui.adsabs.harvard.edu/abs/2019MNRAS.486.2679R} {486, 2679}

\bibitem[\protect\citeauthoryear{{Rizzo}, {Vegetti}, {Powell}, {Fraternali},
  {McKean}, {Stacey}  \& {White}}{{Rizzo} et~al.}{2020}]{Rizzo_2020}
{Rizzo} F.,  {Vegetti} S.,  {Powell} D.,  {Fraternali} F.,  {McKean} J.~P.,
  {Stacey} H.~R.,   {White} S.~D.~M.,  2020, \mn@doi [Nature]
  {10.1038/s41586-020-2572-6}, \href
  {https://ui.adsabs.harvard.edu/abs/2020Natur.584..201R} {584, 201}

\bibitem[\protect\citeauthoryear{{Robotham} et~al.,}{{Robotham}
  et~al.}{2012}]{Robotham_2012}
{Robotham} A.~S.~G.,  et~al., 2012, \mn@doi [MNRAS]
  {10.1111/j.1365-2966.2012.21332.x}, \href
  {https://ui.adsabs.harvard.edu/abs/2012MNRAS.424.1448R} {424, 1448}

\bibitem[\protect\citeauthoryear{{Roshan}, {Banik}, {Ghafourian}, {Thies},
  {Famaey}, {Asencio}  \& {Kroupa}}{{Roshan}
  et~al.}{2021a}]{Roshan_2021_disc_stability}
{Roshan} M.,  {Banik} I.,  {Ghafourian} N.,  {Thies} I.,  {Famaey} B.,
  {Asencio} E.,   {Kroupa} P.,  2021a, \mn@doi [MNRAS] {10.1093/mnras/stab651},
  \href {https://ui.adsabs.harvard.edu/abs/2021MNRAS.503.2833R} {503, 2833}

\bibitem[\protect\citeauthoryear{{Roshan}, {Ghafourian}, {Kashfi}, {Banik},
  {Haslbauer}, {Cuomo}, {Famaey}  \& {Kroupa}}{{Roshan}
  et~al.}{2021b}]{Roshan_2021_bar_speed}
{Roshan} M.,  {Ghafourian} N.,  {Kashfi} T.,  {Banik} I.,  {Haslbauer} M.,
  {Cuomo} V.,  {Famaey} B.,   {Kroupa} P.,  2021b, \mn@doi [MNRAS]
  {10.1093/mnras/stab2553}, \href
  {https://ui.adsabs.harvard.edu/abs/2021MNRAS.508..926R} {508, 926}

\bibitem[\protect\citeauthoryear{{Ruiz-Lara}, {Gallart}, {Bernard}  \&
  {Cassisi}}{{Ruiz-Lara} et~al.}{2020}]{Ruiz_2020}
{Ruiz-Lara} T.,  {Gallart} C.,  {Bernard} E.~J.,   {Cassisi} S.,  2020, \mn@doi
  [Nature Astronomy] {10.1038/s41550-020-1097-0}, \href
  {https://ui.adsabs.harvard.edu/abs/2020NatAs...4..965R} {4, 965}

\bibitem[\protect\citeauthoryear{{Salaris} \& {Weiss}}{{Salaris} \&
  {Weiss}}{2002}]{Salaris_2002}
{Salaris} M.,  {Weiss} A.,  2002, \mn@doi [A\&A] {10.1051/0004-6361:20020554},
  \href {https://ui.adsabs.harvard.edu/abs/2002A\&A...388..492S} {388, 492}

\bibitem[\protect\citeauthoryear{{Salomon}, {Ibata}, {Reyl{\'e}}, {Famaey},
  {Libeskind}, {McConnachie}  \& {Hoffman}}{{Salomon}
  et~al.}{2021}]{Salomon_2021}
{Salomon} J.~B.,  {Ibata} R.,  {Reyl{\'e}} C.,  {Famaey} B.,  {Libeskind}
  N.~I.,  {McConnachie} A.~W.,   {Hoffman} Y.,  2021, \mn@doi [MNRAS]
  {10.1093/mnras/stab2253}, \href
  {https://ui.adsabs.harvard.edu/abs/2021MNRAS.507.2592S} {507, 2592}

\bibitem[\protect\citeauthoryear{{Samuel}, {Wetzel}, {Chapman}, {Tollerud},
  {Hopkins}, {Boylan-Kolchin}, {Bailin}  \& {Faucher-Gigu{\`e}re}}{{Samuel}
  et~al.}{2021}]{Samuel_2021}
{Samuel} J.,  {Wetzel} A.,  {Chapman} S.,  {Tollerud} E.,  {Hopkins} P.~F.,
  {Boylan-Kolchin} M.,  {Bailin} J.,   {Faucher-Gigu{\`e}re} C.-A.,  2021,
  \mn@doi [MNRAS] {10.1093/mnras/stab955}, \href
  {https://ui.adsabs.harvard.edu/abs/2021MNRAS.504.1379S} {504, 1379}

\bibitem[\protect\citeauthoryear{{Sanders}}{{Sanders}}{1998}]{Sanders_1998}
{Sanders} R.~H.,  1998, \mn@doi [MNRAS] {10.1046/j.1365-8711.1998.01459.x},
  \href {https://ui.adsabs.harvard.edu/abs/1998MNRAS.296.1009S} {296, 1009}

\bibitem[\protect\citeauthoryear{{Sanders}}{{Sanders}}{2019}]{Sanders_2019}
{Sanders} R.~H.,  2019, \mn@doi [MNRAS] {10.1093/mnras/stz353}, \href
  {https://ui.adsabs.harvard.edu/abs/2019MNRAS.485..513S} {485, 513}

\bibitem[\protect\citeauthoryear{{Santos-Santos}, {Dom{\'\i}nguez-Tenreiro}  \&
  {Pawlowski}}{{Santos-Santos} et~al.}{2020}]{Isabel_2020}
{Santos-Santos} I.~M.,  {Dom{\'\i}nguez-Tenreiro} R.,   {Pawlowski} M.~S.,
  2020, \mn@doi [MNRAS] {10.1093/mnras/staa3130}, \href
  {https://ui.adsabs.harvard.edu/abs/2020MNRAS.499.3755S} {499, 3755}

\bibitem[\protect\citeauthoryear{{Santos-Santos}, {Fattahi}, {Sales}  \&
  {Navarro}}{{Santos-Santos} et~al.}{2021}]{Isabel_2021}
{Santos-Santos} I. M.~E.,  {Fattahi} A.,  {Sales} L.~V.,   {Navarro} J.~F.,
  2021, \mn@doi [MNRAS] {10.1093/mnras/stab1020}, \href
  {https://ui.adsabs.harvard.edu/abs/2021MNRAS.504.4551S} {504, 4551}

\bibitem[\protect\citeauthoryear{{Savino}, {Koch}, {Prudil}, {Kunder}  \&
  {Smolec}}{{Savino} et~al.}{2020}]{Savino_2020}
{Savino} A.,  {Koch} A.,  {Prudil} Z.,  {Kunder} A.,   {Smolec} R.,  2020,
  \mn@doi [A\&A] {10.1051/0004-6361/202038305}, \href
  {https://ui.adsabs.harvard.edu/abs/2020A\&A...641A..96S} {641, A96}

\bibitem[\protect\citeauthoryear{{Schrabback} et~al.,}{{Schrabback}
  et~al.}{2021}]{Schrabback_2021}
{Schrabback} T.,  et~al., 2021, \mn@doi [A\&A] {10.1051/0004-6361/202037670},
  \href {https://ui.adsabs.harvard.edu/abs/2021A\&A...646A..73S} {646, A73}

\bibitem[\protect\citeauthoryear{{Sellwood}, {Shen}  \& {Li}}{{Sellwood}
  et~al.}{2019}]{Sellwood_2019}
{Sellwood} J.~A.,  {Shen} J.,   {Li} Z.,  2019, \mn@doi [MNRAS]
  {10.1093/mnras/stz1145}, \href
  {https://ui.adsabs.harvard.edu/abs/2019MNRAS.486.4710S} {486, 4710}

\bibitem[\protect\citeauthoryear{{Shao}, {Cautun}, {Frenk}, {Grand},
  {G{\'o}mez}, {Marinacci}  \& {Simpson}}{{Shao} et~al.}{2018}]{Shao_2018}
{Shao} S.,  {Cautun} M.,  {Frenk} C.~S.,  {Grand} R.~J.~J.,  {G{\'o}mez} F.~A.,
   {Marinacci} F.,   {Simpson} C.~M.,  2018, \mn@doi [MNRAS]
  {10.1093/mnras/sty343}, \href
  {http://adsabs.harvard.edu/abs/2018MNRAS.476.1796S} {476, 1796}

\bibitem[\protect\citeauthoryear{{Sharma}, {Salucci}, {Harrison},
  {\VAN{Van}{Van}{van}}~de Ven  \& {Lapi}}{{Sharma} et~al.}{2021}]{Sharma_2021}
{Sharma} G.,  {Salucci} P.,  {Harrison} C.~M.,  {\VAN{Van}{Van}{van}}~de Ven
  G.,   {Lapi} A.,  2021, \mn@doi [MNRAS] {10.1093/mnras/stab249}, \href
  {https://ui.adsabs.harvard.edu/abs/2021MNRAS.503.1753S} {503, 1753}

\bibitem[\protect\citeauthoryear{{Shelest} \& {Lelli}}{{Shelest} \&
  {Lelli}}{2020}]{Shelest_2020}
{Shelest} A.,  {Lelli} F.,  2020, \mn@doi [A\&A] {10.1051/0004-6361/202038184},
  \href {https://ui.adsabs.harvard.edu/abs/2020A\&A...641A..31S} {641, A31}

\bibitem[\protect\citeauthoryear{{Skordis} \& {Z{\l}o{\'s}nik}}{{Skordis} \&
  {Z{\l}o{\'s}nik}}{2019}]{Skordis_2019}
{Skordis} C.,  {Z{\l}o{\'s}nik} T.,  2019, \mn@doi [Physical Review D]
  {10.1103/PhysRevD.100.104013}, \href
  {https://ui.adsabs.harvard.edu/abs/2019PhRvD.100j4013S} {100, 104013}

\bibitem[\protect\citeauthoryear{{Skordis} \& {Z{\l}o{\'s}nik}}{{Skordis} \&
  {Z{\l}o{\'s}nik}}{2021}]{Skordis_2021}
{Skordis} C.,  {Z{\l}o{\'s}nik} T.,  2021, \mn@doi [Physical Review Letters]
  {10.1103/PhysRevLett.127.161302}, \href
  {https://ui.adsabs.harvard.edu/abs/2021PhRvL.127p1302S} {127, 161302}

\bibitem[\protect\citeauthoryear{{Snaith}, {Haywood}, {Di Matteo}, {Lehnert},
  {Combes}, {Katz}  \& {G{\'o}mez}}{{Snaith} et~al.}{2015}]{Snaith_2015}
{Snaith} O.,  {Haywood} M.,  {Di Matteo} P.,  {Lehnert} M.~D.,  {Combes} F.,
  {Katz} D.,   {G{\'o}mez} A.,  2015, \mn@doi [A\&A]
  {10.1051/0004-6361/201424281}, \href
  {https://ui.adsabs.harvard.edu/abs/2015A\&A...578A..87S} {578, A87}

\bibitem[\protect\citeauthoryear{{Sofue}}{{Sofue}}{2015}]{Sofue_2015}
{Sofue} Y.,  2015, \mn@doi [PASJ] {10.1093/pasj/psv042}, \href
  {https://ui.adsabs.harvard.edu/abs/2015PASJ...67...75S} {67, 75}

\bibitem[\protect\citeauthoryear{{Sohn}, {Patel}, {Fardal}, {Besla},
  {\VAN{Van}{Van}{van}}~der Marel, {Geha}  \& {Guhathakurta}}{{Sohn}
  et~al.}{2020}]{Sohn_2020}
{Sohn} S.~T.,  {Patel} E.,  {Fardal} M.~A.,  {Besla} G.,
  {\VAN{Van}{Van}{van}}~der Marel R.~P.,  {Geha} M.,   {Guhathakurta} P.,
  2020, \mn@doi [ApJ] {10.3847/1538-4357/abaf49}, \href
  {https://ui.adsabs.harvard.edu/abs/2020arXiv200806055S} {901, 43}

\bibitem[\protect\citeauthoryear{{Stanway} \& {Eldridge}}{{Stanway} \&
  {Eldridge}}{2018}]{Stanway_2018}
{Stanway} E.~R.,  {Eldridge} J.~J.,  2018, \mn@doi [MNRAS]
  {10.1093/mnras/sty1353}, \href
  {https://ui.adsabs.harvard.edu/abs/2018MNRAS.479...75S} {479, 75}

\bibitem[\protect\citeauthoryear{{Stott} et~al.,}{{Stott}
  et~al.}{2016}]{Stott_2016}
{Stott} J.~P.,  et~al., 2016, \mn@doi [MNRAS] {10.1093/mnras/stw129}, \href
  {https://ui.adsabs.harvard.edu/abs/2016MNRAS.457.1888S} {457, 1888}

\bibitem[\protect\citeauthoryear{{Tepper-Garc{\'\i}a}, {Bland-Hawthorn}  \&
  {Li}}{{Tepper-Garc{\'\i}a} et~al.}{2020}]{Thor_2020}
{Tepper-Garc{\'\i}a} T.,  {Bland-Hawthorn} J.,   {Li} D.,  2020, \mn@doi
  [MNRAS] {10.1093/mnras/staa317}, \href
  {https://ui.adsabs.harvard.edu/abs/2020MNRAS.493.5636T} {493, 5636}

\bibitem[\protect\citeauthoryear{{Teyssier}}{{Teyssier}}{2002}]{Teyssier_2002}
{Teyssier} R.,  2002, \mn@doi [A\&A] {10.1051/0004-6361:20011817}, \href
  {http://adsabs.harvard.edu/abs/2002A\%26A...385..337T} {385, 337}

\bibitem[\protect\citeauthoryear{{Teyssier}, {Chapon}  \&
  {Bournaud}}{{Teyssier} et~al.}{2010}]{Teyssier_2010}
{Teyssier} R.,  {Chapon} D.,   {Bournaud} F.,  2010, \mn@doi [ApJL]
  {10.1088/2041-8205/720/2/L149}, \href
  {https://ui.adsabs.harvard.edu/abs/2010ApJ...720L.149T} {720, L149}

\bibitem[\protect\citeauthoryear{{Thomas}, {Famaey}, {Ibata}, {L{\"u}ghausen}
  \& {Kroupa}}{{Thomas} et~al.}{2017}]{Thomas_2017}
{Thomas} G.~F.,  {Famaey} B.,  {Ibata} R.,  {L{\"u}ghausen} F.,   {Kroupa} P.,
  2017, \mn@doi [A\&A] {10.1051/0004-6361/201730531}, \href
  {https://ui.adsabs.harvard.edu/abs/2017A\&A...603A..65T} {603, A65}

\bibitem[\protect\citeauthoryear{{Tiret} \& {Combes}}{{Tiret} \&
  {Combes}}{2007}]{Tiret_2007}
{Tiret} O.,  {Combes} F.,  2007, \mn@doi [A\&A] {10.1051/0004-6361:20066446},
  \href {https://ui.adsabs.harvard.edu/abs/2007A\%26A...464..517T} {464, 517}

\bibitem[\protect\citeauthoryear{{Tiret} \& {Combes}}{{Tiret} \&
  {Combes}}{2008}]{Tiret_2008_gas}
{Tiret} O.,  {Combes} F.,  2008, \mn@doi [A\&A] {10.1051/0004-6361:200809357},
  \href {https://ui.adsabs.harvard.edu/abs/2008A\%26A...483..719T} {483, 719}

\bibitem[\protect\citeauthoryear{{Toomre}}{{Toomre}}{1964}]{Toomre_1964}
{Toomre} A.,  1964, \mn@doi [ApJ] {10.1086/147861}, \href
  {http://adsabs.harvard.edu/abs/1964ApJ...139.1217T} {139, 1217}

\bibitem[\protect\citeauthoryear{{Tully}, {Libeskind}, {Karachentsev},
  {Karachentseva}, {Rizzi}  \& {Shaya}}{{Tully}
  et~al.}{2015}]{Tully_2015_Cen_A}
{Tully} R.~B.,  {Libeskind} N.~I.,  {Karachentsev} I.~D.,  {Karachentseva}
  V.~E.,  {Rizzi} L.,   {Shaya} E.~J.,  2015, \mn@doi [ApJL]
  {10.1088/2041-8205/802/2/L25}, \href
  {https://ui.adsabs.harvard.edu/abs/2015ApJ...802L..25T} {802, L25}

\bibitem[\protect\citeauthoryear{{\VAN{Van}{Van}{van}}~der Marel \&
  {Kallivayalil}}{{\VAN{Van}{Van}{van}}~der Marel \&
  {Kallivayalil}}{2014}]{Van_der_Marel_2014}
{\VAN{Van}{Van}{van}}~der Marel R.~P.,  {Kallivayalil} N.,  2014, \mn@doi [ApJ]
  {10.1088/0004-637X/781/2/121}, \href
  {http://adsabs.harvard.edu/abs/2014ApJ...781..121V} {781, 121}

\bibitem[\protect\citeauthoryear{{\VAN{Van}{Van}{van}}~der Marel, {Fardal},
  {Besla}, {Beaton}, {Sohn}, {Anderson}, {Brown}  \&
  {Guhathakurta}}{{\VAN{Van}{Van}{van}}~der Marel
  et~al.}{2012}]{Van_der_Marel_2012}
{\VAN{Van}{Van}{van}}~der Marel R.~P.,  {Fardal} M.,  {Besla} G.,  {Beaton}
  R.~L.,  {Sohn} S.~T.,  {Anderson} J.,  {Brown} T.,   {Guhathakurta} P.,
  2012, \mn@doi [ApJ] {10.1088/0004-637X/753/1/8}, \href
  {https://ui.adsabs.harvard.edu/abs/2012ApJ...753....8V} {753, 8}

\bibitem[\protect\citeauthoryear{{\VAN{Van}{Van}{van}}~der Marel, {Fardal},
  {Sohn}, {Patel}, {Besla}, {del Pino}, {Sahlmann}  \&
  {Watkins}}{{\VAN{Van}{Van}{van}}~der Marel et~al.}{2019}]{Van_der_Marel_2019}
{\VAN{Van}{Van}{van}}~der Marel R.~P.,  {Fardal} M.~A.,  {Sohn} S.~T.,  {Patel}
  E.,  {Besla} G.,  {del Pino} A.,  {Sahlmann} J.,   {Watkins} L.~L.,  2019,
  \mn@doi [ApJ] {10.3847/1538-4357/ab001b}, \href
  {https://ui.adsabs.harvard.edu/abs/2019ApJ...872...24V} {872, 24}

\bibitem[\protect\citeauthoryear{{Vasiliev}}{{Vasiliev}}{2018}]{Vasiliev_2018}
{Vasiliev} E.,  2018, \mn@doi [MNRAS] {10.1093/mnrasl/sly168}, \href
  {https://ui.adsabs.harvard.edu/abs/2018MNRAS.481L.100V} {481, L100}

\bibitem[\protect\citeauthoryear{{Wang}, {Hammer}, {Rejkuba}, {Crnojevi{\'c}}
  \& {Yang}}{{Wang} et~al.}{2020}]{Wang_2020}
{Wang} J.,  {Hammer} F.,  {Rejkuba} M.,  {Crnojevi{\'c}} D.,   {Yang} Y.,
  2020, \mn@doi [MNRAS] {10.1093/mnras/staa2508}, \href
  {https://ui.adsabs.harvard.edu/abs/2020MNRAS.498.2766W} {498, 2766}

\bibitem[\protect\citeauthoryear{{Weisz}, {Dolphin}, {Skillman}, {Holtzman},
  {Gilbert}, {Dalcanton}  \& {Williams}}{{Weisz} et~al.}{2014}]{Weisz_2014}
{Weisz} D.~R.,  {Dolphin} A.~E.,  {Skillman} E.~D.,  {Holtzman} J.,  {Gilbert}
  K.~M.,  {Dalcanton} J.~J.,   {Williams} B.~F.,  2014, \mn@doi [ApJ]
  {10.1088/0004-637X/789/2/147}, \href
  {http://adsabs.harvard.edu/abs/2014ApJ...789..147W} {789, 147}

\bibitem[\protect\citeauthoryear{{Wetzstein}, {Naab}  \& {Burkert}}{{Wetzstein}
  et~al.}{2007}]{Wetzstein_2007}
{Wetzstein} M.,  {Naab} T.,   {Burkert} A.,  2007, \mn@doi [MNRAS]
  {10.1111/j.1365-2966.2006.11360.x}, \href
  {http://adsabs.harvard.edu/abs/2007MNRAS.375..805W} {375, 805}

\bibitem[\protect\citeauthoryear{{Widmark}, {Laporte}, {de Salas}  \&
  {Monari}}{{Widmark} et~al.}{2021}]{Widmark_2021}
{Widmark} A.,  {Laporte} C.~F.~P.,  {de Salas} P.~F.,   {Monari} G.,  2021,
  \mn@doi [A\&A] {10.1051/0004-6361/202141466}, \href
  {https://ui.adsabs.harvard.edu/abs/2021A\&A...653A..86W} {653, A86}

\bibitem[\protect\citeauthoryear{{Wittenburg}, {Kroupa}  \&
  {Famaey}}{{Wittenburg} et~al.}{2020}]{Wittenburg_2020}
{Wittenburg} N.,  {Kroupa} P.,   {Famaey} B.,  2020, \mn@doi [ApJ]
  {10.3847/1538-4357/ab6d73}, \href
  {https://ui.adsabs.harvard.edu/abs/2020ApJ...890..173W} {890, 173}

\bibitem[\protect\citeauthoryear{{Yang}, {Hammer}, {Jiao}  \&
  {Pawlowski}}{{Yang} et~al.}{2022}]{Yang_2022}
{Yang} Y.,  {Hammer} F.,  {Jiao} Y.,   {Pawlowski} M.,  2022, \mn@doi [MNRAS]
  {10.1093/mnras/stac644}, \href
  {https://ui.adsabs.harvard.edu/abs/2022MNRAS.512.4171Y} {512, 4171}

\bibitem[\protect\citeauthoryear{{Yu} \& {Liu}}{{Yu} \& {Liu}}{2018}]{Yu_2018}
{Yu} J.,  {Liu} C.,  2018, \mn@doi [MNRAS] {10.1093/mnras/stx3204}, \href
  {http://adsabs.harvard.edu/abs/2018MNRAS.475.1093Y} {475, 1093}

\bibitem[\protect\citeauthoryear{{Zhao}, {Li}  \& {Bienaym{\'e}}}{{Zhao}
  et~al.}{2010}]{Zhao_2010}
{Zhao} H.,  {Li} B.,   {Bienaym{\'e}} O.,  2010, \mn@doi [Physical Review D]
  {10.1103/PhysRevD.82.103001}, \href
  {http://adsabs.harvard.edu/abs/2010PhRvD..82j3001Z} {82, 103001}

\bibitem[\protect\citeauthoryear{{Zhao}, {Famaey}, {L{\"u}ghausen}  \&
  {Kroupa}}{{Zhao} et~al.}{2013}]{Zhao_2013}
{Zhao} H.,  {Famaey} B.,  {L{\"u}ghausen} F.,   {Kroupa} P.,  2013, \mn@doi
  [A\&A] {10.1051/0004-6361/201321879}, \href
  {http://adsabs.harvard.edu/abs/2013A\%26A...557L...3Z} {557, L3}

\makeatother
\end{thebibliography}

\begin{appendix}

\section{Setting up the particle masses in the simulation}
\label{Particle_mass_setup}

An important new aspect of the present \textsc{por} simulations is that we need to consider the outer parts of the simulated discs in much greater detail, since we expect these regions to be the original source material for the SPs \citepalias{Banik_Ryan_2018}. Material at an initial galactocentric distance of ${\approx 50}$~kpc would be very poorly resolved with a feasible number of equal mass particles, so we need to vary the particle masses in our disc templates. This differs from previous \textsc{por} simulations \citep[e.g.][]{Banik_2020_M33}. We therefore adjust the \textsc{dice} algorithm as described below.

To maintain an exponential surface density profile for each component, we require that the particle mass $m$ at some location is
\begin{eqnarray}
	m \left( x \right) ~\propto~ \frac{x \exp \left( -x \right)}{P \left( x \right)} \, ,
	\label{Disc_particle_mass}
\end{eqnarray}
where $P \, dx$ is the probability of a randomly chosen particle having some particular $x \pm dx/2$, with $x$ being the galactocentric radius in units of the exponential scale length for the considered component. For different regions to be resolved similarly well despite the exponential profile, $P$ should be roughly constant. $P$ can theoretically have any non-zero value at any $x$ in the range we wish to model, because this can always be compensated by adjusting $m$ so as to satisfy Equation \ref{Disc_particle_mass} and maintain the surface density profile. We found that setting $P$ to a constant everywhere leads to disastrous numerical effects at low $x$ because here we would need $m \to 0$. Since the central regions of a disc are most unstable to self-gravity, it is essential to minimize artificial relaxation effects here. This can be done if all particles have a similar mass in the central few scale lengths. Our compromise solution is
\begin{eqnarray}
	P ~\propto~ \tanh \left( \alpha x \right) \, .
	\label{alpha_IB_trick}
\end{eqnarray}
Setting the parameter $\alpha = 3$ keeps $m$ nearly constant over as large a range of $x$ as possible.

We assign each particle a scaled radius $x$ using a Monte Carlo technique to realize the distribution
\begin{eqnarray}
	P \left( x \right) ~=~ \frac{\alpha \tanh \left( \alpha x \right)}{ \ln \cosh \left( \alpha x_{\text{max}} \right)} \, ,
\end{eqnarray}
where $x_{\text{max}} \approx 25$ is the maximum normalized extent of the considered component. The azimuthal angle is drawn randomly from the range $\left( 0 - 2\mathrm{\pi} \right)$, while vertically we use a $\sech^2$ profile. Both the radial and the vertical probability distributions can be integrated analytically, simplifying the code. Particle masses are then assigned using Equation \ref{Disc_particle_mass}. To ensure the correct normalization, we multiply the masses of all particles in the considered component by a common factor. These adjustments to our modified \textsc{dice} are enabled by setting the flag $Equal\_mass\_particles$ to 0, while using the default value of 1 causes the particles to have an equal mass within each component \citep[as done in][]{Banik_2020_M33}.

In this way, we are able to adequately resolve the outer parts of the discs while minimizing numerical relaxation effects in their central regions. Relaxation is much less important in the outskirts, where the very weak self-gravity means that any particle is essentially just a test particle orbiting the central region. This is probably why radial variations in $m$ in the outskirts of each disc did not cause numerical difficulties.

\section{Initial gas disc thickness}
\label{Initial_hg}

\begin{figure}
	\centering
	\includegraphics[width = 8.5cm] {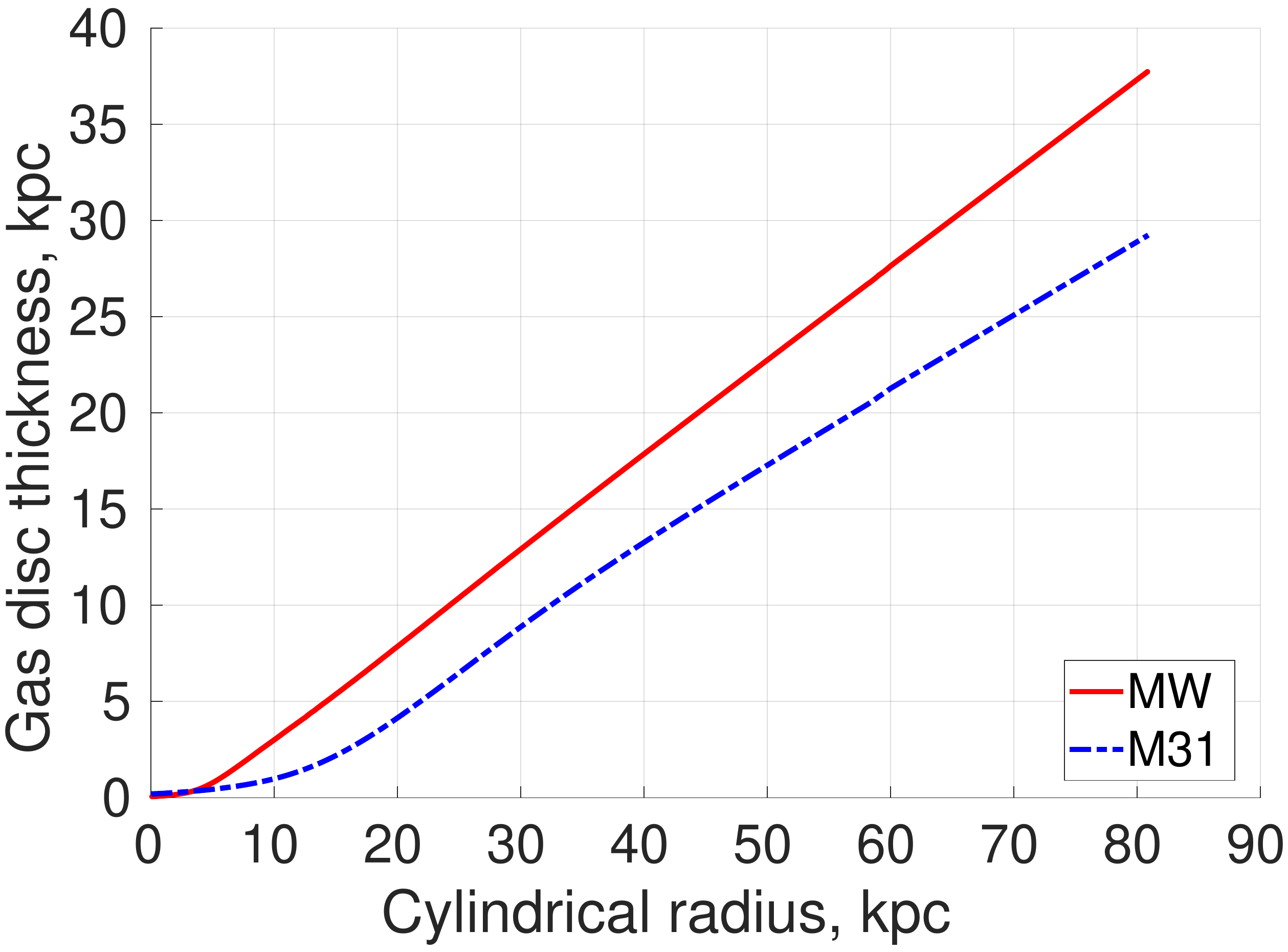}
	\caption{Initial $\sech^2$ thickness of the MW (red) and M31 (blue) gas discs as a function of cylindrical polar radius, as found by our adapted \textsc{dice} algorithm (Section \ref{Disc_templates}).}
	\label{Initial_gas_thickness}
\end{figure}

At each radius within the MW or M31, the vertical profile of the gas disc follows a $\sech^2$ law with characteristic thickness as shown in Figure \ref{Initial_gas_thickness}. The thickness profile is similar to that used by \citet{Banik_2020_M33} in their 100 kK model of M33 (see their figure 3).

\section{Extracting the MW-M31 trajectory from the simulation}
\label{Barycentre_extraction_on_the_fly}

We avoid the need to save simulation snapshots except that of interest by writing each galaxy's centre of mass position and velocity to a text file at every coarse time-step of the \textsc{por} simulation. For concreteness, we focus on how this is calculated for the MW $-$ the same procedure is used for M31. While only a limited analysis is possible `on the fly', this is sufficient to gain a reasonably good understanding of the MW-M31 trajectory as a basis for a more detailed analysis.

We exploit the fact that each particle has a unique identifier, allowing us to find the barycentre of all particles that were initially part of the MW. We then find their root mean square (rms) distance from the MW barycentre in both position and velocity, which we denote $r_{\text{rms}}$ and $v_{\text{rms}}$, respectively. The ratio between the two is used to define a characteristic time-scale
\begin{eqnarray}
	t_{\text{rms}} ~\equiv~ \frac{r_{\text{rms}}}{v_{\text{rms}}} \, .
	\label{t_rms}
\end{eqnarray}
We then iteratively repeat the calculation of all these quantities, each time considering only particles whose 6D position is within some threshold distance $r_{\text{max}}$ from the barycentre identified at the previous step. Differences in velocity are converted to equivalent distances using $t_{\text{rms}}$.

After the first iteration, we consider all particles when finding the MW barycentre, since this could in principle be affected by material that has been accreted from M31. We set $r_{\text{max}}$ to 14.1\% of the MW-M31 separation as calculated on the first iteration, i.e. by considering only particles that were initially part of each galaxy. For safety, we also restrict $r_{\text{max}}$ to the range $10-70$~kpc. The convergence condition is that the 6D position of the MW barycentre should move by $<0.01 \, r_{\text{rms}}$ between successive iterations, with the above-mentioned conversion between velocities and distances. We do not consider the gas as this would be extremely difficult to analyse while the simulation is running, but nonetheless obtain very accurate results using particles alone. The gas is considered when analysing the desired simulation snapshot in more detail.

\section{Tidal debris orbital pole distributions}
\label{Appendix_zmax}

\begin{figure*}
	\centering
	\includegraphics[width = 8.5cm] {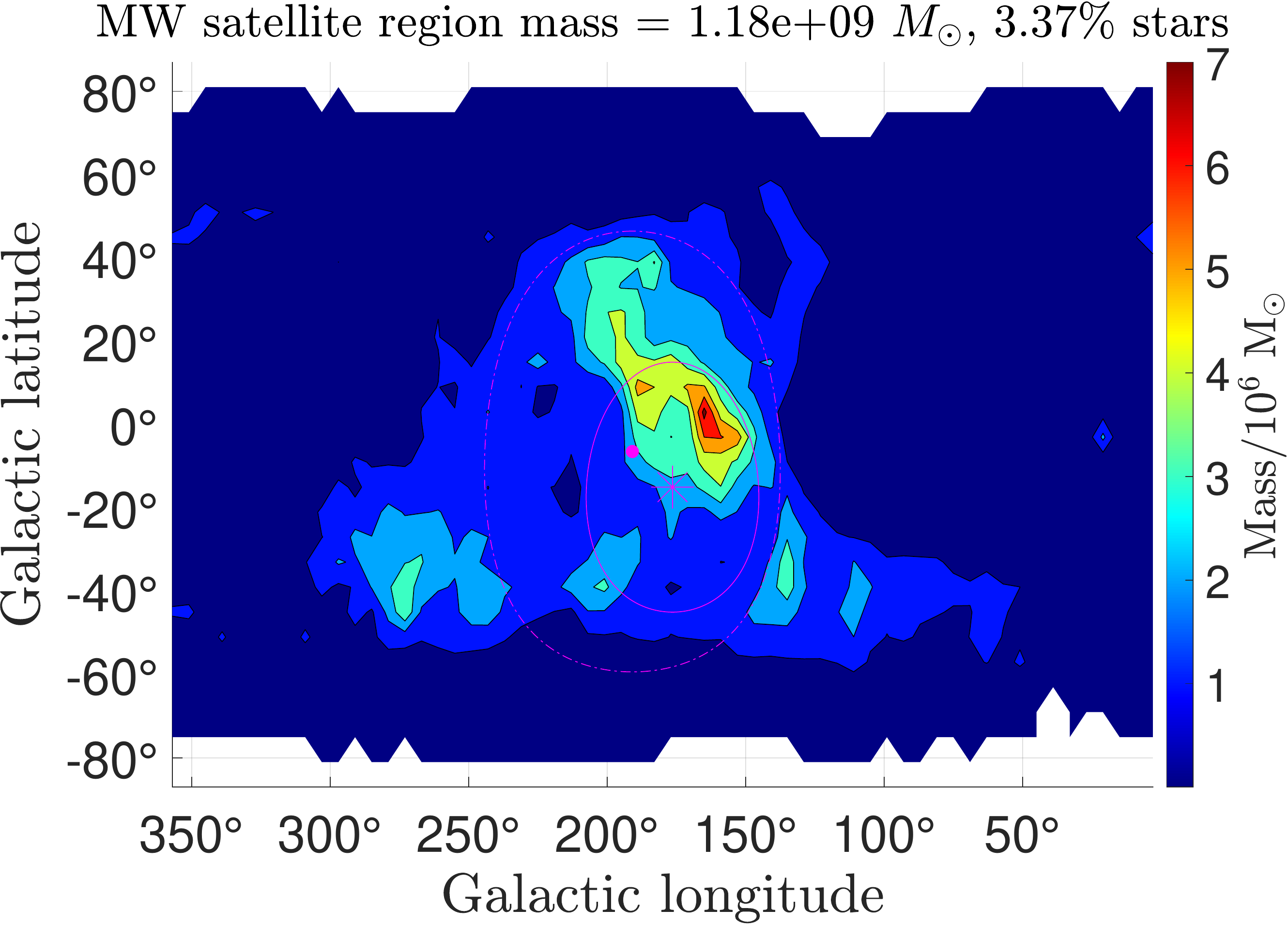}
	\includegraphics[width = 8.5cm] {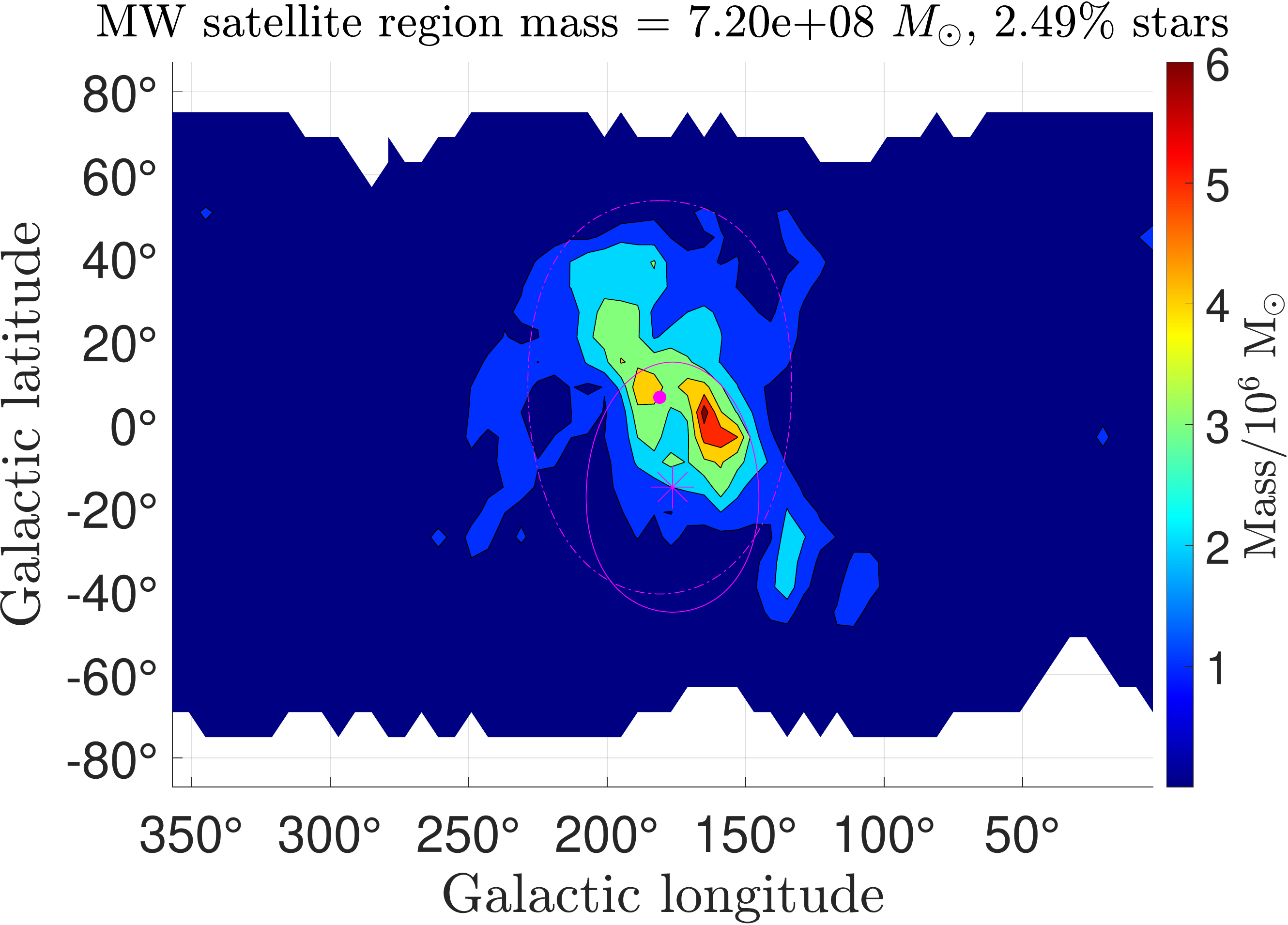}
	\hspace{0.3cm}
	\includegraphics[width = 8.5cm] {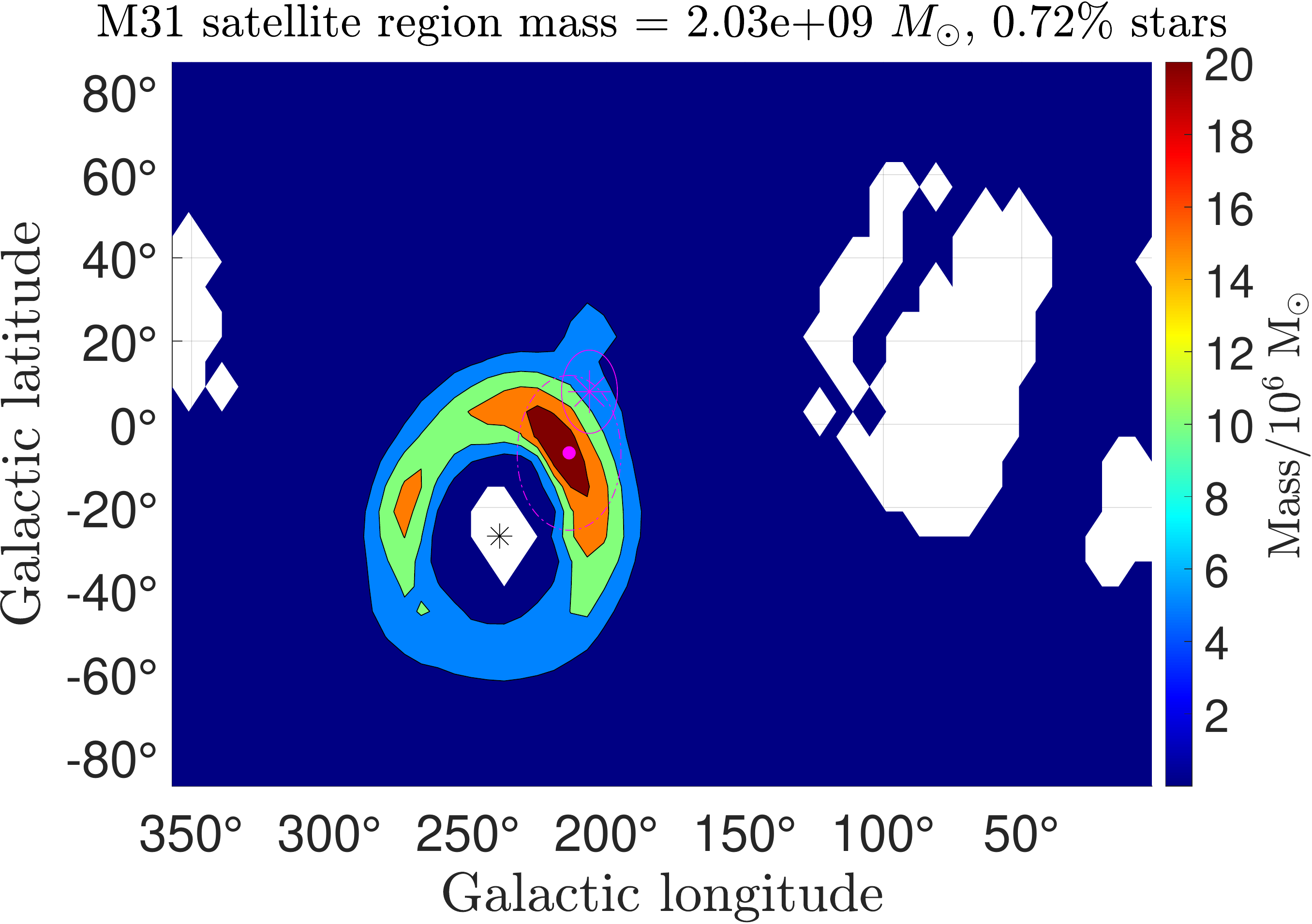}
	\includegraphics[width = 8.5cm] {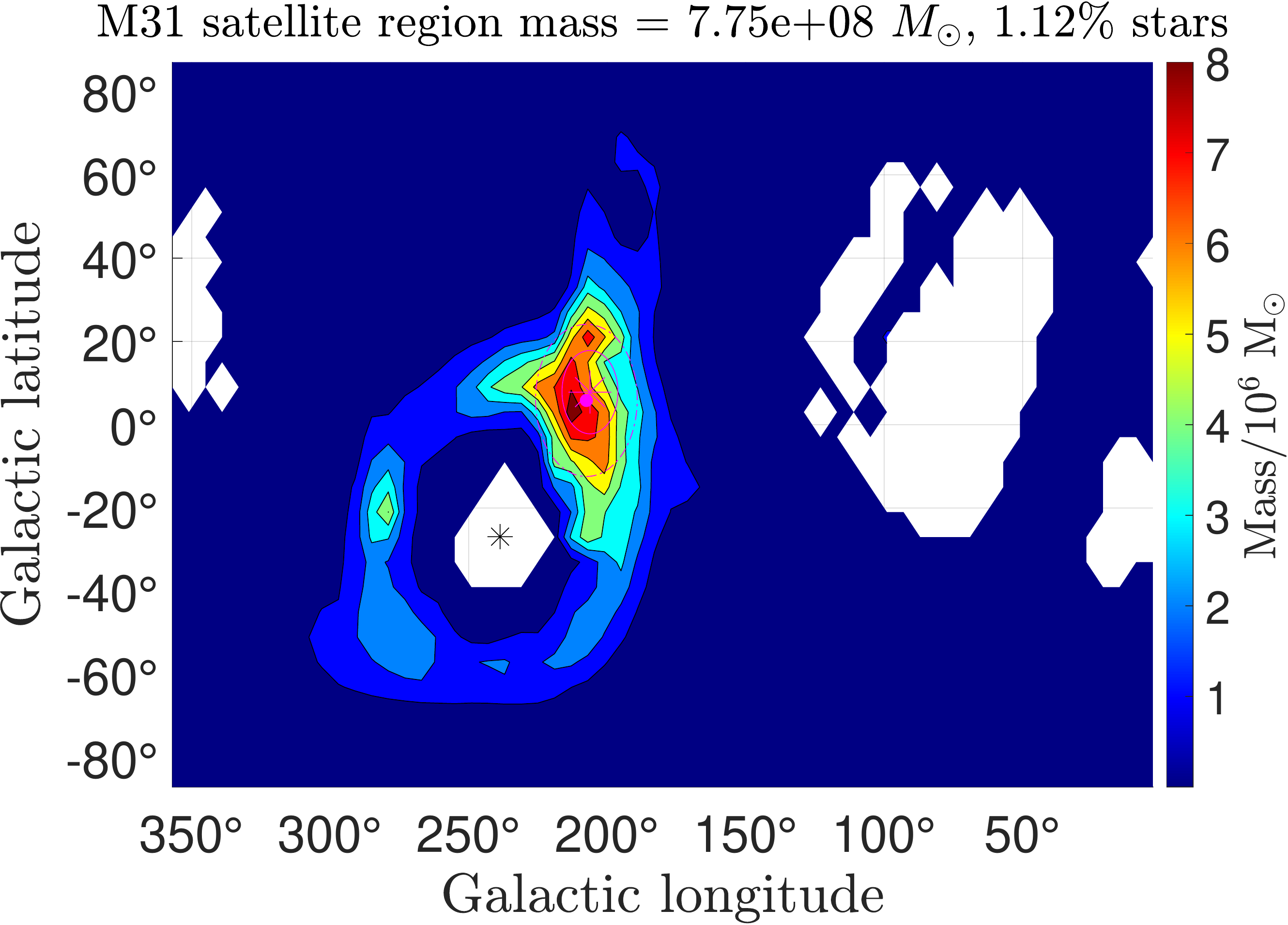}
	\caption{The orbital pole distribution of material in the satellite region of the MW (\emph{top}) and M31 (\emph{bottom}), shown similarly to Figures~\ref{MW_SP_50} and \ref{M31_SP_50} but with an excluded region up to $z_{\text{max}} = 40$ kpc (\emph{left}) or $z_{\text{max}} = 60$ kpc (\emph{right}) to remove the disc remnant. The exact choice of $z_{\text{max}}$ has little effect on our results.}
	\label{SP_40_60}
\end{figure*}

%\begin{figure}
%	\centering
%	\includegraphics[width = 8.5cm] {MW_SP_40}
%	\includegraphics[width = 8.5cm] {MW_SP_60}
%	\caption{The orbital pole distribution of material in the MW satellite region, shown similarly to Figure \ref{MW_SP_50} but with an excluded region up to $z_{\text{max}} = 40$ kpc (\emph{top}) or $z_{\text{max}} = 60$ kpc (\emph{bottom}) to remove the MW disc remnant. The exact choice of $z_{\text{max}}$ has little effect on our results.}
%	\label{MW_SP_40_60}
%\end{figure}
%
%\begin{figure}
%	\centering
%	\includegraphics[width = 8.5cm] {M31_SP_40}
%	\includegraphics[width = 8.5cm] {M31_SP_60}
%	\caption{The orbital pole distribution of material in the M31 satellite region, shown similarly to Figure \ref{M31_SP_50} but with an excluded region up to $z_{\text{max}} = 40$ kpc (\emph{top}) or 60 kpc (\emph{bottom}) to remove the M31 disc remnant. The exact choice of $z_{\text{max}}$ has little effect on our results.}
%	\label{M31_SP_40_60}
%\end{figure}

In Section \ref{Debri_h_hat_distribution}, we presented the orbital pole distributions of material in the MW and M31 satellite regions. This required us to exclude the disc remnant, which we defined as all material within 250 kpc of the respective galaxy's barycentre and up to $z_{\text{max}} = 50$ kpc from its disc plane. To assess the impact of varying the adopted $z_{\text{max}}$, we use Figure \ref{SP_40_60} to show the MW satellite region orbital pole distribution if instead $z_{\text{max}} = 40$ kpc (top left panel) or 60 kpc (top right panel). The bottom panels of Figure \ref{SP_40_60} show the corresponding results for M31. It is clear that for both galaxies, the preferred SP orbital pole direction and the orbital pole dispersion are little affected by the choice of $z_{\text{max}}$.

\section{A higher resolution simulation}
\label{Higher_resolution_simulation}

\begin{figure}
	\centering
	\includegraphics[width = 8.5cm] {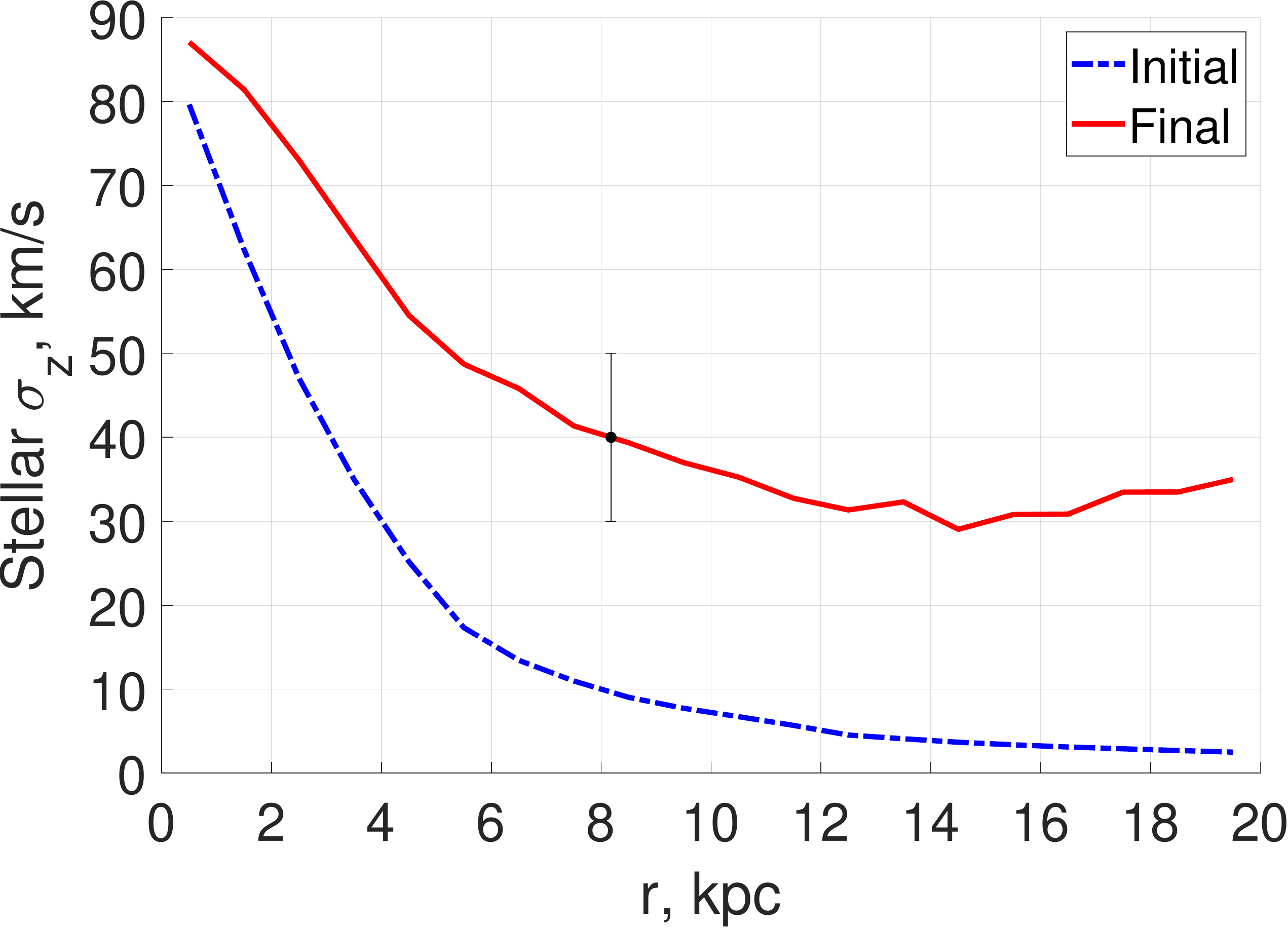}
	\caption{Evolution of the Galactic $\sigma_z$ profile, shown similarly to the bottom right panel of Figure \ref{MW_disc_figures} but for a higher resolution simulation in which we allow up to 13 levels of refinement rather than 12. The result now matches the data point at $40 \pm 10$~km/s for the Solar circle \citep{Yu_2018}.}
	\label{MW_sigma_z_profile_evolution_465kK_lx13}
\end{figure}

To check the numerical consistency of our results, we rerun the \textsc{por} simulation with the maximum number of refinement levels raised from 12 to 13 (Section \ref{PoR_settings}). This reduces the size of the smallest gas cell from 1.5~kpc to 0.75~kpc. The results remain very similar, except for a small difference to the $\sigma_z$ profile of the MW. We therefore show this in Figure \ref{MW_sigma_z_profile_evolution_465kK_lx13}. The main difference is in the central region, presumably due to the improved resolution. $\sigma_z$ is now higher than its initial value at all radii. It also has a much steeper decline away from the disc centre. This makes the shape of the $\sigma_z$ profile more similar to that of M31 (Figure \ref{M31_disc_figures}), which changes little due to the higher resolution (not shown). However, $\sigma_z$ is still higher for M31 than for the MW at low radii. This is also true further out, where the improved resolution slightly reduces the Galactic $\sigma_z$, leading to better agreement with observations \citep{Yu_2018}. A dynamically hotter stellar disc for M31 compared to the MW therefore seems to be a robust feature of our model, which also accounts for the opposite situation with their SPs.

\end{appendix}

\bsp
\label{lastpage}
\end{document}